\PassOptionsToPackage{table}{xcolor} 
\documentclass{article}

\usepackage[accepted]{icml2023}
\usepackage[utf8]{inputenc} %
\usepackage[T1]{fontenc}    %
\usepackage{microtype}
\usepackage{booktabs} %
\usepackage{wrapfig}
\usepackage{amsmath}
\usepackage{amssymb}
\usepackage{mathtools}
\usepackage{amsthm}
\usepackage{bm}
\usepackage{booktabs}       %
\usepackage{amsfonts}       %
\usepackage{nicefrac}       %
\usepackage{microtype}      %
\usepackage{tabularx}
\usepackage{color}
\usepackage{titlesec}
\usepackage{graphicx}
\usepackage{subfigure}
\usepackage[subfigure]{tocloft}
\usepackage{booktabs} %
\usepackage{siunitx}
\usepackage{wrapfig}
\usepackage[para,online,flushleft]{threeparttable}
\usepackage{tikz}
    \usetikzlibrary{fadings}
    \usetikzlibrary{patterns}
    \usetikzlibrary{shadows.blur}
    \usetikzlibrary{shapes}
\usepackage{pdflscape}
\usepackage{tabularx}
\usepackage[toc,page]{appendix}
\usepackage[withpage, smaller, nohyperlinks]{acronym} 
\usepackage{enumitem,amssymb}
\usepackage{pifont}
\usepackage[para,online,flushleft]{threeparttable}
\usepackage[final]{pdfpages}
\usepackage{hyperref}
\usepackage[capitalize,noabbrev]{cleveref}
\usepackage{threeparttable}

\definecolor{jkublue}{RGB}{0, 100, 138}

\hypersetup{
   colorlinks=true,
   linkcolor=jkublue,
   citecolor=jkublue,
   urlcolor=black
}

\newcommand{\stoptocwriting}{ 
  \addtocontents{toc}{\protect\setcounter{tocdepth}{-5}}}
\newcommand{\resumetocwriting}{ 
  \addtocontents{toc}{\protect\setcounter{tocdepth}{\arabic{tocdepth}}}}

\usepackage{color}
\usepackage{amsmath}
\usepackage{bm}

\newcommand\Ba{\bm{a}}
\newcommand\Bb{\bm{b}}

\newcommand\Bf{\bm{f}}
\newcommand\Bg{\bm{g}}
\newcommand\Bh{\bm{h}}

\newcommand\Bm{\bm{m}}

\newcommand\Bv{\bm{v}}
\newcommand\Bw{\bm{w}}

\newcommand\By{\bm{y}}

\newcommand\BM{\bm{M}}

\newcommand\BW{\bm{W}}

\definecolor{lightblue}{RGB}{32, 194, 217}
\definecolor{jku_red}{RGB}{217, 92, 76}
\definecolor{jku_blue}{RGB}{0, 132, 187}
\definecolor{jku_green}{RGB}{91, 167, 85} %
\definecolor{jku_yellow}{RGB}{241, 188, 63}
\definecolor{jku_cyan}{RGB}{79,176,191}
\definecolor{jku_grey}{RGB}{125,130,140}
\definecolor{jku_lightgreen}{RGB}{191,206,82}
\definecolor{jku_violett}{RGB}{174,97,157}

\theoremstyle{plain}

\theoremstyle{definition}

\theoremstyle{remark}

\icmltitlerunning{Enhancing Activity Prediction Models in Drug Discovery with the 
Ability to Understand Human Language}

\begin{document}

\twocolumn[
\icmltitle{Enhancing Activity Prediction Models in Drug Discovery with the 
Ability to Understand Human Language}

\icmlsetsymbol{equal}{*}
\begin{icmlauthorlist}
\icmlauthor{Philipp Seidl}{jku}
\icmlauthor{Andreu Vall}{jku}
\icmlauthor{Sepp Hochreiter}{jku,iarai}
\icmlauthor{Günter Klambauer}{jku}
\end{icmlauthorlist}

\icmlaffiliation{jku}{Insitute for Machine Learning, Johannes Kepler University, Linz, Austria}
\icmlaffiliation{iarai}{IARAI, Vienna, Austria}

\icmlcorrespondingauthor{Günter Klambauer}{klambauer@ml.jku.at}

\icmlkeywords{Machine Learning, ICML}

\vskip 0.3in
]

\printAffiliationsAndNotice{} %

\stoptocwriting
\begin{abstract} %
Activity and property prediction models are the central workhorses in drug discovery and materials sciences, but currently they 
have to be trained or fine-tuned for new tasks. 
Without training or fine-tuning, 
scientific language models could be used for such low-data tasks
through their announced zero- and few-shot capabilities. 
However, their predictive quality at 
activity prediction is lacking. 
In this work, we envision a novel type of 
activity prediction model that is able to adapt to new 
prediction tasks at inference time, via 
understanding textual information describing the task.
To this end, we propose a new architecture with separate modules for 
chemical and natural language inputs, and a contrastive pre-training 
objective on data from large biochemical databases. 
In extensive experiments, we show that our method CLAMP yields improved predictive performance on few-shot learning benchmarks 
and zero-shot problems in drug discovery. 
We attribute the advances of our method to the 
modularized architecture and 
to our pre-training objective.
\end{abstract}

\begin{figure}[ht]
    \includegraphics[width=0.85\columnwidth]{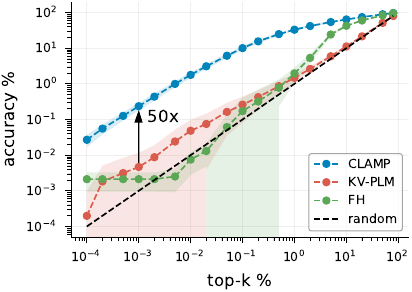}
    \vspace*{-3mm}
    \caption{Performance at retrieving active molecules from a chemical database of 1M molecules in a zero-shot setting. Our method CLAMP enriches active molecules for unseen assays by 50x over the next best method KV-PLM \citep{zeng2022kvplm}, the best-performing scientific language model.  Shaded areas represent standard deviation across 2,543 prediction tasks.\label{fig:compound_retrieval}}
\end{figure}

\begin{figure*}[ht]
    \centering
    \includegraphics[width=0.85\textwidth]{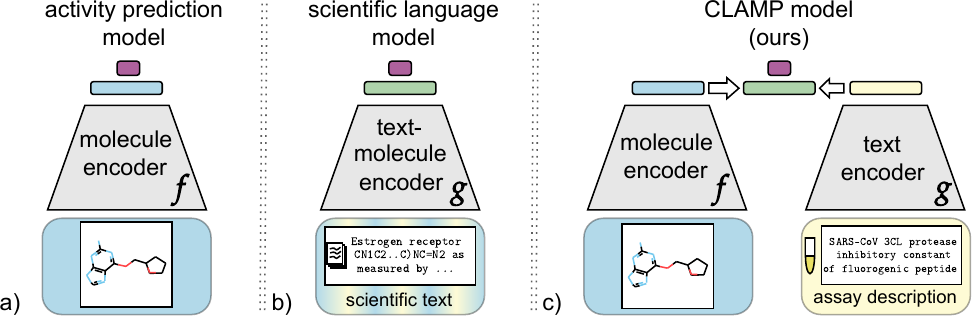}
    \caption{Comparison of approaches for activity prediction. \textbf{a)}
    Activity prediction models use the chemical structure as input and use 
    a molecule encoder to obtain embeddings. \textbf{b)} Scientific language 
    models (SLMs) are able to process biomedical texts that contain both human language
    and chemical structure. As an example, the displayed text 
    contains a string representation of a molecule \texttt{CN12..C)NC=N2}. 
    SLMs tokenize those strings and treat tokens representing chemical 
    structures in the same way as tokens for natural language.
    \textbf{c)} Our model, called CLAMP, uses separate encoders for 
    chemical and natural language data and embeds them into a joint 
    embedding space. This allows CLAMP to predict activities for new
    wet-lab procedures, i.e. bioassays, that are described in human language.
    \label{fig:approaches}}
\end{figure*}

\section{Introduction}
\label{sec:introduction}
\paragraph{Activity and property prediction models are the
main workhorses in computational drug discovery, and hence 
are roughly analogous to language models in natural language processing (NLP) 
and image classification models in computer vision (CV).}
The task to predict chemical, macroscopic properties or biological activity of 
a molecule based on its chemical structure is a decade-old, central
problem in natural sciences \citep{hansch1962correlation,hansch1969quantitative}. 
Machine learning methods have been regularly used to learn these relations between
the chemical structure and the properties based on measurement or simulated data
since at least the early 90s \citep{king1993new}. With the advent of 
Deep Learning (DL) in drug discovery \citep{lusci2013deep, dahl2014multi,unterthiner2014deep,chen2018rise, hochreiter2018machine}
many different \emph{molecule encoders} \citep{xu2019efficient,mayr2018large,gilmer2017neural} have been 
suggested that obtain embeddings from chemical structures which 
are used to predict activities and properties. 
Activity prediction models are used to select or rank molecules for further 
biological testing \citep{melville2009machine,unterthiner2014deep} or for
flagging or removing molecules with unwanted properties \citep{mayr2016deeptox}.
In connection with generative 
models for molecules \citep{segler2018generating,gomez2018automatic}, 
activity prediction models usually serve as a reward function when 
the molecule structure should be optimized toward a particular 
objective \citep{sanchez2018inverse,olivecrona2017molecular}. 
The combination of activity prediction and generative models has brought
a strong speed-up to early phases of drug discovery \citep{zhavoronkov2019deep}.
As a central tool for drug discovery, activity prediction models
are analogous to language models in NLP as well as 
to image classification models used in computer vision.

\paragraph{Molecule encoders extract relevant features 
from chemical structures and are trained on bioactivity data.} 
Activity prediction models based 
on DL use different low-level or initial descriptions of the
chemical structure, such as the molecular graph \citep{scarselli2008graph,merkwirth2005automatic,kipf2016semi,gilmer2017neural},
the string-representation SMILES \citep{weininger1988smiles,mayr2018large},
chemical fingerprints or descriptors \citep{dahl2014multi,unterthiner2014deep,mayr2016deeptox}, 
or a combination of those \citep{yang2019analyzing}. 
While there have been several 
successes with such DL architectures, such as graph neural
networks (GNNs) \citep{scarselli2008graph,gilmer2017neural}, 
their improvements 
are still disputed and have not been as ground-breaking as for 
vision and language \citep{jiang2021could,bender2021artificial,sun2022does}.
These activity prediction models are usually trained 
on pairs of molecules and activity labels 
from biological experiments, so-called 
\emph{biological assays} or \emph{bioassays}. 
Bioassays are often wet-lab procedures 
involving several chemical 
and biological processing steps, such
as growing cell lines and administering chemical agents. 
Because the labels for the training data points, 
called \emph{bioactivities}, are 
highly time- and cost-intensive to acquire, 
there has been a considerable interest 
in being able to efficiently train activity prediction
models on few data points. 
The recently suggested benchmarking dataset FS-Mol \citep{stanley2021fs},
provides as few as 16 labeled molecules for an 
activity prediction task, such that methods must be able to efficiently 
transfer knowledge from other tasks. Although there would also be substantial 
information about the activity prediction tasks available 
in the form of natural language \citep{kim2019pubchem}, 
the textual description of the biological experiment in the wet-lab,
current activity prediction models cannot use this information (Fig.~\ref{fig:approaches}a).
These models require measurement data from that activity prediction task 
or bioassay on which they are trained or fine-tuned. 
Therefore current activity prediction models cannot perform zero-shot
activity prediction \citep{larochelle2008zero,wu2022memorizing} 
and have limited predictive quality in few-shot settings \citep{stanley2021fs}
(Sec.~\ref{sec:relatedwork}). 

\paragraph{Scientific language models (SLMs) can utilize both natural language
and chemical structure but are suboptimal activity predictors.}
Large language models have demonstrated great zero- and few-shot 
capabilities \citep{brown2020language,wei2021finetuned} and 
they have brought a paradigm-shift for 
NLP \citep{sun2022paradigm}. 
Some of these large language models have been also trained on scientific
literature \citep{taylor2022galactica,beltagy2019scibert, singhal2022MedPALM, edwards2022translation}, 
and concretely on biomedical texts \citep{zeng2022kvplm}, 
which also contain limited amounts of chemical structures.
The SLMs Galactica \citep{taylor2022galactica} and KV-PLM \citep{zeng2022deep} 
tokenize the SMILES representations of chemical structures 
and embed those chemical tokens in the same embedding space as language tokens. 
Therefore, these SLMs can in principle be used to perform zero-shot activity 
prediction based on the textual description of 
the bioassay (Fig.~\ref{fig:approaches}b).
However, SLMs still under-perform at activity prediction 
\citep{taylor2022galactica, zeng2022kvplm} (Fig.~\ref{fig:compound_retrieval} 
and Sec.~\ref{sec:experiments}), 
which we attribute to two reasons: 
a) they are using a sub-optimal molecule encoder, and 
b) they are trained on overly limited training data. 
Concerning a), there has been substantial work by the scientific community 
on finding effective molecule encoders \citep{gilmer2017neural, wang2022evaluating, zhu2022unified2d3d, fang2022geometry, sun2022does, liu2021contrastive, abdel2022mfbert, benjamin2022graph, he2022masked, rong2020grover, chilingaryan2022bartsmiles, winter2019cddd, maziarka2020molecule, maziarka2021relative, huang2021therapeutics}
(Appendix~\ref{appsec:representation}).
In comparative studies, the encoder that is implicitly used by the SLMs, i.e., 
tokenization of SMILES strings with subsequent attention-layers,
does not appear as one of the best encoders \citep{xu2019efficient,mayr2018large,jiang2021could}. 
Concerning b), 
biomedical texts only contain few tens of thousands of chemical structures, 
while chemical databases contain hundreds of 
millions of chemical structures and bioactivities \citep{kim2019pubchem},
which are not used to train SLMs. 
In summary, we hypothesize
that choosing an effective molecule encoder and utilizing chemical 
databases as training or pre-training data could lead to improved 
activity prediction. 

\paragraph{We propose a modularized architecture with a separate molecule 
and language encoder and a contrastive learning objective.}
In order to i) use an effective molecule encoder, 
ii) be able to pre-train on data from chemical databases, and
iii) to enhance activity prediction models with the ability to utilize 
human language, we propose an architecture with two separate modules.
The first module is a molecule encoder and 
the second module is a text encoder, 
that are contrastively pre-trained across these
two data modalities (Fig.~\ref{fig:approaches}\textbf{c}). 
Cross-modal contrastive learning \citep{zhang2020contrastive} and especially Contrastive 
Language-Image Pre-training (CLIP) \citep{radford2021learning} 
has strongly impacted several areas of computer vision and NLP. 
CLIP has brought a tremendous improvement for example for generative models 
\citep{ramesh2022hierarchical}, 
retrieval systems \citep{borgeaud2022retro},
and zero- and few-shot prediction \citep{radford2021learning}.
One aspect of these successes is that CLIP is modularized: 
it uses both an effective language encoder \citep{vaswani2017attention, devlin2019bert, brown2020language} and 
an effective vision encoder \citep{he2016deep}. The learning objective of
CLIP enables the interaction of these two encoders and a common 
embedding space of images and language. Furthermore, the
success of CLIP rests on the availability of large datasets
of pairs of images and text captions \citep{schuhmann2022laion}.
Both of these aspects, 
the interaction with predictive or generative models 
through natural language and the availability of a large dataset
of pairs of modalities, could also be beneficial for
machine learning systems in drug discovery. At inference time, such 
a system would be able to acquire new knowledge about a prediction task
by accessing the textual description of the bioassay procedure, 
and thus to predict the activity of molecules without adjusting weights 
or re-training. This ability could be considered as understanding
the bioassay procedure described by human language.
The possibility to pre-train this architecture on large
chemical databases that contain hundreds of millions of chemical structures
paired with textual descriptions of the bioassays, 
offers an opportunity to train encoders  
that provide rich representations \citep{radford2021learning}.

\paragraph{Our proposed approach unlocks large chemical databases for 
pre-training.}
SLMs are pre-trained on datasets such as 
ChEBI-20 \citep{edwards2021text2mol} and 
ChEBI-22 \citep{liu2022multi}, which
comprise only few tens of thousands of molecules.
Galactica \citep{taylor2022galactica} is additionally 
trained on the chemical structure 
of 2M and Grover \citep{rong2020grover} on 10M  
molecules, however, without associated biological information.
In contrast to these pre-training datasets, 
chemical databases, such as PubChem \citep{kim2019pubchem}
and ChEMBL \citep{gaulton2012chembl} contain orders
of magnitude more molecules with associated 
biological information than biomedical texts. 
The chemical database PubChem contains 114M 
chemical structures \citep{kim2019pubchem} and $\sim$300M
bioactivity measurements. Additionally, these chemical databases
contain textual descriptions of the bioassays that 
were used to determine the bioactivity of those molecules
\citep{kim2019pubchem}. A bioactivity datapoint
represents a numeric or binary outcome of bioassay measurement
of a molecule, and hence a label for a molecule-text pair. 
We hypothesize that the chemical databases comprise information that can 
be leveraged for pre-training cross-modal contrastive learning methods in 
drug discovery. The amount of information contained 
in chemical databases could lead to improved molecule 
encoders and richer representations. 
To investigate this, we construct a large-scale, open, 
dataset of chemical structures of molecules 
and natural language descriptions of bioassays, 
together with bioactivity measurements from PubChem.

\paragraph{The zero-shot problem in drug discovery is equivalent to the library design problem.}
In drug discovery, bioassays take the central role to 
determine the biological 
properties of a small molecule, such as inhibitory 
activity on a drug target in a wet-lab test.
A drug target describes a protein whose activity is modulated by a small molecule, 
whereas a bioassay can measure multiple biological interactions not only constrained to a single protein.
New bioassays are often developed
with the aim to screen a large library of molecules
for a particular activity on a drug target. At this 
initial phase, when a new bioassay has been developed, 
the \emph{library design problem} emerges in all drug
discovery projects \citep{nicolaou2013multi}. 
The library design problem concerns how to select
molecules to be screened without previous 
experience about the new bioassay \citep{hajduk2011question, dandapani2012selecting,irwin2006good}, 
and hence this constitutes a zero-shot prediction problem.
A good selection of molecules will lead to 
a high number of active molecules, which can
potentially be further developed into a drug.
Therefore, this initial selection of molecules critically
determines the success of a drug discovery project
and is usually both time- and cost-intensive.
The drug discovery process could be made more effective
by improving the selection of molecules to be tested in 
a newly developed bioassay (Sec.~\ref{sec:relatedwork}), 
which could be tackled with activity prediction models
that understand the description of the bioassay procedure.
Therefore, we aim at enhancing activity prediction models 
with the ability to utilize human language. %

In summary, our \textbf{contributions} are the following:  
\begin{itemize} \itemsep0em 
\item We propose a new architecture for activity prediction that is able to condition on the textual description of the prediction task.
\item In contrast to almost all previous approaches, 
we suggest the use of 
separate modules for chemical and natural language data.
\item We propose a contrastive pre-training objective on 
information contained in chemical databases as training data.
This data contains orders of magnitudes more chemical structures 
than contained in biomedical texts
\item We show that our approach allows for zero-shot activity prediction,
yields transferable representations, and 
improves predictive performance on few-shot benchmarks 
and zero-shot experiments. 
\item From a more general perspective, our results show how ML models
in application domains can be enhanced with an inferface with human language (Sec.~\ref{sec:discussion}).
\end{itemize}

\section{Problem setting: the zero-shot activity prediction 
in drug discovery}
\label{sec:problem}

\textbf{Single-task bioactivity prediction.} Bioactivity prediction 
has been usually considered as a classical supervised, binary 
prediction prediction task. For a given bioassay or drug target, 
a machine learning model $\hat y = g(m)$ can be trained 
on a set of available measurement pairs of molecules and 
activity labels $\{(m_1,y_1),\ldots,(m_N,y_N)\}$, where $m_n \in \mathcal M$ is a representation
of a molecule from the chemical space $\mathcal M$ and $y_n \in \{0,1\}$
is a binary activity label.

\textbf{Multi-task bioactivity prediction.} 
The problem has also been treated as a 
multi-task learning problem \citep{unterthiner2014deep,dahl2014multi,
ramsundar2015massively,mayr2016deeptox,mayr2018large},
in which several types of activity labels are available
for a molecule $\{(m_1,\By_1),\ldots,(m_N,\By_N)\}$, where 
$\By_n \in \{0,1\}^K$ are vectors 
containing activity values for $K$ different bioassays 
or drug targets. The advantage 
of multi-task learning over single-task 
is that a learned molecule encoder $\Bm=\Bf(m)$ 
can be shared across prediction tasks. However, 
multi-task deep neural networks (MT-DNN) 
cannot be used meaningfully for zero-shot transfer learning, 
when predictions should 
be made for a new bioassay 
for which no training data is available. 

\textbf{Zero-shot bioactivity prediction.} 
To allow for zero-shot predictions of new bioassays, 
for which no training data
is available, a textual representation $a$ of the bioassay,
which represents the prediction task, can be used.
Thus, computational methods are allowed to use 
both a molecule representation $m$ and 
a bioassay representation $a \in \mathcal A$ from 
the space $\mathcal A$ of textual description of biossays as input. 
To train such models, the training data can be considered as triplets
$\{(m_1,a_1,y_1),\ldots, (m_N,a_N,y_N)\}$, where $y_n \in \{0,1\} \ \forall n$  
is a binary activity label, from which the models should learn
to provide a prediction $\hat y$ based on a new input molecule $m^*$
and a new input bioassay $a^*$.

\section{Contrastive Language-Assay-Molecule Pre-training (CLAMP)}

\begin{figure*}[ht]
    \centering
    \includegraphics[width=0.85\textwidth]{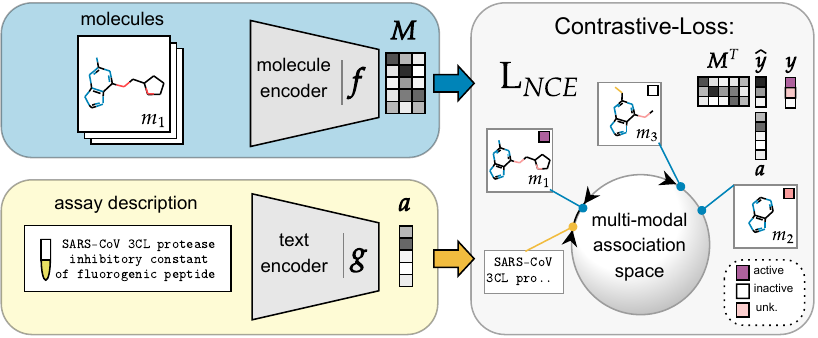}
    
    \caption{Schematic overview of our approach. CLAMP learns to associate active molecules with their corresponding assay descriptions. The stacked molecule embeddings $\BM=[m_1,m_2,m_3]$ are contrasted against the bioassay
    embedding $\Ba$. Similar representations in the association space indicate 
    that molecule $m_1$ is active, while $m_3$ is inactive on 
    bioassay $a$.
    \label{fig:overview}}
\end{figure*}

\textbf{Model architecture and objective.}
Our method uses 
a trainable \emph{molecule encoder} $f:\mathcal M \mapsto \mathbb{R}^d$
to obtain molecule embeddings $\Bm=\Bf(m)$ 
and a trainable 
\emph{text encoder} $g:\mathcal A \mapsto \mathbb{R}^d$ to obtain bioassay embeddings $\Ba=\Bg(a)$. 
We assume that the embeddings are layer-normalized \citep{ba2016layer}.
CLAMP also comprises a 
scoring function $k(\Bm,\Ba)$ that should return high values if 
a molecule $\Bm$ is active on a bioassay $\Ba$ and low values otherwise.
The contrastive learning approach equips our model with the potential
for zero-shot transfer learning, that is, supplying meaningful 
predictions for unseen bioassays.  

The CLAMP model has the following structure:
\begin{align}
\label{eq:similarity}
    \hat y = k(\Bm,\Ba) = k(\Bf(m),\Bg(a)) ,
\end{align}
where $\hat y$ is the predicted activity.
$k(.,.)$ is a score function that should approximate
the targeted distribution $p(y=1 \mid  \Bm,\Ba)$. In practice, we
use the following: $k(\Bm,\Ba)=\frac{\exp (\tau^{-1} \Bm^T\Ba)}{\exp(\tau^{-1} \Bm^T\Ba)+1}$, 
where $\tau^{-1}$ can 
either be a hyperparameter in the range of $1/\sqrt{d}$ or 
a learned parameter (Appendix~\ref{appsec:clamp}).

The objective of our model is to minimize the following
contrastive loss function  with respect to $\Bw$ and 
$\Bv$ \citep{gutmann2010noise,mikolov2013distributed,lopez2021supervised,jiang2019transferable,zang2021scehr,zhai2023sigmoid}: %
\begin{align}
\label{eq:loss}
    \mathrm{L}_{\mathrm{NCE}}=-\frac{1}{N}\sum_{n=1}^N \nonumber & y_n\log(k(\Bf_{\Bw}(m_n),\Bg_{\Bv}(a_n)))+ \\
                       & (1-y_n)\log(1-k(\Bf_{\Bw}(m_n),\Bg_{\Bv}(a_n)),
\end{align}
where $\Bf_{\Bw}(.)$ and $\Bg_{\Bv}(.)$ 
are neural networks with adjustable weights $\Bw$ and $\Bv$, 
respectively. $\{(m_1,a_1,y_1),\ldots, (m_N,a_N,y_N)\}$ is 
the training data set of molecule-text-activity triplets (Sec.~\ref{sec:problem}).

The contrastive loss function encourages molecules that are active on 
a bioassay to have similar embeddings to the embedding of the given bioassay, 
whereas inactive molecules should have embeddings that are dissimlar to it.
In contrast to our approach, in which we have 
access to many labeled pairs, recent prominent 
contrastive learning approaches \citep{radford2021learning,chen2020simple}
only have access to pairs without label. These methods
contrast the matched pair against generated un-matched pairs.
Another difference to these methods is that other 
contrastive learning methods have access to representations of all
classes, whereas in our setting of zero-shot transfer learning 
of bioactivity tasks, only a representation of the positive class, 
but not of the negative class, is available.

\textbf{Encoders.}%
Since there are many possible architectures both for the 
molecule encoder \citep{xu2019efficient} as well as for 
the text encoder, we performed a study in which we assessed
different molecule encoders (Appendix~\ref{appsec:clamp}
and~\ref{appsec:encoder_study}). 
\emph{Molecule encoder.} Briefly, for the molecule encoder we tested 
graph- \citep{kipf2016semi}, SMILES- \citep{mayr2018large}, 
and descriptor-based fully-connected 
architectures \citep{unterthiner2014deep}.
We found that descriptor-based fully-connected 
networks as encoders yielded the best
performance on a validation set, which is 
in accordance with recent results on 
few- and zero-shot drug discovery \citep{stanley2021fs, jiang2021could, schimunek2023contextenriched}.
\emph{Text encoder.} For the text encoder input, we experimented with
BioBERT \citep{lee2020biobert}, Sentence-T5 \citep{ni2021sentencet5} based on T5 \citep{raffel2020exploring}, KV-PLM \citep{zeng2022kvplm},
Galactica \citep{taylor2022galactica}, CLIP text-encoder \citep{radford2021learning}, 
and Latent Semantic Analysis (LSA) \citep{deerwester1990indexing} representations 
of the text. We also consider combinations of these representations.
Surprisingly, LSA works well in combinations 
with language models, which we attribute 
to the specific characteristics of the language 
used to describe bioassays 
(Sec.~\ref{sec:discussion}). 

\textbf{Training and hyperparameters.}
We train the CLAMP architecture from scratch using the 
AdamW \citep{loshchilov2017adamW} optimizer to minimize
the objective Eq.~\eqref{eq:loss}, and for most cases, 
a learning rate of 5e-5 is used. The main hyperparameters
are the size $d$ of the embedding dimension, 
the number of layers and neurons of the molecule encoder,
the initial assay presentation 
as well as the initial molecule representation. 
These hyperparameters are 
selected on a validation set using manual tuning 
(Sec.~\ref{appsec:hyperparams}).

\section{Related work}
\label{sec:relatedwork}

\textbf{Scientific language models.}
Our work is related to scientific language models (SLM) that are able 
to process chemical inputs. Large language models 
such as Galactica \citep{taylor2022galactica}, and
KV-PLM \citep{zeng2022deep} are trained with 
the usual masking objective \citep{devlin2019bert}. 
MolT5 \citep{edwards2022translation} is using a special %
objective \citep{raffel2020exploring} and 
also fine-tunes on molecule caption generation.
Typically the SLMs' input tokens represent chemical structures or sub-structures. 
For example, KV-PLM tokenizes SMILES \citep{weininger1988smiles} strings. 
The pre-training is done on large sets of scientific \citep{taylor2022galactica} or 
biomedical literature \citep{zeng2022deep} which contain
both natural language and chemical structures. 

\textbf{Activity and property prediction models and few- and zero-shot drug discovery.}
There is an immense body of works on activity and property 
prediction models, 
such that we find it useful refer to 
survey articles \citep{muratov2020qsar,lo2018machine,walters2020applications,hochreiter2018machine}.
Since the advent of Deep Learning methods in drug discovery,
activity and property prediction models have been strongly improved
with respect to predictive quality and thus ranking and
selection of molecules with desired activity
\citep{dahl2014multi,unterthiner2014deep,chen2018rise,klambauer2019machine,yang2019analyzing,walters2021critical}. 
Usually several tens of active and inactive 
molecules are necessary to train models with a good
predictive quality \citep{mayr2018large, yang2019analyzing,sturm2020industry}.
To this end, recent efforts have been undertaken to make 
Deep Learning models more efficient with respect to the
necessary training data \citep{altae2017low,nguyen2020meta, stanley2021fs}, 
an area of research which is called \emph{few-shot learning} or 
\emph{low-resource drug discovery}. There are approaches on learning to use 
protein representations of the drug target together with the chemical structure
\citep{van2011proteochemometric} or to use physical simulations to dock molecules
to a protein \citep{dias2008molecular}, both of which allow
for zero-shot activity prediction if the protein target is known. However, 
these approaches restrict zero-shot activity prediction to the space of bioassays that
focus on a particular drug target and exclude all types of functional or toxic activities.

\textbf{Cross-modal contrastive learning methods and pre-training 
strategies in drug discovery.}
\begin{figure}
    \centering
    \includegraphics[width=0.85\columnwidth]{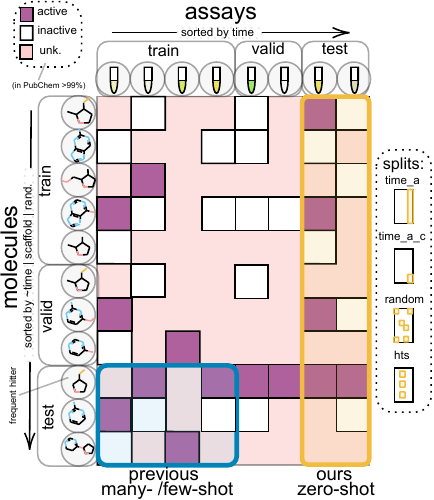}
    \caption{\textbf{Predictive ability of methods:} Rows 
    represent different molecules and columns different 
    bioassays or prediction tasks.
    While previous methods, e.g. multi-task 
    deep networks,
    make predictions
    on unseen molecules and known bioassays 
    (blue box), CLAMP allows
    to make predictions for molecules on unseen bioassays (yellow box).
    \label{fig:datasplit}
    }
\end{figure}
Our work is motivated by the advances that cross-modal contrastive 
learning brought \citep{radford2021learning,furst2022cloob} and 
is related to cross-modal contrastive learning methods in drug discovery. 
Usually, one of the data modalities in drug discovery is the chemical 
structure of the molecule, which is encoded by a molecule encoder. 
\citet{chang2022molecular} contrasts a molecular property vector with a SMILES encoder. %
\citet{guo2022multilingual} contrasts the IUPAC International Chemical Identifier (InChi) with a SMILES encoder. %
\citet{zhu2022unified2d3d} contrasts different representations including SMILES, FP, 3D-geometry, of molecules with each other.
The second modality could be 
natural language \citep{zeng2022deep, edwards2021text2mol}
but also other modalities
such as chemical reaction-templates \citep{seidl2022improving}, 
microscopy images \citep{sanchez2022contrastive}, 
proteins \citep{li2022contrastive} 
or different molecular representations have been suggested. 
\citet{edwards2021text2mol} contrasts molecules with textual descriptions and 
constructs a dataset of 33k corresponding molecule 
and text descriptions termed ChEBI-20. %
\citet{liu2022multi_modal} contrast molecules with a corresponding text 
from proposed dataset PubChemSTM with 280K text-molecule pairs. %
Further related work in Sec.~\ref{appsec:relatedwork}.

\section{Experiments and Results}
\label{sec:experiments}

To demonstrate the effectiveness of our method, we performed three sets of experiments: 
\textbf{a)~Zero-shot transfer learning:} In this experiment, we test the ability 
of methods to predict the activity of molecules on a bioassay of which only 
a textual description is available.  
This setting assesses whether models are able to acquire new knowledge on the 
prediction task from the textual bioassay description. 
The usual activity prediction models (Fig.~\ref{fig:approaches}) 
cannot perform this task, but SLMs have shown some capabilities. 
\textbf{b)~Representation learning:} To check whether the learned 
molecule representations of different methods are rich 
and transferable, we perform linear
probing on a variety of benchmarking datasets \citep{wu2018moleculenet}.
\textbf{c)~Retrieval and library design:}
In this third experiment,
we demonstrate the use of our method as a retrieval system. Based on the
bioassay representation, molecules can be retrieved from 
a chemical database and then ranked for wet-lab testing.
We supply several additional experiments in Sec.~\ref{appsec:extended}.

\begin{table*}[ht]
    \centering
    \caption{Zero-shot results of different methods on four
    different datasets. Green cells indicate the highest values and values within yellow cells are within the standard-deviation to the highest value. %
    Error bars represent standard deviations across the five training re-runs (if computationally feasible). 
    Results are reported in \%. 
    \label{tab:zeroshot}}
    \small
    \begin{threeparttable}

\begin{tabular}{lll|lllllll}
\toprule
metric& dataset &    split & random & GAL 125M$^\dagger$ &   KV-PLM$^\dagger$  & 1-NN &  soft-NN &  FH &CLAMP \\
\midrule
$\Delta$AP &  \cellcolor{jku_grey!10}FS-Mol & \cellcolor{jku_grey!10}default & \cellcolor{jku_grey!10}${01.57^{\pm{0.3}}}$ & \cellcolor{jku_grey!10}${01.44^{\pm{0.0}}}$ & \cellcolor{jku_grey!10}${01.84^{\pm{0.0}}}$ &   \cellcolor{jku_grey!10}${14.68^{\pm{0.7}}}$ & \cellcolor{jku_grey!10}${13.81^{\pm{1.8}}}$ &   \cellcolor{jku_grey!10}${18.50^{\pm{0.2}}}$ &  \cellcolor{jku_grey!10}\cellcolor{jku_green!20}${19.37^{\pm{0.2}}}$ \\
     & PubChem &     hts & ${00.01^{\pm{0.0}}}$ & ${00.00^{\pm{0.0}}}$ & ${00.10^{\pm{0.0}}}$ &   ${01.20^{\pm{0.1}}}$ & ${02.04^{\pm{0.4}}}$ &   ${03.10^{\pm{0.1}}}$ &  \cellcolor{jku_green!20}${08.43^{\pm{0.1}}}$ \\
 &  &  \cellcolor{jku_grey!10}time\_a & \cellcolor{jku_grey!10}${02.13^{\pm{0.3}}}$ & \cellcolor{jku_grey!10}${01.39^{\pm{0.0}}}$ & \cellcolor{jku_grey!10}${03.57^{\pm{0.0}}}$ &   \cellcolor{jku_grey!10}${12.96^{\pm{1.0}}}$ & \cellcolor{jku_grey!10}${05.67^{\pm{0.7}}}$ &   \cellcolor{jku_grey!10}${10.23^{\pm{0.5}}}$ &  \cellcolor{jku_grey!10}\cellcolor{jku_green!20}${14.77^{\pm{0.3}}}$ \\
 & &time\_a\_c & ${04.39^{\pm{0.5}}}$ & ${04.20^{\pm{0.0}}}$ & ${07.99^{\pm{0.0}}}$ & \cellcolor{jku_yellow!20}${11.11^{\pm{0.3}}}$ & ${06.99^{\pm{2.8}}}$ &   ${10.35^{\pm{0.9}}}$ &  \cellcolor{jku_green!20}${11.67^{\pm{0.6}}}$ \\
\midrule
 
    AUROC &  \cellcolor{jku_grey!10}FS-Mol & \cellcolor{jku_grey!10}default & \cellcolor{jku_grey!10}${50.24^{\pm{0.4}}}$ & \cellcolor{jku_grey!10}${50.50^{\pm{0.0}}}$ & \cellcolor{jku_grey!10}${50.56^{\pm{0.0}}}$ &   \cellcolor{jku_grey!10}${64.69^{\pm{0.8}}}$ & \cellcolor{jku_grey!10}${63.92^{\pm{1.9}}}$ &   \cellcolor{jku_grey!10}${68.22^{\pm{0.2}}}$ &  \cellcolor{jku_grey!10}\cellcolor{jku_green!20}${69.26^{\pm{0.2}}}$ \\
    & PubChem &     hts & ${49.92^{\pm{0.2}}}$ & ${49.32^{\pm{0.0}}}$ & ${49.65^{\pm{0.0}}}$ &   ${67.92^{\pm{0.8}}}$ & ${68.41^{\pm{0.9}}}$ &   ${73.48^{\pm{0.4}}}$ &  \cellcolor{jku_green!20}${73.83^{\pm{0.3}}}$ \\
&  &  \cellcolor{jku_grey!10}time\_a & \cellcolor{jku_grey!10}${50.08^{\pm{0.5}}}$ & \cellcolor{jku_grey!10}${47.05^{\pm{0.0}}}$ & \cellcolor{jku_grey!10}${54.92^{\pm{0.0}}}$ &   \cellcolor{jku_grey!10}${66.53^{\pm{0.6}}}$ & \cellcolor{jku_grey!10}${57.85^{\pm{1.7}}}$ &   \cellcolor{jku_grey!10}${66.77^{\pm{1.5}}}$ &  \cellcolor{jku_grey!10}\cellcolor{jku_green!20}${68.66^{\pm{0.5}}}$ \\
    &  &time\_a\_c & ${49.91^{\pm{0.4}}}$ & ${48.04^{\pm{0.0}}}$ & ${57.00^{\pm{0.0}}}$ &   ${61.98^{\pm{0.4}}}$ & ${55.06^{\pm{6.3}}}$ &   ${61.65^{\pm{0.8}}}$ &  \cellcolor{jku_green!20}${63.66^{\pm{0.4}}}$ \\
    \bottomrule
\end{tabular}
\begin{tablenotes}
\end{tablenotes}

\end{threeparttable}

\raggedright
\footnotesize{$^\dagger$ for the SLMs, we chose a 
single model provided by the authors. Training re-runs are computationally infeasible (see Sec.~\ref{appsec:computationl}).} 
\end{table*}

\textbf{Metrics.} We compare different
methods for their ability to rank active molecules higher
than inactive molecules. Due to the imbalanced nature of the prediction
tasks, usually the frequency of inactive molecules is much higher 
than that of active molecules. 
We use delta average precision ($\Delta$AP)
as main metric \citep{stanley2021fs} in 
Experiments \textbf{a)} and \textbf{b)} and the 
enrichment factor (EF) \citep{friesner2004glide}
as metric for experiment \textbf{c)} (see Sec.~\ref{appsec:metric}),

\subsection{Zero-Shot Transfer}
All methods have to predict the activity of molecules for new 
activity prediction tasks based on the textual description of the task
without any available labels for that task. This setting
represents a zero-shot problem in drug discovery.

\textbf{Datasets.} 
To this end, we use the benchmarking dataset FS-Mol \citep{stanley2021fs} and propose a zero-shot mode. Further, we construct three different 
splits of the PubChem \citep{kim2019pubchem} database 
(Fig.~\ref{fig:datasplit}): 
the "hts" split contains high-throughput assays \citep{laufkotter2019combining},
the "time\_a" split contains assays sorted by time. We test on 4,201 assays 
from 2013-11 to 2018-05. For the "time\_a\_c" split, we additionally sort 
compounds by time and test on new unseen molecules from new assays.

\textbf{Methods compared.}
We compared the method \textbf{CLAMP}, the following baseline methods and competitor methods:
\textbf{1-nearest-neighbour (1-NN)}: this method uses textual
description of the given bioassay, to identify the most similar
bioassay in the training set. Then the predictions for this training
set assay are used for the given prediction task. The predictions
are based on a MT-DNN. 
\textbf{soft-nearest-neighbour (soft-NN)}: Uses the same approach 
as 1-NN only that the predictions of the training set assay are weighted by 
their similarity to the given assay. 
The last baseline is the so-called \textbf{Frequent hitters (FH)} model \citep{schimunek2023contextenriched}: 
this model predicts the general 
activity of a molecule across all prediction task. This is a strong 
baseline, since there are molecule that often test positive in 
any bioassay \citep{baell2010new}. 
Another category of methods that are able to perform zero-shot
activity predictions are SLMs that are able to process chemical structures:  
The SLM \textbf{Galactica (GAL 125M)} has been 
included with an appropriate text prompt (Appendix~\ref{appsec:galactica}) 
and also the SLM \textbf{KV-PLM}. We use the 
publicly available SLMs that were 
pre-trained on their suggested text corpora. 
For details, see Appendix~\ref{appsec:zero_shot_experiment}.

\textbf{Results.} The results are shown in Tab.~\ref{tab:zeroshot}. 
CLAMP significantly 
outperforms all other methods with respect to $\Delta$AP (paired Wilcoxon test) 
except the FH models on the "hts" and "time\_a\_c" splits. %
Notably, the SLMs do not reach the predictive performance of the FH baseline that 
completely ignores the textual information.

\subsection{Representation Learning}
\label{sec:representation_learning}
\begin{table*}[ht]
    \noindent
    \centering
    \caption{Linear probing results of different methods 
    with respect to $\Delta$AP. 
    Green cells indicate the highest values in a category of tasks
    and areas in yellow cells within the 
    standard-deviation to the maximum value. 
    Because of the low variability of training re-runs of a linear probing model, the error bars represents standard-deviation obtained through bootstrap resampling.
    Rank-avg represents the mean rank over all assays. Methods are
    assigned to categories (cat): self-supervised learning methods (SSL), scientific 
    lanuage models (SLM), and chemical descriptors or fingerprints (FP)
    \label{tab:linear_probing_wo_gal}}
    \small
    \fontsize{9pt}{9pt}
    \resizebox{\textwidth}{!}{
    \begin{tabular}{llcccccccc||l}
\toprule
 dataset& &BACE &   BBBP &   ClinTox &HIV &  SIDER & Tox21 &ToxCast & Tox21-10k &  \\
 split&&scaffold & scaffold &scaffold & scaffold & scaffold &scaffold & scaffold & 
 original & \\
 \# of assays & cat &  1 &  1 &  2 & 1  & 27  &  12 &  617 & 68  & rank-avg  \\
\midrule
       \cellcolor{jku_grey!10}CLAMP & \cellcolor{jku_grey!10}ours& \cellcolor{jku_grey!10}\cellcolor{jku_yellow!20}${27.47^{\pm{\scriptscriptstyle4}}}$ & \cellcolor{jku_grey!10}\cellcolor{jku_yellow!20}${16.47^{\pm{\scriptscriptstyle4}}}$ &  \cellcolor{jku_grey!10}${11.05^{\pm{\scriptscriptstyle6}}}$ &  \cellcolor{jku_grey!10}\cellcolor{jku_green!20}${28.49^{\pm{\scriptscriptstyle4}}}$ &  \cellcolor{jku_grey!10}\cellcolor{jku_green!20}${08.96^{\pm{\scriptscriptstyle4}}}$ & \cellcolor{jku_grey!10}\cellcolor{jku_green!20}${23.35^{\pm{\scriptscriptstyle6}}}$ &  \cellcolor{jku_grey!10}\cellcolor{jku_green!20}${09.44^{\pm{\scriptscriptstyle5}}}$ &  \cellcolor{jku_grey!10}\cellcolor{jku_green!20}${51.27^{\pm{\scriptscriptstyle10}}}$ & \cellcolor{jku_grey!10}\cellcolor{jku_green!20}${02.85^{\pm{\scriptscriptstyle2}}}$ \\
                              Grover &  SSL &                                                 ${21.74^{\pm{\scriptscriptstyle4}}}$ &                         \cellcolor{jku_yellow!20}${16.76^{\pm{\scriptscriptstyle4}}}$ &                          ${22.74^{\pm{\scriptscriptstyle8}}}$ &                                                   ${13.58^{\pm{\scriptscriptstyle3}}}$ &                         \cellcolor{jku_yellow!20}${05.41^{\pm{\scriptscriptstyle4}}}$ &                                                  ${12.88^{\pm{\scriptscriptstyle5}}}$ &                         \cellcolor{jku_yellow!20}${05.22^{\pm{\scriptscriptstyle3}}}$ &                         \cellcolor{jku_yellow!20}${42.23^{\pm{\scriptscriptstyle10}}}$ &                        \cellcolor{jku_yellow!20}${05.08^{\pm{\scriptscriptstyle3}}}$ \\
\cellcolor{jku_grey!10}Mc+RDKc & \cellcolor{jku_grey!10}FP &                          \cellcolor{jku_grey!10}${23.87^{\pm{\scriptscriptstyle4}}}$ & \cellcolor{jku_grey!10}\cellcolor{jku_yellow!20}${18.39^{\pm{\scriptscriptstyle4}}}$ &  \cellcolor{jku_grey!10}${17.75^{\pm{\scriptscriptstyle7}}}$ & \cellcolor{jku_grey!10}\cellcolor{jku_yellow!20}${25.57^{\pm{\scriptscriptstyle4}}}$ & \cellcolor{jku_grey!10}\cellcolor{jku_yellow!20}${08.43^{\pm{\scriptscriptstyle4}}}$ &                          \cellcolor{jku_grey!10}${13.96^{\pm{\scriptscriptstyle5}}}$ & \cellcolor{jku_grey!10}\cellcolor{jku_yellow!20}${05.18^{\pm{\scriptscriptstyle4}}}$ & \cellcolor{jku_grey!10}\cellcolor{jku_yellow!20}${42.45^{\pm{\scriptscriptstyle10}}}$ &                          \cellcolor{jku_grey!10}${05.64^{\pm{\scriptscriptstyle3}}}$ \\
                                CDDD & SSL &                                                 ${17.51^{\pm{\scriptscriptstyle5}}}$ &                          \cellcolor{jku_green!20}${20.56^{\pm{\scriptscriptstyle4}}}$ & \cellcolor{jku_green!20}${33.82^{\pm{\scriptscriptstyle7}}}$ &                                                   ${12.29^{\pm{\scriptscriptstyle3}}}$ &                         \cellcolor{jku_yellow!20}${05.30^{\pm{\scriptscriptstyle4}}}$ &                                                  ${10.34^{\pm{\scriptscriptstyle4}}}$ &                                                   ${04.82^{\pm{\scriptscriptstyle3}}}$ &                                                    ${36.32^{\pm{\scriptscriptstyle9}}}$ &                                                  ${05.79^{\pm{\scriptscriptstyle3}}}$ \\
  \cellcolor{jku_grey!10}BARTSmiles &  \cellcolor{jku_grey!10}SSL &\cellcolor{jku_grey!10}\cellcolor{jku_green!20}${29.16^{\pm{\scriptscriptstyle3}}}$ & \cellcolor{jku_grey!10}\cellcolor{jku_yellow!20}${17.09^{\pm{\scriptscriptstyle4}}}$ &  \cellcolor{jku_grey!10}${10.94^{\pm{\scriptscriptstyle6}}}$ &                           \cellcolor{jku_grey!10}${07.88^{\pm{\scriptscriptstyle2}}}$ & \cellcolor{jku_grey!10}\cellcolor{jku_yellow!20}${05.24^{\pm{\scriptscriptstyle4}}}$ &                          \cellcolor{jku_grey!10}${10.34^{\pm{\scriptscriptstyle4}}}$ & \cellcolor{jku_grey!10}\cellcolor{jku_yellow!20}${05.07^{\pm{\scriptscriptstyle3}}}$ &                            \cellcolor{jku_grey!10}${30.40^{\pm{\scriptscriptstyle9}}}$ &                          \cellcolor{jku_grey!10}${05.79^{\pm{\scriptscriptstyle3}}}$ \\
                              KV-PLM &    SLM &                                                ${24.44^{\pm{\scriptscriptstyle4}}}$ &                         \cellcolor{jku_yellow!20}${17.61^{\pm{\scriptscriptstyle4}}}$ &                          ${07.08^{\pm{\scriptscriptstyle4}}}$ &                                                   ${06.20^{\pm{\scriptscriptstyle2}}}$ &                                                   ${04.61^{\pm{\scriptscriptstyle4}}}$ &                                                  ${09.90^{\pm{\scriptscriptstyle4}}}$ &                                                   ${04.53^{\pm{\scriptscriptstyle3}}}$ &                                                   ${30.92^{\pm{\scriptscriptstyle10}}}$ &                                                  ${05.88^{\pm{\scriptscriptstyle3}}}$ \\
      \cellcolor{jku_grey!10}MFBERT & \cellcolor{jku_grey!10}SSL &                         \cellcolor{jku_grey!10}${14.89^{\pm{\scriptscriptstyle4}}}$ & \cellcolor{jku_grey!10}\cellcolor{jku_yellow!20}${19.77^{\pm{\scriptscriptstyle4}}}$ &  \cellcolor{jku_grey!10}${11.84^{\pm{\scriptscriptstyle7}}}$ &                           \cellcolor{jku_grey!10}${06.09^{\pm{\scriptscriptstyle1}}}$ & \cellcolor{jku_grey!10}\cellcolor{jku_yellow!20}${06.93^{\pm{\scriptscriptstyle4}}}$ &                          \cellcolor{jku_grey!10}${08.46^{\pm{\scriptscriptstyle4}}}$ &                           \cellcolor{jku_grey!10}${04.72^{\pm{\scriptscriptstyle3}}}$ &                            \cellcolor{jku_grey!10}${25.61^{\pm{\scriptscriptstyle8}}}$ &                          \cellcolor{jku_grey!10}${06.20^{\pm{\scriptscriptstyle3}}}$ \\
                          Graphormer &  SSL &                                                 ${20.22^{\pm{\scriptscriptstyle4}}}$ &                                                   ${08.67^{\pm{\scriptscriptstyle4}}}$ &                          ${04.51^{\pm{\scriptscriptstyle4}}}$ &                                                   ${06.87^{\pm{\scriptscriptstyle2}}}$ &                         \cellcolor{jku_yellow!20}${05.71^{\pm{\scriptscriptstyle4}}}$ &                                                  ${08.16^{\pm{\scriptscriptstyle4}}}$ &                                                   ${04.10^{\pm{\scriptscriptstyle3}}}$ &                                                    ${29.50^{\pm{\scriptscriptstyle9}}}$ &                                                  ${06.55^{\pm{\scriptscriptstyle3}}}$ \\
      \cellcolor{jku_grey!10}Morgan & \cellcolor{jku_grey!10}FP &\cellcolor{jku_grey!10}\cellcolor{jku_yellow!20}${27.79^{\pm{\scriptscriptstyle4}}}$ &                           \cellcolor{jku_grey!10}${16.06^{\pm{\scriptscriptstyle4}}}$ &  \cellcolor{jku_grey!10}${23.86^{\pm{\scriptscriptstyle8}}}$ &                           \cellcolor{jku_grey!10}${16.23^{\pm{\scriptscriptstyle4}}}$ & \cellcolor{jku_grey!10}\cellcolor{jku_yellow!20}${06.97^{\pm{\scriptscriptstyle4}}}$ &                          \cellcolor{jku_grey!10}${09.69^{\pm{\scriptscriptstyle4}}}$ &                           \cellcolor{jku_grey!10}${04.28^{\pm{\scriptscriptstyle3}}}$ &                           \cellcolor{jku_grey!10}${36.07^{\pm{\scriptscriptstyle10}}}$ &                          \cellcolor{jku_grey!10}${06.64^{\pm{\scriptscriptstyle3}}}$ \\
                               MolT5 & SLM &                                               ${11.51^{\pm{\scriptscriptstyle4}}}$ &                                                   ${16.03^{\pm{\scriptscriptstyle4}}}$ &                          ${10.06^{\pm{\scriptscriptstyle6}}}$ &                                                   ${15.92^{\pm{\scriptscriptstyle3}}}$ &                                                   ${02.91^{\pm{\scriptscriptstyle3}}}$ &                                                  ${06.20^{\pm{\scriptscriptstyle3}}}$ &                                                   ${03.53^{\pm{\scriptscriptstyle3}}}$ &                                                    ${15.65^{\pm{\scriptscriptstyle6}}}$ &                                                  ${07.36^{\pm{\scriptscriptstyle3}}}$ \\
      \cellcolor{jku_grey!10}MolCLR & \cellcolor{jku_grey!10}SSL &                           \cellcolor{jku_grey!10}${15.59^{\pm{\scriptscriptstyle4}}}$ &                           \cellcolor{jku_grey!10}${13.01^{\pm{\scriptscriptstyle4}}}$ &  \cellcolor{jku_grey!10}${00.99^{\pm{\scriptscriptstyle3}}}$ &                           \cellcolor{jku_grey!10}${02.57^{\pm{\scriptscriptstyle1}}}$ & \cellcolor{jku_grey!10}\cellcolor{jku_yellow!20}${06.13^{\pm{\scriptscriptstyle4}}}$ &                          \cellcolor{jku_grey!10}${06.08^{\pm{\scriptscriptstyle3}}}$ &                           \cellcolor{jku_grey!10}${02.43^{\pm{\scriptscriptstyle2}}}$ &                            \cellcolor{jku_grey!10}${13.21^{\pm{\scriptscriptstyle5}}}$ &                          \cellcolor{jku_grey!10}${08.20^{\pm{\scriptscriptstyle3}}}$ \\
\bottomrule
\end{tabular}
    }
\end{table*}

In this experiment, we use linear probing \citep{alain2016understanding}
to assess how robust and transferable the embeddings 
of different encoders are.

\textbf{Datasets.} We use the 
MoleculeNet \citep{wu2018moleculenet} benchmarking datasets, 
BACE, BBBP, ClinTox, 
HIV, SIDER, Tox21 and ToxCast. 
Additionally, we compare methods on Tox21-10k \citep{richard2021tox21_10k}.
We remove %
all downstream test-set
molecules from the pre-training dataset 
(Sec.~\ref{appsec:data_overlap}).

\textbf{Methods compared.} %
Molecular 
encoders are pre-trained in their proposed way. 
\textbf{CLAMP} was pre-trained on PubChem with a 
random split and we removed all test-set-molecules that 
are contained in the downstream tasks to avoid data leakage
(Appendix~\ref{appsec:data_overlap}). 
We included several baseline encoders, that extract substructures
from molecules, that is, \textbf{Morgan} fingerprints
 \citep{morgan1965generation}
of length 1024, and a combination of those with
chemical descriptors \citep{landrum2013rdkit}
\textbf{Mc+RDKc} of length 8192.
Furthermore, the following \emph{molecule encoders} that were
pre-trained in self-supervised fashion were assessed: 
\textbf{Grover} \citep{rong2020grover}, a graph transformer, 
\textbf{CDDD} \citep{winter2019cddd}, 
a SMILES-LSTM based autoencoder, 
the SMILES-Tranformers \textbf{BARTSmiles} \citep{chilingaryan2022bartsmiles}, 
\textbf{Graphormer} \citep{ying2021graphormer}, 
\textbf{MFBERT} \citep{abdel2022mfbert},
and \textbf{MolCLR} \citep{wang2021molclr}, 
a contrastively pre-trained GNN.
We also use the \textbf{SLMs} 
\textbf{KV-PLM} \citep{zeng2022kvplm}, 
\textbf{MolT5} \citep{edwards2022translation}
and \textbf{Galactica}\footnote{The model cannot be evaluated for all datasets, due to test-set measurements being present in pre-training. Valid results can be found in Tab.~\ref{tab:linear_probing_with_gal}} \citep{taylor2022galactica}.

\textbf{Results.} Tab.~\ref{tab:linear_probing_wo_gal} displays the results
of different methods on the linear probing experiments
with respect to $\Delta$AP. 
CLAMP performs best on average and 
significantly (paired Wilcoxon test; all $p$-values $<$1e-10) 
outperforms all other methods with respect to $\Delta AP$ 
across prediction tasks. 
Our method yields the best performing representations in 
5 of the 8 datasets, 
and the datasets on which CLAMP is not the best method, it is within the standard-deviation.
Notably, CLAMP strongly improves predictive performance on %
on ToxCast from a $\Delta AP$ of $05.22^{\pm 3}$ to $09.44^{\pm 5}$,
which is an average over 617 prediction tasks.

\subsection{Retrieval and library design}
\label{sec:retrieval}

Here we consider a retrieval task, in which
molecules from a chemical database must be ranked based on 
a given bioassay that represents the query. Molecules
that are active on the given bioassay should be ranked high. 
The enrichment-factor (EF) is used as metric to evaluate 
this type of retrieval tasks \citep{truchon2007vs_metric}. 
The EF calculates how much a given method 
improves the top-$k$ accuracy over a random ordering for a given $k$. 

\textbf{Dataset.}
We use the PubChem dataset with assay based temporal split time\_a.
For molecule retrieval, we chose the assay-based
temporal split time\_a (Appendix~\ref{appsec:dataset}).
For the chemical databases we consider two sizes: 
1M or 10k molecules. Molecules have been selected in order of their PubChem compound-ID (CID).
To obtain robust estimates, %
we consider assays with more than 100 active molecules and 
report the average over assays. 
This results in 190 assays for the 10k molecules setting 
and 2,543 assays for the 1M molecule setting for testing.

\textbf{Methods compared.}
The methods \textbf{CLAMP} and \textbf{FH} are trained on the time\_a split and 
chosen based on validation $\Delta AP$.
For FH baseline the ranking for the molecules remains 
the same regardless of the assay.
We benchmarked against \textbf{KV-PLM} and \textbf{Galactica}.
Evaluating Galactica on the 1M benchmark is
computationally infeasible (Appendix~\ref{appsec:computationl})
because for each combination %
a full forward-pass has to be performed. 
At the 10k setting, it takes $\sim$13sec for KV-PLM\footnote{at batchsize of 2048, including encoding}, 
$\sim$9sec for CLAMP \footnote{64 cores, batch size of 2048, including encoding} and $\sim$19h for GAL\footnote{batch size of 256,  $\sim$2.9 years at 10M setting} for 190 assays.

\textbf{Results.}
We find that CLAMP enriches active molecules for unseen assays 
by more than 10-fold in the case of 10k, 
and by more than 250-fold in the case of 1M molecules. 
KV-PLM \citep{zeng2022kvplm}, the best-
performing scientific language model is outperformed by 50x (Fig.~\ref{fig:compound_retrieval}).
Fig.~\ref{fig:compound_retr}
shows a comparison between the methods across different top-$k$
accurracies. CLAMP consistently outperforms all other methods.

\section{Discussion}
\label{sec:discussion}
\emph{Conclusions.} 
Our proposed contrastive learning method 
CLAMP exhibits the best performance at zero-shot prediction 
drug discovery tasks on several large datasets. The
pre-trained molecule-encoder of CLAMP yields transferable 
representations. Our results also point out that, although 
the scientific language models can in principle be used 
for zero-shot activity prediction, 
they are not performing well at this task 
and are computationally demanding (Sec.~\ref{sec:retrieval}). 
\emph{Limitations and Future work.} 
Currently, our approach is mostly limited by computational complexity, since both hyperparameter and model selection is computationally demanding. We leave it to future work, to expand on the search for different encoders in conjunction with our approach.
The models can perform activity prediction, but is not able to generate molecules. However, the embedding space of CLAMP is a prime candidate for conditional generation as it is analogous to the CLIP latent space \citep{ramesh2022hierarchical}.
Chemical dosage, which affects the assay outcome, as with many other approaches, is not considered. It may also struggle with negations and grammatical nuances, resulting in inaccurate predictions.
As for all ML methods, the predictive ability of CLAMP can decrease outside the chemical and bioassay space of the training data and suffers from biases that are present in chemical databases.
\emph{Broader impact.} See Sec.~\ref{appsec:broader}.\\

\textbf{Data and software availability.} 
Python code and instructions to reproduce the results are provided
as Supplementary Material and will be available at \url{https://github.com/ml-jku/clamp} \\

\textbf{Acknowledgments}
The ELLIS Unit Linz, the LIT AI Lab, the Institute for Machine Learning, are supported by the Federal State Upper Austria. IARAI is supported by Here Technologies. We thank the projects AI-MOTION (LIT-2018-6-YOU-212), DeepFlood (LIT-2019-8-YOU-213), Medical Cognitive Computing Center (MC3), INCONTROL-RL (FFG-881064), PRIMAL (FFG-873979), S3AI (FFG-872172), DL for GranularFlow (FFG-871302), EPILEPSIA (FFG-892171), AIRI FG 9-N (FWF-36284, FWF-36235), ELISE (H2020-ICT-2019-3 ID: 951847), Stars4Waters (HORIZON-CL6-2021-CLIMATE-01-01). We thank Audi.JKU Deep Learning Center, TGW LOGISTICS GROUP GMBH, Silicon Austria Labs (SAL), FILL Gesellschaft mbH, Anyline GmbH, Google, ZF Friedrichshafen AG, Robert Bosch GmbH, UCB Biopharma SRL, Merck Healthcare KGaA, Verbund AG, GLS (Univ. Waterloo) Software Competence Center Hagenberg GmbH, T\"{U}V Austria, Frauscher Sensonic and the NVIDIA Corporation.

\bibliography{main_icml}
\bibliographystyle{icml2023}

 \cleardoublepage

\resumetocwriting
 
\appendix
\onecolumn

\tableofcontents

\renewcommand{\thefigure}{A\arabic{figure}}
\renewcommand{\thetable}{A\arabic{table}}
\renewcommand{\theequation}{A\arabic{equation}}

\section{Appendix}

\subsection{Further related work}
\label{appsec:relatedwork}

\paragraph{Zero-shot learning problems.}
From the perspective of machine-learning, 
the described problem represents
a zero-data or zero-shot prediction 
task \citep{chang2008importance,larochelle2008zero, farhadi2009describing,palatucci2009zero},
for which several methods 
in the area of computer vision and natural language 
processing have been developed \citep{wang2019survey}.
The setting is that no training data are available 
and only a description of the classes or tasks are
provided, which in our case is the textual 
description of the bioassay. In contrast to zero-shot
problems in computer vision where a description of 
each class is available, in the drug discovery setting only
a description of the positive class is available. 
Contrastive learning for
zero-data problems has recently been exemplified 
with the ConVIRT \citep{zhang2020contrastive} or the CLIP algorithm \citep{radford2021learning},
in which representations of natural images and language
are learned.

\paragraph{Proteo-chemometric and molecular docking.}
Several efforts have been devoted to 
being able to make predictions for new biological 
targets, such as proteins. The set 
of proteochemometric methods \citep{van2011proteochemometric}
use information about the protein, such as its
1D structure, and combine it with information 
about the molecule. Molecular docking methods
use the 3D structure of the protein and search 
for a conformation of a ligand that fits into
a binding pocket \citep{pagadala2017software,meng2011molecular}.
However, many bioassays are not focused on 
a target, but rather measure a general effect, 
such as a toxic response or cell proliferation,
which limits or prohibits the use of proteochemic or docking methods.

\paragraph{Recommender systems.}
The zero-data problem has earlier been identified 
by the recommender systems and matrix factorization
research community as cold-start problem \citep{schein2002methods}.
The cold-start problem is how to 
to provide good recommendations for novel users or items.
Remedies for the cold-start problem of recommender systems
exploit similarities of initial descriptions between users and items
\citep{lika2014facing}. Contrastive learning has recently
been suggested to learn the similarities between users 
and items \citep{liu2021contrastive,zhou2020contrastive}.
From the perspective of recommender systems, our method
CLAMP can be understood as having to suggest molecules
representing items for a new bioassay representing a new user.

\paragraph{Selection strategies for bioassay screening.}
Our approach is related to works on 
different strategies for selecting a molecule library
for wet-lab testing. 
A prominent approach is high-throughput screening, 
in which large parts of physically available molecules 
are screened at high-throughput \citep{hajduk2011question}.
This is possible if the bioassay can be performed in high 
throughput, the wet-lab facilities and a large molecule 
library are available. High-throughput screening has been
seen as a strong improvement in drug discovery. 
Naturally, many computational methods have also been suggested
to first virtually screen \citep{shoichet2004virtual} chemical
libraries and then perform bioassay screening on the top-ranked
molecules. 
Data-driven strategies, such as machine learning and 
Deep Learning, have brought a strong improvement of
virtual screening methods. 
However, data-driven strategies are not possible for new bioassays \citep{ain2015machine}
since no data is available, and no actives or inactives are known. 
To ameliorate this central problem, 
practitioners and scientists have resorted to using information 
from similar bioassays, facilitated by efforts to semantically structure
the information about bioassays \citep{visser2011bioassay}. 
However, this type of information has not been integrated into 
machine learning approaches yet. 
In summary, while data-driven virtual screening strategies have been 
shown to be highly effective, it is currently unclear 
how those approaches could be used for designing
libraries for newly developed bioassays.

\paragraph{Information from the textual description of 
the bioassay can be leveraged with contrastive learning.}
Despite the lack of known active and inactive molecules for 
novel bioassays, 
there is information available that could potentially 
be used for machine learning: the textual description 
of the bioassay. For each bioassay, the procedure in 
the wet-lab, their endpoint, and the substrate, are 
usually described in textual form. There have even been 
efforts to semantically describe such bioassays using an
ontology \citep{visser2011bioassay}. %

\subsection{Datasets}
\label{appsec:dataset}
\label{appsec:data}
Here we provide an overview of the datasets used in this work. 
The summary statistics can be found in Tab.~\ref{tab:datasets}.

\begin{table*}[t]
    \centering
    \caption{Main datasets summary statistics overview. 
    \label{tab:datasets}}
    \small
    \begin{tabular}{lrrr} %
\toprule
 & PubChem & PubChem HTS &   FS-Mol \\% &   HIV &   ToxCast &SIDER &  ... \\
\midrule
  Source & PubChem & PubChem &ChEMBL27 \\% & Molec.Net  & Molec.Net   &Molec.Net   &  \\
  \cellcolor{jku_grey!10}\# measurements & \cellcolor{jku_grey!10}223,219,241 & \cellcolor{jku_grey!10}143,886,653 & \cellcolor{jku_grey!10}501,366 \\% &\cellcolor{jku_grey!10}41,127 & \cellcolor{jku_grey!10}5,291,392 &   \cellcolor{jku_grey!10}38,529 & \cellcolor{jku_grey!10} \\
 \# molecules &   2,120,811 & 715,231 & 240,465 \\% &41,127 & 8,576 &1,427 &  \\
\cellcolor{jku_grey!10}\# assays &  \cellcolor{jku_grey!10}21,002 & \cellcolor{jku_grey!10}582 &   \cellcolor{jku_grey!10}5,135 \\%  & \cellcolor{jku_grey!10}1 &   \cellcolor{jku_grey!10}617 &   \cellcolor{jku_grey!10}27 & \cellcolor{jku_grey!10} \\
 \% of assays with only one class &   74.54 & 1.37 &  0.00 \\% & 0.00 &   0.00 & 0.00 &  \\
\cellcolor{jku_grey!10}Mean \# compounds / assay &   \cellcolor{jku_grey!10}10,628.48 &  \cellcolor{jku_grey!10}247,227.93 &   \cellcolor{jku_grey!10}96.32 \\% &  \cellcolor{jku_grey!10}41,127.00 &  \cellcolor{jku_grey!10}8,576.00 & \cellcolor{jku_grey!10}1,427.00 & \cellcolor{jku_grey!10} \\
  Median \# compounds / assay &   35.00 &  304,804.00 &   46.00  \\% & 41,127.00 &  8,576.00 & 1,427.00 &  \\
\cellcolor{jku_grey!10}\% active &\cellcolor{jku_grey!10}1.51 &\cellcolor{jku_grey!10}0.70 &   \cellcolor{jku_grey!10}46.48 \\% &  \cellcolor{jku_grey!10}3.51 &  \cellcolor{jku_grey!10}2.39 &\cellcolor{jku_grey!10}56.76 & \cellcolor{jku_grey!10} \\
   \% density & 0.50 &   34.57 & 0.04 \\% & 100.00  & 100.00 &   100.00 &  \\
 \cellcolor{jku_grey!10}Mean \% active per assay &   \cellcolor{jku_grey!10}79.46 &\cellcolor{jku_grey!10}1.04 &   \cellcolor{jku_grey!10}47.17 \\% &  \cellcolor{jku_grey!10}3.51 &  \cellcolor{jku_grey!10}2.39 &\cellcolor{jku_grey!10}56.76 & \cellcolor{jku_grey!10} \\
   Median \% active per assay &  100.00 & 0.42 &   48.84 \\% &  3.51 &  1.28 &66.29 &  \\
\bottomrule
\end{tabular}
\end{table*}

\subsubsection{An open, large-scale dataset for zero-shot drug discovery 
derived from PubChem}
We constructed a large public dataset extracted from 
PubChem \citep{kim2019pubchem,preuer2018frechet}, an open chemistry database, and the largest collection of readily available chemical data. We take assays ranging from 2004 to 2018-05. It initially comprises 224,290,250 records of molecule-bioassay activity, corresponding to 2,120,854 unique molecules and 21,003 unique bioassays. We find that some molecule-bioassay pairs have multiple activity records, which may not all agree. We reduce every molecule-bioassay pair to exactly one activity measurement by applying majority voting. Molecule-bioassay pairs with ties are discarded. This step yields our final bioactivity dataset, which features 223,219,241 records of molecule-bioassay activity, corresponding to 2,120,811 unique molecules and 21,002 unique bioassays ranging from AID 1 to AID 1259411. Molecules range up to CID 132472079.
The dataset has 3 different splitting schemes which are further described in 
Sec.~\ref{appsec:data_splits}.

\subsubsection{FS-Mol}
The FS-Mol dataset \citep{stanley2021fs} 
has been constructed based on ChEMBL \citep{gaulton2012chembl}
with the focus on providing a few-shot 
learning dataset to the research community. 
The dataset comprises  $\sim$240k molecules and
5k prediction tasks, roughly equivalent to bioassays. 
In the original form of FS-Mol, for each prediction 
task a small training set (support-set) of 16, 32, 64, or 128
molecules together with binary activity labels are 
available for few-shot learning.

We extend the dataset with textual descriptions sourced from ChEMBL. 
We further use this dataset in a zero-shot setting,
where we only have the text description of the prediction task available 
(examples in Tab.~\ref{tab:assay_descriptions}).
For the standard setting in few-shot learning, with 
a training set of 16 molecules, the prediction metrics 
of few-shot learning methods are in the range of 
a $\Delta$AP of $20$-$25$ \%
\citep{schimunek2023contextenriched}. Notably, CLAMP 
reaches a $\Delta$AP of $19.4$\% without any training data for 
that task.

\subsubsection{Downstream datasets}
In this work, we further test on a range of datasets in the domains of biophysics as well as physiology.
We mainly focus on datasets from the MoleculeNet benchmark 
\citep{wu2018moleculenet}. 
Despite their small size and a limited number of tasks, these
datasets are widely used to assess and compare graph neural networks \citep{rong2020grover}.
The benchmark contains the following datasets:

\begin{table*}[tb]
    \centering
    \caption{Downstream datasets overview}
    \footnotesize
    \begin{tabular}{llllllllll}
\toprule
                                            & \multicolumn{9}{l}{Downstream} \\
                                                 &                             BACE &                             BBBP &                          ClinTox &                               HIV &                            SIDER &                            Tox21 &                           ToxCast &                        Tox21-10k &                          Tox21og \\
\midrule
          \cellcolor{jku_grey!10}\# measurements &    \cellcolor{jku_grey!10}1,513 &    \cellcolor{jku_grey!10}2,039 &    \cellcolor{jku_grey!10}2,956 &    \cellcolor{jku_grey!10}41,127 &   \cellcolor{jku_grey!10}38,529 &   \cellcolor{jku_grey!10}93,972 & \cellcolor{jku_grey!10}5,291,392 &  \cellcolor{jku_grey!10}402,885 &  \cellcolor{jku_grey!10}108,372 \\
                                     \# compounds &                            1,513 &                            1,975 &                            1,459 &                            41,127 &                            1,427 &                            7,831 &                             8,576 &                            7,659 &                            8,968 \\
                \cellcolor{jku_grey!10}\# assays &        \cellcolor{jku_grey!10}1 &        \cellcolor{jku_grey!10}1 &        \cellcolor{jku_grey!10}2 &         \cellcolor{jku_grey!10}1 &       \cellcolor{jku_grey!10}27 &       \cellcolor{jku_grey!10}12 &       \cellcolor{jku_grey!10}617 &       \cellcolor{jku_grey!10}68 &       \cellcolor{jku_grey!10}12 \\
                 \% of assays with only one class &                             0.00 &                             0.00 &                             0.00 &                              0.00 &                             0.00 &                             0.00 &                              0.00 &                             0.00 &                             0.00 \\
\cellcolor{jku_grey!10}Mean \# compounds / assay & \cellcolor{jku_grey!10}1,513.00 & \cellcolor{jku_grey!10}1,975.00 & \cellcolor{jku_grey!10}1,459.00 & \cellcolor{jku_grey!10}41,127.00 & \cellcolor{jku_grey!10}1,427.00 & \cellcolor{jku_grey!10}7,831.00 &  \cellcolor{jku_grey!10}8,576.00 & \cellcolor{jku_grey!10}5,924.78 & \cellcolor{jku_grey!10}7,482.67 \\
                      Median \# compounds / assay &                         1,513.00 &                         1,975.00 &                         1,459.00 &                         41,127.00 &                         1,427.00 &                         7,831.00 &                          8,576.00 &                         5,963.00 &                         7,531.00 \\
                \cellcolor{jku_grey!10}\% active &    \cellcolor{jku_grey!10}45.67 &    \cellcolor{jku_grey!10}76.51 &    \cellcolor{jku_grey!10}50.61 &      \cellcolor{jku_grey!10}3.51 &    \cellcolor{jku_grey!10}56.76 &     \cellcolor{jku_grey!10}6.24 &      \cellcolor{jku_grey!10}2.39 &     \cellcolor{jku_grey!10}5.10 &     \cellcolor{jku_grey!10}7.27 \\
                                       \% density &                           100.00 &                           103.24 &                           101.30 &                            100.00 &                           100.00 &                           100.00 &                            100.00 &                            77.36 &                           100.70 \\
 \cellcolor{jku_grey!10}Mean \% active per assay &    \cellcolor{jku_grey!10}45.67 &    \cellcolor{jku_grey!10}76.51 &    \cellcolor{jku_grey!10}50.61 &      \cellcolor{jku_grey!10}3.51 &    \cellcolor{jku_grey!10}56.76 &     \cellcolor{jku_grey!10}6.24 &      \cellcolor{jku_grey!10}2.39 &     \cellcolor{jku_grey!10}5.34 &     \cellcolor{jku_grey!10}7.52 \\
                       Median \% active per assay &                            45.67 &                            76.51 &                            50.61 &                              3.51 &                            66.29 &                             4.61 &                              1.28 &                             4.52 &                             5.16 \\
\bottomrule
\end{tabular}
    \label{sec:downstream_dsets}
\end{table*}

\paragraph{BACE.}
BACE contains $\sim$1.5 molecules and their 
bioactivity measurements 
for inhibition of human$\beta$-secretase 1 (BACE-1).
The bioactivity values are an aggregate of scientific literature 
and not from a single bioassay. 
Scaffold splitting is recommended and used.

\paragraph{BBBP.}
The Blood–brain barrier penetration (BBBP) dataset, original form 
\citet{martins2012bbbp},  contains $\sim$2k molecules and activity values 
for whether a molecule is able to pass a highly selective membrane and 
enter into the brain fluid. This is a vital physiological function an has 
implications for drug-design.
Scaffold splitting is recommended and used.
We note that some molecules are duplicates and have conflicting 
measurements, but those remain in the minority.

\paragraph{ClinTox.}
The Clinical Toxicity (ClinTox) \citep{artemov2016clintox, gayvert2016clintox} contains 
two bioactivity prediction tasks: 
(1) FDA-approval and (2) failed clinical trials for toxicity reasons.
The dataset contains $\sim$58k molecules, 
we find that molecules have multiple 
measurements. We also hypothesize that zero-imputation was used for 
unknown values, since the activity matrix does not contain missing values.
Predicting if a molecule will fail clinical trials or whether it be 
approved by the FDA is a difficult task because of the
high variability of the labeling procedure, which arises from 
genetic variability, variability assessment procedure, and environmental
variability.
Random splitting is recommended, but we chose the more challenging 
scaffold splitting scheme.

We found that SMILES-based models performed best on this task 
and hypothesize that there might be an artifact in the 
representations of the SMILES-strings. 
We found that after standardization of the SMILES-strings 
(Sec.~\ref{appsec:data_overlap}) of molecules, the metric 
$\Delta$AP drops significantly, e.g. for CDDD from 33.82 to 8.85\%.
Due to comparability, we report results for the default representation of molecules.

\paragraph{HIV.}
This dataset was introduced by \href{https://wiki.nci.nih.gov/display/NCIDTPdata/AIDS+Antiviral+Screen+Data}{Drug Therapeutics Program AIDS Antiviral Screen}, contains $\sim$40k molecules, and measures evidence of anti-HIV activity.
The original source also contains moderately active molecules, which were classified as inactive by the MoleculeNet authors for this classification task. There are several bioassays in PubChem that are highly related to this assay e.g. AID 179, because inhibitory activity against HIV is 
a highly researched area. We removed standardized molecules from pre-training for the downstream experiments.

\paragraph{SIDER.}
The SIDe Effect Resource (SIDER) dataset has been derived 
from a database of marketed drugs and 
their corresponding adverse drug reactions \citep{kuhn2016sider}.
The MoleculeNet version uses a subset of side effects and contains 27 tasks, which correspond to system organ classes.
Although a random split is recommended, we opt for
the more challenging scaffold splitting scheme.

\paragraph{Tox21.}
A public database assessing toxicity was made available by the 
Toxicology in the 21st Century (Tox21) initiative. 
A subset of the screening data was used in the Tox21 Data Challenge
in 2014 \citep{mayr2016deeptox}. The dataset poses
12 prediction tasks of different toxic effects, such as 
heat-shock response and DNA damage. 
The Tox21 dataset in MoleculeNet made several changes to the original
Challenge dataset, such as zero-imputation for missing data. 
Despite these changes introduced by MoleculeNet, we used the Tox21
as provided with scaffold splitting due to comparability.

\paragraph{ToxCast.}
The ToxCast dataset provides toxicological information for a sizable drug 
library based on in vitro high-throughput screening. This dataset was also 
part of the same program as Tox21. The qualitative outcomes of more 
than 600 studies on 8,615 chemicals are included \citep{richard2016toxcast}.
Random splitting is recommended by MoleculeNet, but again we use 
the more challenging scaffold splitting. 
Note that for the scaffold split, 9 tasks only have 1 class in the test-set. Since the metrics cannot be computed, they are omitted from the mean.

\paragraph{Tox21-10k.}
We also use an extended version of the Tox21 dataset that has 
additional toxicity prediction 
tasks \citep{richard2021tox21_10k, wu2021tox21trade} and 
~408k bioactivity measurements. 
Concretely Tox21-10k has 68 prediction tasks compared to 12 prediction
tasks in the original Tox21 dataset (see above). 
We use the original splitting of the dataset.

\subsubsection{Data Overlap Analysis}
\label{appsec:data_overlap}

To avoid biases in the evaluation procedure on downstream tasks, 
we avoid data leakage between pre-training data and the test-data. Data
leakage might occur if the bioactivity measurements of downstream 
tasks were present in the database on which pre-training is done. 
To this end, we made sure that these bioactivity measurements were removed
from the pre-training data.

We therefore standardized all molecules. We remove Hs, disconnect metal atoms, and normalize the molecule. The molecules are re-ionized and we correct for valence information as well as for ring information. We adapted the  \href{https://github.com/greglandrum/RSC_OpenScience_Standardization_202104/blob/main/MolStandardize\%20pieces.ipynb}{following steps} suggested by the RDKit \citep{landrum2013rdkit} author. 
The InChiKeys have been calculated to identify the molecules between the datasets.
All molecules that overlap with 
the downstream test set were removed from the pre-training dataset.
For example, 1,740 molecules were removed from the PubChem pre-train dataset based on the 784 molecules in Tox21-test set.

A total of 11,828 unique molecules were removed from pre-training. %

\subsubsection{Data splitting procedures}
\label{appsec:data_splits}

\paragraph{Assay and Molecule temporal split time\_a\_c.}
We conduct a temporal split \citep{sheridan2013time} to simulate
the situation in which new molecules and bioassays 
have to be predicted. We approximate this effect by assuming 
that new molecules and bioassays receive increasingly 
larger identifiers \citep{kim2019pubchem}. 
We split both the unique PubChem Compound identifiers (CIDs) and 
the unique PubChem bioassay identifiers (AIDs) into the oldest 60\%, 
the following 20\%, and the most recent 20\%. 
Then, we take the bioactivity records corresponding to the 60\% 
oldest molecules and bioassays for training, 
the bioactivity records corresponding 
to the following 20\% of molecules and bioassays for validation, 
and the bioactivity records corresponding to the 20\% 
most recent molecules and bioassays for testing (Fig.~\ref{fig:datasplit}).
We train on assays from the first assay deposited in 2004 up to AID 602432 from 2012-03,
validate on assays up to AID 602433 
from 2013-11 and test assays from AID 602434 to AID 1259411 
from 2018-05. 
The training dataset contains 11,661 assays and 143M bioactivity measurements, 
the validation dataset  2,959 assays and 8M bioactivity measurements, 
and test 3,933 assays and 1.2M bioactivity measurements.

Our temporal splitting setting for both molecules and bioassays 
represents a challenging prediction problem. 
Molecule-bioassay pairs corresponding to older molecules tested 
on newer bioassays, will not be included in any of the splits. 
Thus, molecules that are well characterized by measurements
on bioactivity tasks in the training datasets are removed 
from evaluation. We favor this difficult setting of 
both new molecules and bioassays splits,  
to assess the evaluation of whether our proposed 
method both generalize to novel molecules and bioassays.

\paragraph{Assay temporal split time\_a.}
We again split assays in the same manner as in time\_a\_c, 
but we test and validate also on "known" molecules, that is, 
molecules that have been entered into PubChem up to this time.
These known molecules might be characterized by bioactivity
measurements on bioassays included in the training datasets
and thus might be easier to predict. 
The size of the test-set increases drastically from 
1.2M up to 11M measurements with 4,201 assays that can be used 
for testing. This split is also effective and 
applicable to the real situation of testing new assays.

\paragraph{High Throughput Screening hts-split.}
High throughput screenings (HTS) are bioassays that 
can test a large scale of molecules in an automated manner \citep{mayr2009novel}.
\citet{laufkotter2019combining} suggested a dataset based on PubChem based around HTS data. 
We adapt the same split as suggested to test zero-shot capabilities. 
Additionally, we benchmark against the proposed methods 
HTSFP and BaSH \citep{laufkotter2019combining}, 
two fingerprint-variants 
based on HTS data (Sec.~\ref{appsec:hts_benchmarking}).

\paragraph{Scaffold split.}
Chemical datasets are characterized by strong biases, such 
as the compound series bias \citep{mayr2018large}. The compound
series bias arises from the fact that molecules are often generated
by adding functional groups to a scaffold molecule. This leads
to clusters of highly similar molecules being present in a chemical 
database. As a further consequence, performance estimation on 
a test set of molecules that were randomly selected from the dataset
usually overestimated the predictive quality \citep{mayr2018large}. 
To counter this bias, the data is often split by chemical 
clusters or scaffolds, called "scaffold split" \citep{wu2018moleculenet}. 

We use the scaffold split for several downstream datasets. 
This splitting procedure is one of the standard settings 
in MoleculeNet \citep{wu2018moleculenet,bemis1996properties}. 
We use the exact split provided by the framework, for comparability, 
if not stated otherwise.

\paragraph{FS-Mol split.}
The FS-Mol split corresponds to the data splits 
in \citep{stanley2021fs}. The version that is used is fsmol-0.1.  FS-Mol splits the 
data into training, validation, and test tasks, which
corresponds to splitting the bioassays. Bioassays 
are not split by a temporal procedure as in 
our PubChem splits (see above), but are randomly
distributed across training and test (stratified by protein class). We evaluate
the compared methods on the test tasks of FS-Mol, but in a zero-shot setting, where no support-set molecules are drawn.

\subsubsection{Assay Description}
\paragraph{Processing.}
For PubChem, we chose to use the title of the bioassay and
the descriptions provided by the database \citep{kim2019pubchem}.
The descriptions from FS-Mol have been added based on their ChEMBL ID, 
and multiple attributes have been used, such as: 
assay-organism, confidence-description or assay\_type, if present.
The textual descriptions for MoleculeNet datasets have been used in the zero-shot out-
of-domain experiments found in Sec.~\ref{appsec:zero_shot_experiment}. 
We hand-crafted the textual descriptions from information from multiple sources. 
Currently, a field of research called prompt engineering is devoted on how to optimize such task descriptions,
often called text prompts \citep{liu2022design}. 
In our case, only one prompt per task is used, and we leaf it 
to further work to optimize the assay descriptions.

\begin{figure}[htb]
     \centering
    \includegraphics[width=\columnwidth]{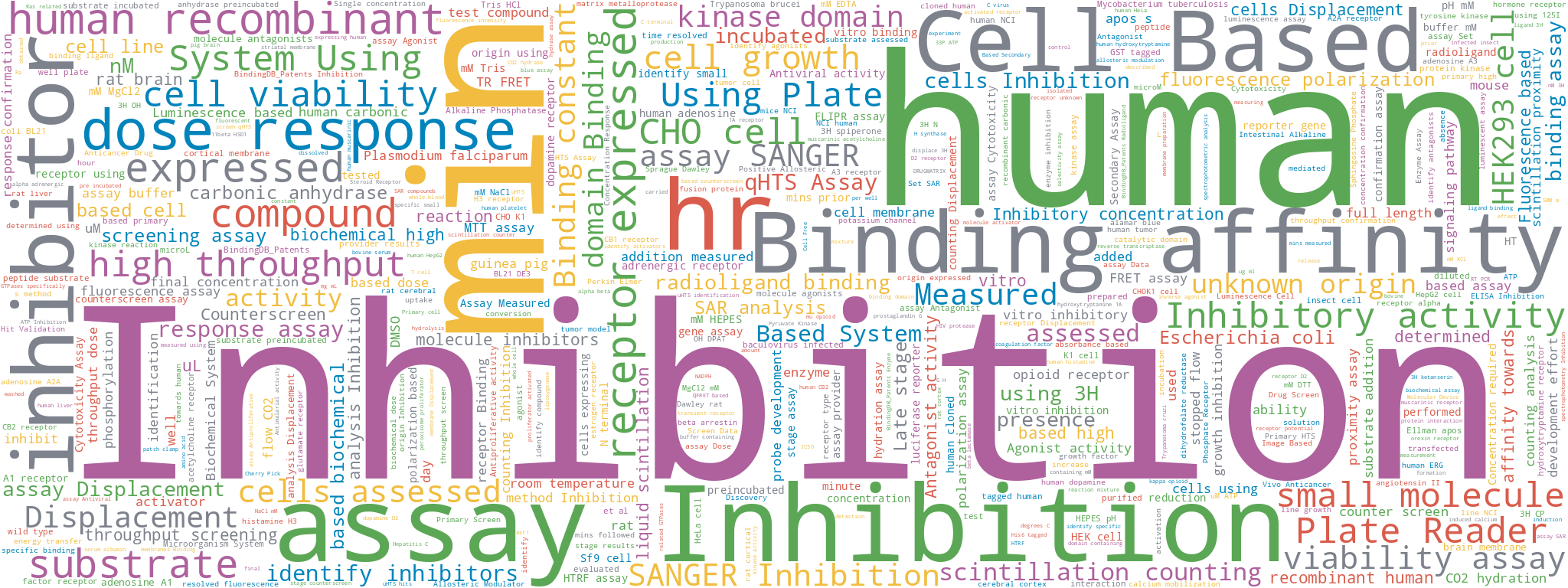}
     \caption{Wordcloud of PubChem Assay descriptions.}
     \label{fig:appendix:wordcloud}
\end{figure}

\paragraph{Examples for assay descriptions.}
In this section, we provide examples of bioassay descriptions, 
which we sampled randomly from PubChem, FS-Mol and MoleculeNet
(see Tab.~\ref{tab:assay_descriptions}).
These bioassay descriptions usually 
use very compact language and
contain terminology from molecular biology or chemistry 
and, thus, are very dissimilar
from the language on which language models are usually trained
\citep{devlin2019bert}.
Fig.~\ref{fig:appendix:wordcloud} shows a Wordcloud of assay-descriptions for PubChem.

\begin{table*}[]
    \centering
    \begin{tabular}{lp{\textwidth-1.8cm}}
\toprule
index & PubChem \\
\midrule
 \cellcolor{jku_grey!10}8568 & \cellcolor{jku_grey!10}Antiviral activity against pseudotype HIV1 JRFL infected in human HeLa67 cells expressing CD4/CCR5 after 3 days by luciferase reporter gene assay \\
16956 & SANGER: Inhibition of human MHH-PREB-1 cell growth in a cell viability assay. \\
 \cellcolor{jku_grey!10}8871 &\cellcolor{jku_grey!10}Screen and Counter Screen to Identify Novel Compounds that Selectively Sensitize Mycobacterium Tuberculosis to Beta-lactam Antibiotics \\
 4871 & In vitro inhibitory concentration against HIV-1 reverse transcriptase using rC-dG template primer \\
\cellcolor{jku_grey!10}17467 &\cellcolor{jku_grey!10}A549 Cytotoxicity Assay Measured in Cell-Based System Using Plate Reader - 7071-06\_Inhibitor\_Dose\_DryPowder\_Activity\_Set11 \\
18903 &Inhibition of human Cav1.3 channel in human SH-SY5Y cells assessed as 70 mM K+ induced calcium elevation compound treated 15 mins before stimulus by Fluo-4/AM assay \\
 \cellcolor{jku_grey!10}1194 &\cellcolor{jku_grey!10}Screen for compounds that decrease glutamate induced motor neuron death measured by TUNEL staining for DNA degradation (MNGlu) \\
482 &Dose Response Cell Based Assay for Antagonists of the 5-Hydroxytryptamine Receptor Subtype 1E (5HT1E \\
\cellcolor{jku_grey!10}17241 &\cellcolor{jku_grey!10}HepG2 Cytotoxicity Assay Measured in Cell-Based System Using Plate Reader - 7071-02\_Inhibitor\_Dose\_DryPowder\_Activity\_set2 \\
 9444 & Inhibition of p38 alpha \\
\bottomrule
\end{tabular}

\vspace{2mm}

\begin{tabular}{lp{\textwidth-1.8cm}}
\toprule
     index &  FS-Mol \\
\midrule
2175 &  assay\_organism: Homo sapiens confidence\_description: Direct single protein target assigned relationship\_description: Direct protein target assigned description: PUBCHEM\_BIOASSAY: MITF Measured in Cell-Based System Using Plate Reader - 2084-01\_Inhibitor\_DoseNoFile\_CherryPick\_Activity\_Set4. (Class of assay: confirmatory) [Related pubchem assays (depositor defined):AID488944] target\_chembl\_id: CHEMBL1741165 \\
\cellcolor{jku_grey!10}1850 &   \cellcolor{jku_grey!10} confidence\_description: Homologous single protein target assigned relationship\_description: Homologous protein target assigned description: Allosteric enhancer activity score measured by its ability to stabilize the agonist-receptor-G protein ternary complex at a concentration of 100 uM target\_chembl\_id: CHEMBL226 \\
1246 &  confidence\_description: Homologous single protein target assigned relationship\_description: Homologous protein target assigned description: Inhibition of uridine phosphorylase (UrdPase) from murine liver. target\_chembl\_id: CHEMBL3718 \\
\cellcolor{jku_grey!10}2196 &    \cellcolor{jku_grey!10} assay\_test\_type: In vitro assay\_test\_type: In vitro confidence\_description: Homologous single protein target assigned relationship\_description: Homologous protein target assigned description: Binding constant for TNK2 kinase domain target\_chembl\_id: CHEMBL4599 \\
\bottomrule
\end{tabular}

\vspace{2mm}

\begin{tabular}{lp{\textwidth-1.8cm}}
\toprule
 index & MoleculeNet - BBBP \\
\midrule
     0 & BBBP: Binary labels of blood-brain barrier penetration (permeability). \\
\bottomrule
\end{tabular}

\vspace{2mm}

\begin{tabular}{lp{\textwidth-1.8cm}}
\toprule
  index & MoleculeNet - SIDER \\
\midrule
     18 &   Infections and infestations, drug side effect \\
\cellcolor{jku_grey!10}0 & \cellcolor{jku_grey!10}Hepatobiliary disorders, drug side effect \\
     22 &  Pregnancy, puerperium and perinatal conditions, drug side effect \\
\cellcolor{jku_grey!10}4 &    \cellcolor{jku_grey!10}Investigations, drug side effect \\
\bottomrule
\end{tabular}

\vspace{2mm}

\begin{tabular}{lp{\textwidth-1.8cm}}
\toprule
  index & MoleculeNet - Tox21 \\
\midrule
2 &     qHTS assay to identify small molecule agonists of the androgen receptor (AR) signaling pathway: Summary \\
\cellcolor{jku_grey!10}9 &  \cellcolor{jku_grey!10}qHTS assay for small molecule activators of the heat shock response signaling pathway: Summary \\
0 & qHTS assay to identify small molecule agonists of the androgen  receptor (AR) signaling pathway using the MDA cell line \\
\cellcolor{jku_grey!10}6 & \cellcolor{jku_grey!10}qHTS assay to identify small molecule agonists of the peroxisome proliferator-activated receptor gamma (PPARg) signaling pathway: Summary \\
\bottomrule
\end{tabular}
    \caption{Examples for bioassay descriptions. \label{tab:assay_descriptions}}
    
\end{table*}

\subsection{Metrics}
\label{app:metric}
In this work, we report four performance metrics, which we denote as AUROC, AP, $\Delta$AP, and $\neg$AP. 
AUROC is the area under the ROC curve. AP is the mean average precision, which is an approximation of the area under the precision-recall curve. $\Delta$AP additionally subtracts the base-rate, portion of positive-samples in the test-set from AP for each task. The last metric, which we dub $\neg$AP for negative-class mean average precision, is not standard, but it is very informative. It is simply the mean average precision of the negative class. That is, if the negative class is coded as 0 and the positive class is coded as 1, then NegAVGP is, 
\begin{align}
\neg \text{AP} &= \text{AP} (1 - y, 1 - \hat{y}),
\end{align}
where $\hat y$ are the predictions and $y$ is the true label. 
For retrieval tasks, we use the top-$k$ accuracy as well as the enrichment-factor EF \citep{friesner2004glide}.
Top-$k$ accuracy measures how often the active compound is within the $k$ suggested, ranked by the method.
Top-$k$\% is the top-$k$ accuracy divided by the number of total choices for the method.
EF normalizes top-$k$ accuracy by the score of a random choice. 

\paragraph{Choice of main metric.}
\label{appsec:metric}
Typically AUROC is used to compare different methods \citep{wu2018moleculenet}. It measures the area under the true-positive rate vs the false-positive rate for different thresholds and can be regarded as a general sorting ability of a model. A valuable property of this metric is that a random model will produce an AUROC of 0.5.
However, for imbalanced datasets, AUROC should be used with caution because it weighs the more frequent class higher \citep{zhao2009statistical}.
In drug discovery, there is a bigger emphasis on finding active compounds which are typically rare. 
This is much more reflected by the Area under the Precision-Recall Curve (AUPRC) (aka PRCAUC, equiv. to AP).
AP is highly dependent on the base rate of the task, 
and averaging over many tasks biases towards imbalanced tasks. $\Delta$AP subtracts the base-rate and for a random classifier is at 0.
The metric is also adopted as the main metric in the FS-Mol benchmark (termed $\Delta$AUPRC there) \citep{stanley2021fs}.
Alternatively, BedROC \citep{truchon2007vs_metric} might also be used (see Tab.~\ref{tab:appendix:zeroshot_full}). The metric has additional parameters such as $\alpha$.

\clearpage

\subsection{Details on Contrastive Language-Assay-Molecule Pre-training (CLAMP)}
\label{appsec:clamp}

\subsubsection{Molecule encoders}
Programmatically we handle molecules as
SMILES strings \citep{weininger1988smiles},
and convert them to different other representations 
from which an encoding can be generated.
Several encodings of molecules were considered, which  
we categorize into a) fingerprint-based encoding and 
b) NN-based encoding. 
Encodings can also be combined for modeling. 

\paragraph{Fingerprint-based.}
We use the Python\footnote{\url{https://www.python.org}} API of the 
RDKit\footnote{\url{http://www.rdkit.org}} open-source 
chemoinformatics software to extract fingerprints.
The sparse fingerprint (FP) \textbf{sprsFP} is a concatenation of 
Daylight-like FPs, 
Morgan fingerprints  and MACCS keys. We remove fingerprint features
with low variance across training samples and finally 
end up with a vector of size 2176.
\textbf{Mc+RDKc} is an addition of fingerprint-counts 
from the Morgan fingerprint and the RDKit-FP. We additionally scale the counts by $\ln (1+c)$.

\paragraph{NN-based.}
We included the following NN-based embeddings, which were
pre-trained in a self-supervised fashion: 
\textbf{Grover} \citep{rong2020grover}, a graph transformer, 
\textbf{CDDD} \citep{winter2019cddd}, a SMILES-LSTM based autoencoder, 
\textbf{MFBERT} \citep{abdel2022mfbert} a SMILES-Transformer.
The embeddings from these neural networks are 
projected by a linear layer into 
$d$-dimensional embeddings. 
Further details on the network architecture and the choice of 
the dimension $d$ are provided in Sec.~\ref{appsec:ffnn}.
Results for different embeddings can be seen in Tab.~\ref{tab:compound_modes}.

\subsubsection{Assay encoders}

For each unique bioassay in the bioactivity dataset, 
we retrieve a textual description of the bioassay, which 
includes both the bioassay title and, if available, a more detailed description.
We process each textual description to obtain a 
fix-length bioassay vector. We follow 
two different pipelines, which we then assess separately.

\paragraph{Term-based.}
We pre-train a Latent Semantic Analysis \textbf{LSA} \citep{deerwester1990indexing} 
model on bioassay descriptions. 
In the case of PubChem, we obtain textual descriptions for 1,252,874 bioassays. 
To avoid leaking information, we exclude those corresponding to bioassays present 
in the validation and test splits of the bioactivity dataset. 
To train the LSA model, we first compute a bioassay-term matrix of tf-idf coefficients 
and then compute its truncated SVD decomposition. 
Finally, we extract the LSA embedding vectors for all the bioassay descriptions 
in our bioactivity dataset. The LSA embedding vectors have dimension 
2048 in the case of PubChem and 128 in the case of FS-Mol.
In the case of FS-Mol we additionally obtain different attributes for each assay. 
A collection of attributes is one-hot encoded leading to an encoding vector 
of size 355 termed \textbf{category}. A further simplistic encoding-vector 
in the case of FS-Mol, is a 2-dimensional vector describing if the company \textbf{millipore} 
has conducted the experiment. The FH baseline can be seen as a special case of the 
architecture in which a \textbf{constant} is used as encoding.

\paragraph{NN-based.}
We use pre-trained model instances of several publicly available language models.
We consider a \textbf{BioBERT} model \citep{lee2020biobert},\footnote{\url{https://huggingface.co/dmis-lab/biobert-large-cased-v1.1}} which uses a 
transformer architecture \citep{vaswani2017attention} and 
has been trained on biomedical text corpora. 
Each bioassay description is provided as input 
to BioBERT and we keep the activations at the 
last layer as the bioassay vector and apply mean pooling. 
These are of dimension 1024.
We further use the text-encoder part of \textbf{CLIP} \citep{radford2021learning}, 
which has been trained on a dataset of 400M image, text pairs from 
the internet. The output of the model is 768 dimensional.
Sentence-T5 \textbf{sT5} \citep{ni2021sentencet5} is a transformer variant based on T5 \citep{raffel2020exploring}, 
trained to summarize sentences and for textual similarity. 
We use the XXL variant with 11b parameters available at huggingface \citep{jain2022hugging}.

A feed-forward neural network takes the assay encodings as 
input and projects them to $d$-dimensional embeddings. 
Further details on the network architecture and 
the choice of the dimension $d$ are provided in Sec.~\ref{appsec:hyperparams}.
Results for different assay encodings can be seen in Tab.~\ref{tab:assay_modes}. 
We found a combination of LSA and CLIP to work well in our setting.

\subsubsection{Feed forward neural networks}
\label{appsec:ffnn}

The molecule and the bioassay encoders process their input vectors 
using each a feed-forward neural network. The network architecture 
on each encoder can be different, except for the 
output dimensionality $d$, which maps into the 
shared multi-modal embedding space.

The input and hidden layers of the fully-connected networks have the following structure
\begin{align}
    \Bh^{l+1}&=\text{dropout} \Big( \text{ReLU} \big(\text{norm}(\BW \Bh^{l} + \Bb) \big) \Big),
\end{align}
where $\Bh^{l}$ is the input to the layer and
$\Bh^{l+1}$ is its output. The variables 
$\BW$ and $\Bb$ are learnable weights. The preactivations 
are followed by batch normalization~\citep{ioffe2015batch} or layer 
normalization~\citep{ba2016layer}, a rectified linear unit (ReLU) 
activation function, and dropout~\citep{srivastava2014dropout}. The 
output layer does not have normalization, activation function, nor 
dropout, as it directly serves as the molecule or the bioassay embedding.

\subsubsection{Hyperparameters}
\label{appsec:hyperparams}
Models were selected by conducting a manual hyperparameter search 
(Tab.~\ref{table:hyperparams_explored}). We explored different configurations 
for the embedding dimension $d$, the learning rate, the number of layers, 
the number of hidden units in each layer, and the dropout probability. 
We also experimented with the parameter~$\tau$, 
necessary for the scoring function (Eq.~\ref{eq:similarity}),
being set to 1 or learned, and with using either batch or layer normalization.%

\begin{table*}
\centering

\begin{threeparttable}
   \caption{Hyperparameter settings explored during model selection}
   \label{table:hyperparams_explored}
   
   \begin{tabular}{lp{\textwidth-5cm}}
\toprule
\textbf{Hyperparameter}&\textbf{Explored values} \\
\midrule

  assay encoding &  -, BioBERT, LSA, category, CLIP, millipore, sT5 \\
  \cellcolor{jku_grey!10}batch\_size &   \cellcolor{jku_grey!10}12, 32, 128, 256, 512, 1024, 2048, 4096, 32768 \\
 $\tau$ & 1.0, learned \\
  \cellcolor{jku_grey!10}molecule encoding &\cellcolor{jku_grey!10} CDDD, CLIP, Graphormer, GROVER, MFBERT, Morganc+RDKc, sprsFP \\
 dropout\_hidden &0.05, 0.1, 0.15, 0.2, 0.3, 0.4 \\
 \cellcolor{jku_grey!10}dropout\_input & \cellcolor{jku_grey!10}0.0, 0.1 \\
 embedding\_size &  128,  256,  512, 768, 1024, 2048  \\
 \cellcolor{jku_grey!10}epoch\_max &  \cellcolor{jku_grey!10} 1, 2, 3, 4, 5, 7, 10, 50\\
  hidden\_layers &[1024, 1024], [2048, 1024], [4096, 2048, 1024], [4096, 2048], [4096, 4096, 2048], [4096], [4098, 4098], [512, 512] \\
\cellcolor{jku_grey!10}lr\_factor & \cellcolor{jku_grey!10}  0.95, 1.0 \\
lr\_ini & 0.0001, 1e-05, 2e-05, 3e-05, 4e-05, 5e-05 \\
\cellcolor{jku_grey!10}model\_class &\cellcolor{jku_grey!10}Galactica, FH, KVPLM, MLPLayerNorm, MultitaskMLPLayerNorm, ScaledMLPLayerNorm \\
multitask\_temperature & 1, 100, 10e6, nan \\
\cellcolor{jku_grey!10}nonlinearity & \cellcolor{jku_grey!10}  ReLU \\
optimizer & Adam, AdamW \\
 patience &1, 3, 5, 10 \\
\bottomrule
\end{tabular}
   
   \begin{tablenotes}
   \tiny
        \item[] dropout\_hidden: Dropout rate at the hidden layers
        \item[] dropout\_input: Dropout rate at the input layer
        \item[] lr\_factor: cosine annealing lr-schedule factor
        \item[] lr\_ini: initial learning-rate
        \item[] embedding\_size: association dimension $d$
        \item[] multitask\_temperature: softmax temperature for 1-/soft-NN baseline
        \item[] patience: nr. of epochs to continue training after best optimization score has been reached
        \end{tablenotes}
   \end{threeparttable}
  
\end{table*}

\begin{table*}
   \centering
   \tiny
   \caption{Hyperparameter settings selected based on validation performance.}
   \label{table:hp_selected}
   \begin{threeparttable}
   \begin{tabular}{lp{0.5cm}p{0.5cm}p{0.5cm}p{0.5cm}|p{0.5cm}p{0.5cm}p{0.5cm}p{0.5cm}|p{0.5cm}p{0.5cm}p{0.5cm}p{0.5cm}|p{0.5cm}p{0.5cm}p{0.5cm}p{0.5cm}}
\toprule
                               \textbf{Hyperparameter}& \multicolumn{4}{l}{FSMOL} & \multicolumn{12}{l}{PubChem} \\
                                      & \multicolumn{4}{l}{default} & \multicolumn{4}{l}{hts} & \multicolumn{4}{l}{time\_a} & \multicolumn{4}{l}{time\_a\_c} \\
                                      &                      1-NN  &                                    CLAMP &                     Frequent hitters &                   soft-NN  &                                1-NN  &                                              CLAMP &                     Frequent hitters &                             soft-NN  &                                1-NN  &                                CLAMP &                               Frequent hitters &                   soft-NN  &                                1-NN  &                                              CLAMP &                               Frequent hitters &                             soft-NN  \\
\midrule
   \cellcolor{jku_grey!10}assay encoding &          \cellcolor{jku_grey!10}LSA & \cellcolor{jku_grey!10}clip\& lsa &            \cellcolor{jku_grey!10}- &          \cellcolor{jku_grey!10}LSA &                    \cellcolor{jku_grey!10}LSA & \cellcolor{jku_grey!10}clip\& lsa\&sT5 &            \cellcolor{jku_grey!10}- &                    \cellcolor{jku_grey!10}LSA &                    \cellcolor{jku_grey!10}LSA &    \cellcolor{jku_grey!10}clip\& lsa &                      \cellcolor{jku_grey!10}- &    \cellcolor{jku_grey!10}clip\& lsa &                    \cellcolor{jku_grey!10}LSA & \cellcolor{jku_grey!10}clip\& lsa\&sT5 &                      \cellcolor{jku_grey!10}- &              \cellcolor{jku_grey!10}clip\& lsa \\
                           batch\_size &                               512 &                                   256 &                               512 &                               512 &                                         512 &                                           2,048 &                             1,024 &                                       4,096 &                                         512 &                             1,024 &                                         512 &                             1,024 &                                       1,024 &                                          32,768 &                                         512 &                                         512 \\
\cellcolor{jku_grey!10}molecule encoding & \cellcolor{jku_grey!10}mc+rdkc &     \cellcolor{jku_grey!10}mc+rdkc & \cellcolor{jku_grey!10}mc+rdkc & \cellcolor{jku_grey!10}mc+rdkc & \cellcolor{jku_grey!10}sprsFP &               \cellcolor{jku_grey!10}mc+rdkc & \cellcolor{jku_grey!10}mc+rdkc & \cellcolor{jku_grey!10}sprsFP & \cellcolor{jku_grey!10}sprsFP & \cellcolor{jku_grey!10}mc+rdkc & \cellcolor{jku_grey!10}sprsFP & \cellcolor{jku_grey!10}mc+rdkc & \cellcolor{jku_grey!10}sprsFP &               \cellcolor{jku_grey!10}mc+rdkc & \cellcolor{jku_grey!10}sprsFP & \cellcolor{jku_grey!10}sprsFP \\
                       dropout\_hidden &                                 0.20 &                                     0.40 &                                 0.20 &                                 0.20 &                                           0.20 &                                               0.30 &                                 0.20 &                                           0.20 &                                           0.20 &                                 0.20 &                                           0.20 &                                 0.20 &                                           0.30 &                                               0.30 &                                           05 &                                           0.20 \\
\cellcolor{jku_grey!10}dropout\_input &          \cellcolor{jku_grey!10}0 &              \cellcolor{jku_grey!10}0.1 &          \cellcolor{jku_grey!10}0 &          \cellcolor{jku_grey!10}0 &                    \cellcolor{jku_grey!10}0 &                        \cellcolor{jku_grey!10}0.1 &          \cellcolor{jku_grey!10}0.1 &                    \cellcolor{jku_grey!10}0 &                    \cellcolor{jku_grey!10}0 &          \cellcolor{jku_grey!10}0.1 &                    \cellcolor{jku_grey!10}0 &          \cellcolor{jku_grey!10}0.1 &                    \cellcolor{jku_grey!10}0.1 &                        \cellcolor{jku_grey!10}0.1 &                    \cellcolor{jku_grey!10}0.1 &                    \cellcolor{jku_grey!10}0 \\
                       embedding\_size &                             1,024 &                                   512 &                             1,024 &                             1,024 &                                       1,024 &                                             768 &                             1,024 &                                       1,024 &                                       1,024 &                               768 &                                       1,024 &                               768 &                                         768 &                                             768 &                                         256 &                                       1,024 \\
    \cellcolor{jku_grey!10}epoch\_max &         \cellcolor{jku_grey!10}50 &             \cellcolor{jku_grey!10}50 &         \cellcolor{jku_grey!10}50 &         \cellcolor{jku_grey!10}50 &                   \cellcolor{jku_grey!10}50 &                        \cellcolor{jku_grey!10}50 &         \cellcolor{jku_grey!10}50 &                    \cellcolor{jku_grey!10}50 &                   \cellcolor{jku_grey!10}50 &         \cellcolor{jku_grey!10}50 &                   \cellcolor{jku_grey!10}50 &         \cellcolor{jku_grey!10}50 &                   \cellcolor{jku_grey!10}50 &                        \cellcolor{jku_grey!10}50 &                    \cellcolor{jku_grey!10}50 &                   \cellcolor{jku_grey!10}50 \\
                        hidden\_layers &                         [4096, 2048] &                             [2048, 1024] &                         [4096, 2048] &                         [4096, 2048] &                                   [4096, 2048] &                                       [4096, 2048] &                         [2048, 1024] &                                   [4096, 2048] &                                   [4096, 2048] &                   [4096, 2048, 1024] &                                   [4096, 2048] &                         [4096, 2048] &                                   [4096, 2048] &                                       [4096, 2048] &                                   [4096, 2048] &                                   [4096, 2048] \\
   \cellcolor{jku_grey!10}layer\_norm &         \cellcolor{jku_grey!10}True &             \cellcolor{jku_grey!10}True &        \cellcolor{jku_grey!10}False &         \cellcolor{jku_grey!10}True &                   \cellcolor{jku_grey!10}True &                       \cellcolor{jku_grey!10}True &        \cellcolor{jku_grey!10}False &                   \cellcolor{jku_grey!10}True &                   \cellcolor{jku_grey!10}True &         \cellcolor{jku_grey!10}True &                  \cellcolor{jku_grey!10}False &         \cellcolor{jku_grey!10}True &                   \cellcolor{jku_grey!10}True &                       \cellcolor{jku_grey!10}True &                  \cellcolor{jku_grey!10}False &                   \cellcolor{jku_grey!10}True \\
                            lr\_factor &                                 1 &                                     1 &                                 1 &                                 1 &                                           1 &                                               1 &                                 1 &                                           1 &                                           1 &                                 1 &                                           1 &                                 1 &                                           1 &                                               1 &                                           1 &                                           1 \\
       \cellcolor{jku_grey!10}lr\_ini &        \cellcolor{jku_grey!10}1e-05 &            \cellcolor{jku_grey!10}1e-05 &        \cellcolor{jku_grey!10}1e-05 &        \cellcolor{jku_grey!10}1e-05 &                  \cellcolor{jku_grey!10}1e-05 &                      \cellcolor{jku_grey!10}4e-05 &        \cellcolor{jku_grey!10}5e-05 &                  \cellcolor{jku_grey!10}1e-05 &                  \cellcolor{jku_grey!10}1e-05 &        \cellcolor{jku_grey!10}5e-05 &                  \cellcolor{jku_grey!10}1e-05 &        \cellcolor{jku_grey!10}5e-05 &                  \cellcolor{jku_grey!10}5e-05 &                      \cellcolor{jku_grey!10}5e-05 &                  \cellcolor{jku_grey!10}5e-05 &                  \cellcolor{jku_grey!10}1e-05 \\
                MT\_temperature &                        10e6 &                                      NaN &                                  NaN &                                 1 &                                  10e6 &                                                NaN &                                 1 &                                           1 &                                  10e6 &                                  NaN &                                            NaN &                                 1 &                                  10e6 &                                                NaN &                                            NaN &                                           1 \\
     \cellcolor{jku_grey!10}patience &          \cellcolor{jku_grey!10}3 &              \cellcolor{jku_grey!10}3 &          \cellcolor{jku_grey!10}5 &          \cellcolor{jku_grey!10}3 &                    \cellcolor{jku_grey!10}3 &                        \cellcolor{jku_grey!10}3 &          \cellcolor{jku_grey!10}3 &                    \cellcolor{jku_grey!10}3 &                    \cellcolor{jku_grey!10}3 &          \cellcolor{jku_grey!10}3 &                    \cellcolor{jku_grey!10}3 &          \cellcolor{jku_grey!10}3 &                    \cellcolor{jku_grey!10}3 &                        \cellcolor{jku_grey!10}3 &                    \cellcolor{jku_grey!10}1 &                    \cellcolor{jku_grey!10}3 \\
                               $\tau$ &                                1 &                                     1 &                                1 &                                1 &                                          1 &                                               1 &                                1 &                                          1 &                                          1 &                                 1 &                                          1 &                                1 &                                          1 &                                               1 &                                          1 &                                          1 \\
\cellcolor{jku_grey!10}warmup\_epochs &          \cellcolor{jku_grey!10}2 &              \cellcolor{jku_grey!10}2 &          \cellcolor{jku_grey!10}0 &          \cellcolor{jku_grey!10}2 &                    \cellcolor{jku_grey!10}2 &                        \cellcolor{jku_grey!10}2 &          \cellcolor{jku_grey!10}2 &                    \cellcolor{jku_grey!10}2 &                    \cellcolor{jku_grey!10}2 &          \cellcolor{jku_grey!10}2 &                    \cellcolor{jku_grey!10}2 &          \cellcolor{jku_grey!10}2 &                    \cellcolor{jku_grey!10}2 &                        \cellcolor{jku_grey!10}2 &                    \cellcolor{jku_grey!10}1 &                    \cellcolor{jku_grey!10}2 \\
\bottomrule
\end{tabular}
      \begin{tablenotes}
   \tiny
        \item[] dropout\_hidden: Dropout rate at the hidden layers
        \item[] dropout\_input: Dropout rate at the input layer
        \item[] lr\_factor: cosine annealing lr-schedule factor
        \item[] lr\_ini: initial learning-rate
        \item[] embedding\_size: association dimension $d$
        \item[] multitask\_temperature: softmax temperature for 1-/soft-NN baseline
        \item[] patience: nr. of epochs to continue training after best optimization score has been reached
        \end{tablenotes}
   \end{threeparttable}
  
\end{table*}

 Model weights were initialized with MSRA~\citep{he2015delving}. 
 For each hyperparameter configuration, we optimized the objective function (Eq.~\ref{eq:loss}) 
 using AdamW \citep{loshchilov2017adamW}. For each hyperparameter configuration, 
 a copy of the model weights achieving the highest $\Delta AP$ on the respective
 validation set over max\_epochs training epochs was stored. 
 Upon analysis of the obtained validation metrics, we selected the final models~(Tab.~\ref{table:hp_selected}).

\subsection{Details on compared methods and baselines}
\paragraph{1-NN and soft-NN.}

We propose baselines, which could be considered two variants 
of multi-task deep networks (MT-DNN), for the purpose of making activity predictions for novel bioassays. Our baselines could, however, also be 
considered as new methods, as, to the best of our knowledge, they have not been suggested before.  
In both cases, for a target novel bioassay, we compute the 
cosine similarity between its text encoding and those of all the training 
bioassays, thus obtaining a vector of textual similarities. 
The first baseline method, 1-nearest neighbour (1-NN), predicts the bioactivity values that 
MT-DNN would predict for the training bioassay most similar to the target novel bioassay. 
If the training set of this reference is vastly different from the molecules that should be predicted, technically predictions can be made, but they will suffer from the usual decreased performance caused by the domain shift. In this scenario also an experienced chemist might find it hard to find a reference experiment. 

The second baseline, soft-nearest neighbours (soft-NN), is 
a smoother version of the first one. The vector of similarities between the 
target novel bioassay and the training bioassays is normalized using the 
softmax function, such that the resulting vector of weights sums up to one. 
Then, soft-NN predicts the weighted average of the values that MT-DNN 
would predict for all the training bioassays.

\paragraph{Frequent hitters (FH).}
A further baseline is the Frequent hitters (FH) model \citep{schimunek2023contextenriched}.
This baseline models the average activity of a molecule, regardless of the assay. 
This model can perform well because there are molecules that 
show up repeatedly as “active” across many different bioassays, and are also termed promiscuous molecules.
Several publications have described the phenomena \citep{roche2002development, gilberg2016highly, senger2016filtering, schuffenhauer2020evolution}, but it is missing as a baseline in publications related to few-and-zero-shot drug discovery \citep{stanley2021fs, chen2022meta}.
Frequent hitter molecules can be related to molecules which interfere 
with the measurement rather than the object of interest, 
so called Pan-Assay Interference Compounds (PAINS) \cite{baell2010new,baell2018seven}.

\paragraph{Hyperparameters.}
We trained a dedicated MT-DNN for each baseline model. Since our training-, validation-, and test-splits are bioassay-wise disjoint, we propose the following training procedure. Each MT-DNN visits the training-set as usual, but it is then evaluated on the (bioassay-wise disjoint) validation set by using its predictions directly as 1-NN or soft-NN. In this way, we can train MT-DNN models for our baselines using exactly the same splits and information that CLAMP used.

Given the results of the hyperparameter search conducted for CLAMP, we conducted a hyperparameter search where we explored different configurations for the learning rate, the number of layers, the number of hidden units in each layer, and the dropout probability~(Tab.~\ref{table:hyperparams_explored}). We set the parameter~$\tau$ to $1$ and used layer normalization.

Model weights were initialized with MSRA~\citep{he2015delving}. For each hyperparameter configuration, we optimized the multi-task masked loss~\citep{mayr2016deeptox} using AdamW~\citep{loshchilov2017adamW} with a batch size of 512 samples. For each hyperparameter configuration, a copy of the model weights achieving the highest validation $\Delta$AP  over epoch\_max training epochs was stored. Upon analysis of the obtained validation metrics, we selected the final models~(Tab.~\ref{table:hp_selected}). 

\newpage

\subsection{Details on the Zero-Shot Transfer experiment (5.1)}
\label{appsec:zero_shot_experiment}

\subsubsection{Details on compared SLMs.}

\paragraph{Galactica (GAL).}
\label{appsec:galactica}
Galactica (GAL) is a SLM, trained on a large scientific corpus of papers. Among other scientific tasks, it was trained on the structure of 20M molecules, as well as on a set of 44 assays from the MoleculeNet benchmark \citep{wu2018moleculenet}. We used the publicly available weights, and test the model. 
Currently, only one of the specifically designed prompts is known, in order to query the model about bioassays. We design our prompt similar to the one suggested in the paper in the form of:

\small{\texttt{"Here is a SMILES formula: [START\_SMILES]" + SMILES + "[END\_SMILES] Will the chemical molecule be active in the assay:" + ASSAY\_DESCRIPTION + "Answer (Yes or No):"}.}

Further, we take the logit at the position of the \texttt{"Yes"} token as ranking between molecules.

\begin{table*}[tb]
    \centering
    \caption{Zero-shot results for different metrics in \%. Green values indicate the highest values and yellow values within the std to the maximum value. Error bars represent standard deviations across the best five runs based on validation $\Delta$AP.
    \label{tab:appendix:zeroshot_full}}
    \small
    \begin{tabular}{lll|lllllll}
\toprule
metric& dataset &    split & random & GAL 125M$^\dagger$ &   KV-PLM$^\dagger$  & 1-NN &  soft-NN &  FH &CLAMP \\
\midrule
$\Delta$AP &  \cellcolor{jku_grey!10}FS-Mol & \cellcolor{jku_grey!10}default & \cellcolor{jku_grey!10}${01.57^{\pm{0.3}}}$ & \cellcolor{jku_grey!10}${01.44^{\pm{0.0}}}$ & \cellcolor{jku_grey!10}${01.84^{\pm{0.0}}}$ &   \cellcolor{jku_grey!10}${14.68^{\pm{0.7}}}$ & \cellcolor{jku_grey!10}${13.81^{\pm{1.8}}}$ &   \cellcolor{jku_grey!10}${18.50^{\pm{0.2}}}$ &  \cellcolor{jku_grey!10}\cellcolor{jku_green!20}${19.37^{\pm{0.2}}}$ \\
     & PubChem &     hts & ${00.01^{\pm{0.0}}}$ & ${00.00^{\pm{0.0}}}$ & ${00.10^{\pm{0.0}}}$ &   ${01.20^{\pm{0.1}}}$ & ${02.04^{\pm{0.4}}}$ &   ${03.10^{\pm{0.1}}}$ &  \cellcolor{jku_green!20}${08.43^{\pm{0.1}}}$ \\
 &  &  \cellcolor{jku_grey!10}time\_a & \cellcolor{jku_grey!10}${02.13^{\pm{0.3}}}$ & \cellcolor{jku_grey!10}${01.39^{\pm{0.0}}}$ & \cellcolor{jku_grey!10}${03.57^{\pm{0.0}}}$ &   \cellcolor{jku_grey!10}${12.96^{\pm{1.0}}}$ & \cellcolor{jku_grey!10}${05.67^{\pm{0.7}}}$ &   \cellcolor{jku_grey!10}${10.23^{\pm{0.5}}}$ &  \cellcolor{jku_grey!10}\cellcolor{jku_green!20}${14.77^{\pm{0.3}}}$ \\
 & &time\_a\_c & ${04.39^{\pm{0.5}}}$ & ${04.20^{\pm{0.0}}}$ & ${07.99^{\pm{0.0}}}$ & \cellcolor{jku_yellow!20}${11.11^{\pm{0.3}}}$ & ${06.99^{\pm{2.8}}}$ &   ${10.35^{\pm{0.9}}}$ &  \cellcolor{jku_green!20}${11.67^{\pm{0.6}}}$ \\
\midrule
 
    AUROC &  \cellcolor{jku_grey!10}FS-Mol & \cellcolor{jku_grey!10}default & \cellcolor{jku_grey!10}${50.24^{\pm{0.4}}}$ & \cellcolor{jku_grey!10}${50.50^{\pm{0.0}}}$ & \cellcolor{jku_grey!10}${50.56^{\pm{0.0}}}$ &   \cellcolor{jku_grey!10}${64.69^{\pm{0.8}}}$ & \cellcolor{jku_grey!10}${63.92^{\pm{1.9}}}$ &   \cellcolor{jku_grey!10}${68.22^{\pm{0.2}}}$ &  \cellcolor{jku_grey!10}\cellcolor{jku_green!20}${69.26^{\pm{0.2}}}$ \\
    & PubChem &     hts & ${49.92^{\pm{0.2}}}$ & ${49.32^{\pm{0.0}}}$ & ${49.65^{\pm{0.0}}}$ &   ${67.92^{\pm{0.8}}}$ & ${68.41^{\pm{0.9}}}$ &   ${73.48^{\pm{0.4}}}$ &  \cellcolor{jku_green!20}${73.83^{\pm{0.3}}}$ \\
&  &  \cellcolor{jku_grey!10}time\_a & \cellcolor{jku_grey!10}${50.08^{\pm{0.5}}}$ & \cellcolor{jku_grey!10}${47.05^{\pm{0.0}}}$ & \cellcolor{jku_grey!10}${54.92^{\pm{0.0}}}$ &   \cellcolor{jku_grey!10}${66.53^{\pm{0.6}}}$ & \cellcolor{jku_grey!10}${57.85^{\pm{1.7}}}$ &   \cellcolor{jku_grey!10}${66.77^{\pm{1.5}}}$ &  \cellcolor{jku_grey!10}\cellcolor{jku_green!20}${68.66^{\pm{0.5}}}$ \\
    &  &time\_a\_c & ${49.91^{\pm{0.4}}}$ & ${48.04^{\pm{0.0}}}$ & ${57.00^{\pm{0.0}}}$ &   ${61.98^{\pm{0.4}}}$ & ${55.06^{\pm{6.3}}}$ &   ${61.65^{\pm{0.8}}}$ &  \cellcolor{jku_green!20}${63.66^{\pm{0.4}}}$ \\
    \midrule
BEDROC &  \cellcolor{jku_grey!10}FS-Mol & \cellcolor{jku_grey!10}default & \cellcolor{jku_grey!10}${48.16^{\pm{0.6}}}$ & \cellcolor{jku_grey!10}${49.12^{\pm{0.0}}}$ & \cellcolor{jku_grey!10}${47.63^{\pm{0.0}}}$ &   \cellcolor{jku_grey!10}${67.53^{\pm{1.0}}}$ & \cellcolor{jku_grey!10}${66.04^{\pm{2.5}}}$ & \cellcolor{jku_grey!10}\cellcolor{jku_yellow!20}${73.30^{\pm{0.5}}}$ &  \cellcolor{jku_grey!10}\cellcolor{jku_green!20}${73.77^{\pm{0.6}}}$ \\
    & PubChem &     hts & ${05.31^{\pm{0.1}}}$ & ${05.21^{\pm{0.0}}}$ & ${05.96^{\pm{0.0}}}$ &   ${18.53^{\pm{1.2}}}$ & ${21.37^{\pm{0.7}}}$ &   ${27.27^{\pm{0.6}}}$ &  \cellcolor{jku_green!20}${32.02^{\pm{0.7}}}$ \\
&  &  \cellcolor{jku_grey!10}time\_a & \cellcolor{jku_grey!10}${34.44^{\pm{0.8}}}$ & \cellcolor{jku_grey!10}${33.23^{\pm{0.0}}}$ & \cellcolor{jku_grey!10}${36.00^{\pm{0.0}}}$ &   \cellcolor{jku_grey!10}${50.67^{\pm{1.1}}}$ & \cellcolor{jku_grey!10}${40.55^{\pm{1.3}}}$ &   \cellcolor{jku_grey!10}${47.51^{\pm{0.7}}}$ &  \cellcolor{jku_grey!10}\cellcolor{jku_green!20}${54.06^{\pm{0.7}}}$ \\
& &time\_a\_c & ${35.18^{\pm{1.6}}}$ & ${35.31^{\pm{0.0}}}$ & ${39.16^{\pm{0.0}}}$ &   ${43.53^{\pm{0.4}}}$ & ${37.06^{\pm{4.3}}}$ &   ${42.73^{\pm{1.8}}}$ &  \cellcolor{jku_green!20}${44.81^{\pm{1.1}}}$ \\
\midrule
  AP &  \cellcolor{jku_grey!10}FS-Mol & \cellcolor{jku_grey!10}default & \cellcolor{jku_grey!10}${48.75^{\pm{0.3}}}$ & \cellcolor{jku_grey!10}${48.48^{\pm{0.0}}}$ & \cellcolor{jku_grey!10}${49.02^{\pm{0.0}}}$ &   \cellcolor{jku_grey!10}${61.86^{\pm{0.7}}}$ & \cellcolor{jku_grey!10}${60.99^{\pm{1.8}}}$ &   \cellcolor{jku_grey!10}${65.68^{\pm{0.2}}}$ &  \cellcolor{jku_grey!10}\cellcolor{jku_green!20}${66.55^{\pm{0.2}}}$ \\
 & PubChem &     hts & ${00.58^{\pm{0.0}}}$ & ${00.57^{\pm{0.0}}}$ & ${00.67^{\pm{0.0}}}$ &   ${01.78^{\pm{0.1}}}$ & ${02.61^{\pm{0.4}}}$ &   ${03.67^{\pm{0.1}}}$ &  \cellcolor{jku_green!20}${09.00^{\pm{0.1}}}$ \\
&  &  \cellcolor{jku_grey!10}time\_a & \cellcolor{jku_grey!10}${35.42^{\pm{0.3}}}$ & \cellcolor{jku_grey!10}${34.68^{\pm{0.0}}}$ & \cellcolor{jku_grey!10}${36.87^{\pm{0.0}}}$ &   \cellcolor{jku_grey!10}${46.26^{\pm{1.0}}}$ & \cellcolor{jku_grey!10}${38.97^{\pm{0.7}}}$ &   \cellcolor{jku_grey!10}${43.53^{\pm{0.5}}}$ &  \cellcolor{jku_grey!10}\cellcolor{jku_green!20}${48.06^{\pm{0.3}}}$ \\
   & &time\_a\_c & ${38.50^{\pm{0.5}}}$ & ${38.31^{\pm{0.0}}}$ & ${42.10^{\pm{0.0}}}$ & \cellcolor{jku_yellow!20}${45.22^{\pm{0.3}}}$ & ${41.10^{\pm{2.8}}}$ &   ${44.46^{\pm{0.9}}}$ &  \cellcolor{jku_green!20}${45.78^{\pm{0.6}}}$ \\
   \midrule
$\neg$AP &  \cellcolor{jku_grey!10}FS-Mol & \cellcolor{jku_grey!10}default & \cellcolor{jku_grey!10}${54.17^{\pm{0.3}}}$ & \cellcolor{jku_grey!10}${54.56^{\pm{0.0}}}$ & \cellcolor{jku_grey!10}${54.38^{\pm{0.0}}}$ &   \cellcolor{jku_grey!10}${67.87^{\pm{0.8}}}$ & \cellcolor{jku_grey!10}${67.24^{\pm{1.8}}}$ &   \cellcolor{jku_grey!10}${71.04^{\pm{0.3}}}$ &  \cellcolor{jku_grey!10}\cellcolor{jku_green!20}${72.22^{\pm{0.2}}}$ \\
   & PubChem &     hts & ${99.43^{\pm{0.0}}}$ & ${99.45^{\pm{0.0}}}$ & ${99.43^{\pm{0.0}}}$ &   ${99.71^{\pm{0.0}}}$ & ${99.68^{\pm{0.0}}}$ &  \cellcolor{jku_green!20}${99.78^{\pm{0.0}}}$ & \cellcolor{jku_yellow!20}${99.77^{\pm{0.0}}}$ \\
 &  &  \cellcolor{jku_grey!10}time\_a & \cellcolor{jku_grey!10}${69.03^{\pm{0.2}}}$ & \cellcolor{jku_grey!10}${68.27^{\pm{0.0}}}$ & \cellcolor{jku_grey!10}${71.45^{\pm{0.0}}}$ &   \cellcolor{jku_grey!10}${75.30^{\pm{0.2}}}$ & \cellcolor{jku_grey!10}${73.07^{\pm{0.8}}}$ &  \cellcolor{jku_grey!10}\cellcolor{jku_green!20}${76.24^{\pm{0.4}}}$ & \cellcolor{jku_grey!10}\cellcolor{jku_yellow!20}${76.19^{\pm{0.3}}}$ \\
  &  &time\_a\_c & ${69.36^{\pm{0.1}}}$ & ${68.36^{\pm{0.0}}}$ & ${72.78^{\pm{0.0}}}$ & \cellcolor{jku_yellow!20}${74.63^{\pm{0.1}}}$ & ${72.05^{\pm{2.1}}}$ &  \cellcolor{jku_green!20}${74.75^{\pm{0.5}}}$ & \cellcolor{jku_yellow!20}${74.67^{\pm{0.5}}}$ \\
\bottomrule
\end{tabular}
\raggedright
    \footnotesize{$^\dagger$ for the SLMs, we chose a single model provided by the authors. Training re-runs are computationally too costly.} 
\end{table*}

\paragraph{KV-PLM.}
The SLM KV-PLM \citep{zeng2022kvplm} is also able to process molecules and text in one model. KV-PLM is a BERT \citep{devlin2019bert} transformer variant and the weights, as well as the tokenizer, are available \href{https://github.com/thunlp/KV-PLM}{online}.
We use the matching score from SMILES input and textual description as ranking for molecules within an assay. As textual description, we use the same assay description source as in CLAMP (see Tab.~\ref{tab:assay_descriptions} for example descriptions).
KV-PLM has not been trained on assay descriptions specifically and tuning the prompt remains future work.

\subsubsection{Additional metrics}

Tab.~\ref{tab:appendix:zeroshot_full} is an extension of Tab.~\ref{tab:zeroshot} with additional metrics. 
CLAMP significantly (paired Wilcoxon test) outperforms all baselines and other methods with respect to both $\Delta$AP and AUROC.
The FH method performs well for $\neg$AP for the three splits in PubChem (not significantly compared to CLAMP). 
We speculate, that it might have learned general attributes about molecules that make them very unlikely to be active in any assay.

\subsection{Details on the representation learning study (5.2)}
We examine the effectiveness of representation learning methods by evaluating the 
ability to produce rich representations for a variety of tasks. 
The quality of a learned representation is a highly debated topic in the field, 
but one important aspect is, that a given representation is linearly separable for a given new task \citep{alain2016understanding}. 

\begin{figure*}[tb]
    \centering
    \includegraphics[width=.8\columnwidth]{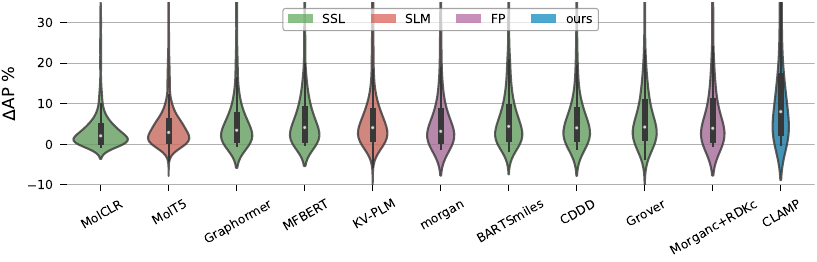}
    \caption{$\Delta AP$ violin-plot over all tasks within the downstream datasets. 
    SSL represents self-supervised methods, SLM scientific language models, 
    FP fingerprint baselines.
    \label{fig:lin_probing_violin}}
\end{figure*}

We use linear probing performance as a proxy for representation quality \citep{alain2016understanding,radford2021learning}. 
We precompute the representations for several methods, and run a 
logistic regression model for 1500 iterations with balanced class weights, 
L2 regularization and lbfgs solver. 
We compute a Logistic Regression model for each method and 
assay, and compute the metrics as mean performance over assays.
Tab.~\ref{tab:linear_probing_wo_gal} in the main manuscript shows 
the $\Delta$AP results and Fig.~\ref{fig:lin_probing_violin} shows a 
violin-plot representing the $\Delta$AP distribution 
for each method and per assay.

\subsubsection{Further metrics for results}
Tab.~\ref{tab:linear_probing_appendix} is an extension 
of Tab.~\ref{tab:linear_probing_wo_gal} with the AUROC metric. 
CLAMP significantly (paired Wilcoxon test) outperforms 
all baselines and other methods with respect to AUROC.

\begin{table*}[t]
    \noindent
    \centering
    \caption{Linear probing results of different methods 
    with respect to $\Delta$AP, including Galactica. 
    Green cells indicate the highest values in a category of tasks
    and areas in yellow cells within the 
    standard deviation to the maximum value. 
    Because of the low variability of training re-runs of a linear probing model, the error bars represent standard deviation obtained through bootstrap resampling.
    Rank-avg represents the mean rank over all assays. Methods are
    assigned to categories (cat): self-supervised learning methods (SSL), scientific 
    language models (SLM), and chemical descriptors or fingerprints (FP)
    \label{tab:linear_probing_with_gal}}
    \small
    \fontsize{9pt}{9pt}
    \resizebox{\textwidth}{!}{
    \begin{tabular}{llllllllll||l}
\toprule
 dataset& & BACE &   BBBP &   ClinTox &HIV &  SIDER & Tox21 &ToxCast & Tox21-10k &  \\
 split&&scaffold & scaffold &scaffold & scaffold & scaffold &scaffold & scaffold & 
 original & \\
 \# of assays & cat &  1 &  1 &  2 & 1  & 27  &  12 &  617 & 68  & rank-avg  \\
\midrule
   \cellcolor{jku_grey!10}CLAMP & \cellcolor{jku_grey!10}ours& \cellcolor{jku_grey!10}\cellcolor{jku_yellow!20}${27.47^{\pm{\scriptscriptstyle4}}}$ & \cellcolor{jku_grey!10}${16.47^{\pm{\scriptscriptstyle4}}}$ &  \cellcolor{jku_grey!10}${11.05^{\pm{\scriptscriptstyle6}}}$ &  \cellcolor{jku_grey!10}\cellcolor{jku_green!20}${28.49^{\pm{\scriptscriptstyle4}}}$ &  \cellcolor{jku_grey!10}\cellcolor{jku_green!20}${08.96^{\pm{\scriptscriptstyle4}}}$ & \cellcolor{jku_grey!10}\cellcolor{jku_green!20}${23.35^{\pm{\scriptscriptstyle6}}}$ &  \cellcolor{jku_grey!10}\cellcolor{jku_green!20}${09.44^{\pm{\scriptscriptstyle5}}}$ &  \cellcolor{jku_grey!10}\cellcolor{jku_green!20}${51.27^{\pm{\scriptscriptstyle10}}}$ & \cellcolor{jku_grey!10}\cellcolor{jku_green!20}${02.85^{\pm{\scriptscriptstyle2}}}$ \\
  Grover &  SSL & ${21.74^{\pm{\scriptscriptstyle4}}}$ & ${16.76^{\pm{\scriptscriptstyle4}}}$ &  ${22.74^{\pm{\scriptscriptstyle8}}}$ &   ${13.58^{\pm{\scriptscriptstyle3}}}$ & \cellcolor{jku_yellow!20}${05.41^{\pm{\scriptscriptstyle4}}}$ &  ${12.88^{\pm{\scriptscriptstyle5}}}$ & \cellcolor{jku_yellow!20}${05.22^{\pm{\scriptscriptstyle3}}}$ & \cellcolor{jku_yellow!20}${42.23^{\pm{\scriptscriptstyle10}}}$ &\cellcolor{jku_yellow!20}${05.08^{\pm{\scriptscriptstyle3}}}$ \\
\cellcolor{jku_grey!10}Mc+RDKc & \cellcolor{jku_grey!10}FP &  \cellcolor{jku_grey!10}${23.87^{\pm{\scriptscriptstyle4}}}$ & \cellcolor{jku_grey!10}${18.39^{\pm{\scriptscriptstyle4}}}$ &  \cellcolor{jku_grey!10}${17.75^{\pm{\scriptscriptstyle7}}}$ & \cellcolor{jku_grey!10}\cellcolor{jku_yellow!20}${25.57^{\pm{\scriptscriptstyle4}}}$ & \cellcolor{jku_grey!10}\cellcolor{jku_yellow!20}${08.43^{\pm{\scriptscriptstyle4}}}$ &  \cellcolor{jku_grey!10}${13.96^{\pm{\scriptscriptstyle5}}}$ & \cellcolor{jku_grey!10}\cellcolor{jku_yellow!20}${05.18^{\pm{\scriptscriptstyle4}}}$ & \cellcolor{jku_grey!10}\cellcolor{jku_yellow!20}${42.45^{\pm{\scriptscriptstyle10}}}$ &  \cellcolor{jku_grey!10}${05.64^{\pm{\scriptscriptstyle3}}}$ \\
CDDD & SSL & ${17.51^{\pm{\scriptscriptstyle5}}}$ &  ${20.56^{\pm{\scriptscriptstyle4}}}$ & \cellcolor{jku_green!20}${33.82^{\pm{\scriptscriptstyle7}}}$ &   ${12.29^{\pm{\scriptscriptstyle3}}}$ & \cellcolor{jku_yellow!20}${05.30^{\pm{\scriptscriptstyle4}}}$ &  ${10.34^{\pm{\scriptscriptstyle4}}}$ &   ${04.82^{\pm{\scriptscriptstyle3}}}$ &${36.32^{\pm{\scriptscriptstyle9}}}$ &  ${05.79^{\pm{\scriptscriptstyle3}}}$ \\
  \cellcolor{jku_grey!10}BARTSmiles &  \cellcolor{jku_grey!10}SSL &\cellcolor{jku_grey!10}\cellcolor{jku_green!20}${29.16^{\pm{\scriptscriptstyle3}}}$ & \cellcolor{jku_grey!10}${17.09^{\pm{\scriptscriptstyle4}}}$ &  \cellcolor{jku_grey!10}${10.94^{\pm{\scriptscriptstyle6}}}$ &   \cellcolor{jku_grey!10}${07.88^{\pm{\scriptscriptstyle2}}}$ & \cellcolor{jku_grey!10}\cellcolor{jku_yellow!20}${05.24^{\pm{\scriptscriptstyle4}}}$ &  \cellcolor{jku_grey!10}${10.34^{\pm{\scriptscriptstyle4}}}$ & \cellcolor{jku_grey!10}\cellcolor{jku_yellow!20}${05.07^{\pm{\scriptscriptstyle3}}}$ &\cellcolor{jku_grey!10}${30.40^{\pm{\scriptscriptstyle9}}}$ &  \cellcolor{jku_grey!10}${05.79^{\pm{\scriptscriptstyle3}}}$ \\
  KV-PLM &SLM &${24.44^{\pm{\scriptscriptstyle4}}}$ & ${17.61^{\pm{\scriptscriptstyle4}}}$ &  ${07.08^{\pm{\scriptscriptstyle4}}}$ &   ${06.20^{\pm{\scriptscriptstyle2}}}$ &   ${04.61^{\pm{\scriptscriptstyle4}}}$ &  ${09.90^{\pm{\scriptscriptstyle4}}}$ &   ${04.53^{\pm{\scriptscriptstyle3}}}$ &   ${30.92^{\pm{\scriptscriptstyle10}}}$ &  ${05.88^{\pm{\scriptscriptstyle3}}}$ \\
  \cellcolor{jku_grey!10}MFBERT & \cellcolor{jku_grey!10}SSL & \cellcolor{jku_grey!10}${14.89^{\pm{\scriptscriptstyle4}}}$ & \cellcolor{jku_grey!10}${19.77^{\pm{\scriptscriptstyle4}}}$ &  \cellcolor{jku_grey!10}${11.84^{\pm{\scriptscriptstyle7}}}$ &   \cellcolor{jku_grey!10}${06.09^{\pm{\scriptscriptstyle1}}}$ & \cellcolor{jku_grey!10}\cellcolor{jku_yellow!20}${06.93^{\pm{\scriptscriptstyle4}}}$ &  \cellcolor{jku_grey!10}${08.46^{\pm{\scriptscriptstyle4}}}$ &   \cellcolor{jku_grey!10}${04.72^{\pm{\scriptscriptstyle3}}}$ &\cellcolor{jku_grey!10}${25.61^{\pm{\scriptscriptstyle8}}}$ &  \cellcolor{jku_grey!10}${06.20^{\pm{\scriptscriptstyle3}}}$ \\
  Graphormer &  SSL & ${20.22^{\pm{\scriptscriptstyle4}}}$ &   ${08.67^{\pm{\scriptscriptstyle4}}}$ &  ${04.51^{\pm{\scriptscriptstyle4}}}$ &   ${06.87^{\pm{\scriptscriptstyle2}}}$ & \cellcolor{jku_yellow!20}${05.71^{\pm{\scriptscriptstyle4}}}$ &  ${08.16^{\pm{\scriptscriptstyle4}}}$ &   ${04.10^{\pm{\scriptscriptstyle3}}}$ &${29.50^{\pm{\scriptscriptstyle9}}}$ &  ${06.55^{\pm{\scriptscriptstyle3}}}$ \\
  \cellcolor{jku_grey!10}Morgan & \cellcolor{jku_grey!10}FP &\cellcolor{jku_grey!10}\cellcolor{jku_yellow!20}${27.79^{\pm{\scriptscriptstyle4}}}$ &   \cellcolor{jku_grey!10}${16.06^{\pm{\scriptscriptstyle4}}}$ &  \cellcolor{jku_grey!10}${23.86^{\pm{\scriptscriptstyle8}}}$ &   \cellcolor{jku_grey!10}${16.23^{\pm{\scriptscriptstyle4}}}$ & \cellcolor{jku_grey!10}\cellcolor{jku_yellow!20}${06.97^{\pm{\scriptscriptstyle4}}}$ &  \cellcolor{jku_grey!10}${09.69^{\pm{\scriptscriptstyle4}}}$ &   \cellcolor{jku_grey!10}${04.28^{\pm{\scriptscriptstyle3}}}$ &   \cellcolor{jku_grey!10}${36.07^{\pm{\scriptscriptstyle10}}}$ &  \cellcolor{jku_grey!10}${06.64^{\pm{\scriptscriptstyle3}}}$ \\
   MolT5 & SLM &   ${11.51^{\pm{\scriptscriptstyle4}}}$ &   ${16.03^{\pm{\scriptscriptstyle4}}}$ &  ${10.06^{\pm{\scriptscriptstyle6}}}$ &   ${15.92^{\pm{\scriptscriptstyle3}}}$ &   ${02.91^{\pm{\scriptscriptstyle3}}}$ &  ${06.20^{\pm{\scriptscriptstyle3}}}$ &   ${03.53^{\pm{\scriptscriptstyle3}}}$ &${15.65^{\pm{\scriptscriptstyle6}}}$ &  ${07.36^{\pm{\scriptscriptstyle3}}}$ \\
  \cellcolor{jku_grey!10}MolCLR & \cellcolor{jku_grey!10}SSL &   \cellcolor{jku_grey!10}${15.59^{\pm{\scriptscriptstyle4}}}$ &   \cellcolor{jku_grey!10}${13.01^{\pm{\scriptscriptstyle4}}}$ &  \cellcolor{jku_grey!10}${00.99^{\pm{\scriptscriptstyle3}}}$ &   \cellcolor{jku_grey!10}${02.57^{\pm{\scriptscriptstyle1}}}$ & \cellcolor{jku_grey!10}\cellcolor{jku_yellow!20}${06.13^{\pm{\scriptscriptstyle4}}}$ &  \cellcolor{jku_grey!10}${06.08^{\pm{\scriptscriptstyle3}}}$ &   \cellcolor{jku_grey!10}${02.43^{\pm{\scriptscriptstyle2}}}$ &\cellcolor{jku_grey!10}${13.21^{\pm{\scriptscriptstyle5}}}$ &  \cellcolor{jku_grey!10}${08.20^{\pm{\scriptscriptstyle3}}}$ \\

GAL 6.7B & SLM & ${23.14^{\pm{\scriptscriptstyle4}}}$	& \cellcolor{jku_green!20} ${24.53 ^{\pm{\scriptscriptstyle 4}}}$ &	- $^\dagger$	& ${18.35 ^{\pm{\scriptscriptstyle 4}}}$ &	-$^\dagger$	& ${13.79^{\pm{\scriptscriptstyle 5}}}$	 & -$^\dagger$ & \cellcolor{jku_yellow!20}${42.80 ^{\pm{\scriptscriptstyle 11}}}$ & - \\
\bottomrule
\end{tabular}
    }
    \footnotesize{$^\dagger$ model was pre-trained on a different split which would result in dataset-leakage and an overestimate in performance} 
\end{table*}

\begin{table*}[t]
    \noindent
    \centering
    \caption{Linear probing results of different methods with respect to AUROC. 
    Green cells indicate the highest values in a category of tasks and 
    areas in yellow cells within the standard-deviation to the best value. 
    Because of the low variability of training re-runs of a linear probing model,
    the error bars represent bootstrap standard deviation. 
    Rank-avg represents the mean rank over all assays.
    \label{tab:linear_probing_appendix_auroc_default_split}}
    \label{tab:linear_probing_appendix}
    \small
    \fontsize{10pt}{10pt}
    \begin{tabular}{lllllllll||l}
\toprule
                                   &                                                                                 BACE &                                                                                   BBBP &                                                        ClinTox &                                                                                    HIV &                                                                                  SIDER &                                                                                 Tox21 &                                                                               ToxCast &                                                                             Tox21-10k &                                                                   \\
                                        &                                                                         scaffold &                                                                         scaffold &                                                 scaffold &                                                                         scaffold &                                                                         scaffold &                                                                        scaffold &                                                                        scaffold & split & rank-avg \\
\midrule
       \cellcolor{jku_grey!10}CLAMP &  \cellcolor{jku_grey!10}\cellcolor{jku_green!20}${84.28^{\pm{\scriptscriptstyle3}}}$ &                           \cellcolor{jku_grey!10}${68.22^{\pm{\scriptscriptstyle4}}}$ &   \cellcolor{jku_grey!10}${75.31^{\pm{\scriptscriptstyle8}}}$ & \cellcolor{jku_grey!10}\cellcolor{jku_yellow!20}${76.34^{\pm{\scriptscriptstyle2}}}$ &  \cellcolor{jku_grey!10}\cellcolor{jku_green!20}${65.15^{\pm{\scriptscriptstyle7}}}$ & \cellcolor{jku_grey!10}\cellcolor{jku_green!20}${78.23^{\pm{\scriptscriptstyle4}}}$ & \cellcolor{jku_grey!10}\cellcolor{jku_green!20}${74.00^{\pm{\scriptscriptstyle6}}}$ & \cellcolor{jku_grey!10}\cellcolor{jku_green!20}${90.85^{\pm{\scriptscriptstyle4}}}$ & \cellcolor{jku_grey!10}\cellcolor{jku_green!20}${02.41^{\pm{\scriptscriptstyle2}}}$ \\
                              Grover &                                                   ${78.64^{\pm{\scriptscriptstyle4}}}$ &                                                   ${67.94^{\pm{\scriptscriptstyle4}}}$ & \cellcolor{jku_yellow!20}${89.48^{\pm{\scriptscriptstyle4}}}$ &                          \cellcolor{jku_green!20}${77.52^{\pm{\scriptscriptstyle2}}}$ &                         \cellcolor{jku_yellow!20}${59.63^{\pm{\scriptscriptstyle7}}}$ &                                                  ${68.16^{\pm{\scriptscriptstyle5}}}$ &                                                  ${67.26^{\pm{\scriptscriptstyle7}}}$ &                                                  ${86.62^{\pm{\scriptscriptstyle5}}}$ &                                                  ${04.66^{\pm{\scriptscriptstyle3}}}$ \\
      \cellcolor{jku_grey!10}KV-PLM &                           \cellcolor{jku_grey!10}${79.91^{\pm{\scriptscriptstyle4}}}$ & \cellcolor{jku_grey!10}${69.27^{\pm{\scriptscriptstyle4}}}$ &   \cellcolor{jku_grey!10}${73.10^{\pm{\scriptscriptstyle9}}}$ &                           \cellcolor{jku_grey!10}${69.76^{\pm{\scriptscriptstyle2}}}$ &                           \cellcolor{jku_grey!10}${57.36^{\pm{\scriptscriptstyle7}}}$ &                          \cellcolor{jku_grey!10}${64.74^{\pm{\scriptscriptstyle5}}}$ &                          \cellcolor{jku_grey!10}${66.32^{\pm{\scriptscriptstyle7}}}$ &                          \cellcolor{jku_grey!10}${82.16^{\pm{\scriptscriptstyle5}}}$ &                          \cellcolor{jku_grey!10}${05.32^{\pm{\scriptscriptstyle3}}}$ \\
                                CDDD &                                                   ${76.83^{\pm{\scriptscriptstyle4}}}$ &                          ${72.45^{\pm{\scriptscriptstyle4}}}$ &  \cellcolor{jku_green!20}${92.41^{\pm{\scriptscriptstyle5}}}$ &                                                   ${73.87^{\pm{\scriptscriptstyle3}}}$ &                         \cellcolor{jku_yellow!20}${59.72^{\pm{\scriptscriptstyle7}}}$ &                                                  ${68.55^{\pm{\scriptscriptstyle5}}}$ &                                                  ${64.85^{\pm{\scriptscriptstyle7}}}$ &                                                  ${85.66^{\pm{\scriptscriptstyle5}}}$ &                                                  ${05.52^{\pm{\scriptscriptstyle3}}}$ \\
  \cellcolor{jku_grey!10}BARTSmiles & \cellcolor{jku_grey!10}\cellcolor{jku_yellow!20}${83.21^{\pm{\scriptscriptstyle3}}}$ & \cellcolor{jku_grey!10}${70.90^{\pm{\scriptscriptstyle4}}}$ &   \cellcolor{jku_grey!10}${79.30^{\pm{\scriptscriptstyle8}}}$ &                           \cellcolor{jku_grey!10}${70.91^{\pm{\scriptscriptstyle3}}}$ &                           \cellcolor{jku_grey!10}${57.66^{\pm{\scriptscriptstyle7}}}$ &                          \cellcolor{jku_grey!10}${65.16^{\pm{\scriptscriptstyle5}}}$ &                          \cellcolor{jku_grey!10}${64.98^{\pm{\scriptscriptstyle7}}}$ &                          \cellcolor{jku_grey!10}${83.04^{\pm{\scriptscriptstyle5}}}$ &                          \cellcolor{jku_grey!10}${05.58^{\pm{\scriptscriptstyle3}}}$ \\
                              MFBERT &                                                   ${71.63^{\pm{\scriptscriptstyle4}}}$ &                         ${71.61^{\pm{\scriptscriptstyle3}}}$ &                          ${77.99^{\pm{\scriptscriptstyle10}}}$ &                                                   ${71.12^{\pm{\scriptscriptstyle2}}}$ &                         \cellcolor{jku_yellow!20}${61.14^{\pm{\scriptscriptstyle7}}}$ &                                                  ${63.95^{\pm{\scriptscriptstyle5}}}$ &                                                  ${63.72^{\pm{\scriptscriptstyle7}}}$ &                                                  ${80.62^{\pm{\scriptscriptstyle5}}}$ &                                                  ${06.13^{\pm{\scriptscriptstyle3}}}$ \\
  \cellcolor{jku_grey!10}Graphormer &                           \cellcolor{jku_grey!10}${75.72^{\pm{\scriptscriptstyle4}}}$ &                           \cellcolor{jku_grey!10}${58.71^{\pm{\scriptscriptstyle4}}}$ &   \cellcolor{jku_grey!10}${68.64^{\pm{\scriptscriptstyle9}}}$ &                           \cellcolor{jku_grey!10}${71.61^{\pm{\scriptscriptstyle3}}}$ & \cellcolor{jku_grey!10}\cellcolor{jku_yellow!20}${58.82^{\pm{\scriptscriptstyle7}}}$ &                          \cellcolor{jku_grey!10}${65.48^{\pm{\scriptscriptstyle5}}}$ &                          \cellcolor{jku_grey!10}${62.92^{\pm{\scriptscriptstyle7}}}$ &                          \cellcolor{jku_grey!10}${83.38^{\pm{\scriptscriptstyle5}}}$ &                          \cellcolor{jku_grey!10}${06.38^{\pm{\scriptscriptstyle3}}}$ \\
                               MolT5 &                                                   ${63.46^{\pm{\scriptscriptstyle5}}}$ &                                                   ${65.06^{\pm{\scriptscriptstyle4}}}$ &                           ${73.78^{\pm{\scriptscriptstyle9}}}$ &                                                   ${72.81^{\pm{\scriptscriptstyle3}}}$ &                                                   ${53.20^{\pm{\scriptscriptstyle7}}}$ &                                                  ${63.63^{\pm{\scriptscriptstyle4}}}$ &                                                  ${62.98^{\pm{\scriptscriptstyle7}}}$ &                                                  ${74.99^{\pm{\scriptscriptstyle6}}}$ &                                                  ${06.88^{\pm{\scriptscriptstyle3}}}$ \\
\cellcolor{jku_grey!10}Mc+RDKc & \cellcolor{jku_grey!10}\cellcolor{jku_yellow!20}${81.16^{\pm{\scriptscriptstyle4}}}$ & \cellcolor{jku_grey!10}${69.67^{\pm{\scriptscriptstyle4}}}$ &   \cellcolor{jku_grey!10}${84.81^{\pm{\scriptscriptstyle5}}}$ &                           \cellcolor{jku_grey!10}${72.25^{\pm{\scriptscriptstyle3}}}$ & \cellcolor{jku_grey!10}\cellcolor{jku_yellow!20}${62.41^{\pm{\scriptscriptstyle7}}}$ &                          \cellcolor{jku_grey!10}${69.61^{\pm{\scriptscriptstyle5}}}$ &                          \cellcolor{jku_grey!10}${60.73^{\pm{\scriptscriptstyle8}}}$ &                          \cellcolor{jku_grey!10}${82.72^{\pm{\scriptscriptstyle6}}}$ &                          \cellcolor{jku_grey!10}${06.99^{\pm{\scriptscriptstyle3}}}$ \\
                              Morgan &                         \cellcolor{jku_yellow!20}${80.94^{\pm{\scriptscriptstyle3}}}$ &                                                   ${66.09^{\pm{\scriptscriptstyle4}}}$ &                           ${74.99^{\pm{\scriptscriptstyle9}}}$ &                                                   ${71.31^{\pm{\scriptscriptstyle3}}}$ &                         \cellcolor{jku_yellow!20}${59.00^{\pm{\scriptscriptstyle7}}}$ &                                                  ${64.65^{\pm{\scriptscriptstyle5}}}$ &                                                  ${59.12^{\pm{\scriptscriptstyle8}}}$ &                                                  ${79.99^{\pm{\scriptscriptstyle6}}}$ &                                                  ${07.98^{\pm{\scriptscriptstyle3}}}$ \\
      \cellcolor{jku_grey!10}MolCLR&                           \cellcolor{jku_grey!10}${70.43^{\pm{\scriptscriptstyle5}}}$ &                           \cellcolor{jku_grey!10}${61.07^{\pm{\scriptscriptstyle4}}}$ &  \cellcolor{jku_grey!10}${53.64^{\pm{\scriptscriptstyle12}}}$ &                           \cellcolor{jku_grey!10}${63.87^{\pm{\scriptscriptstyle3}}}$ &                           \cellcolor{jku_grey!10}${56.84^{\pm{\scriptscriptstyle7}}}$ &                          \cellcolor{jku_grey!10}${63.16^{\pm{\scriptscriptstyle4}}}$ &                          \cellcolor{jku_grey!10}${59.23^{\pm{\scriptscriptstyle7}}}$ &                          \cellcolor{jku_grey!10}${73.82^{\pm{\scriptscriptstyle6}}}$ &                          \cellcolor{jku_grey!10}${08.16^{\pm{\scriptscriptstyle3}}}$ \\

    Gal 6.7B & ${78.97^{\pm{\scriptscriptstyle4}}}$	& \cellcolor{jku_green!20}${75.72^{\pm{\scriptscriptstyle3}}}$	& -	& ${74.10^{\pm{\scriptscriptstyle3}}}$	& -	& ${69.14^{\pm{\scriptscriptstyle5}}}$	& -	& ${86.84^{\pm{\scriptscriptstyle5}}}$	& -  \\
      
\bottomrule
\end{tabular}
\end{table*}

\subsubsection{Comparison to literature.}
In this section, 
we present a selection of prior works 
that have also benchmarked on the MoleculeNet dataset.
For a fair comparison, is important to ensure that the same data 
split in referenced prior works is also the one used in our experiments. 
We use the initial scaffold-split from MoleculeNet and 
do not take multiple-scaffold splits as in e.g. Graphormer\citep{ying2021graphormer}.
Furthermore, while many of the prior works report results using the AUROC 
metric, we prefer $\Delta$AP. In our experiments optimize on validation $\Delta$AP as 
the primary metric (Sec.~\ref{app:metric}).
Tab.~\ref{tab:moleculenet_literature} shows our linear probing results 
compared to fine-tuned models from the literature.
CLAMP yields the best performance
of the linear probing methods. 

\begin{table*}[ht]
    \noindent
    \centering
    \caption{MoleculeNet ROC-AUC comparison to literature reported values on scaffold-split for all datasets. LP stands for linear probing, all other methods have been fine-tuned. Standard-Error for methods with LP reports dataset bootstrap variance whereby other methods typically report re-run variance.  Results are reported in \% and methods are sorted by the last column. \label{tab:moleculenet_literature}}
    
    \fontsize{9pt}{9pt}
    \tiny
    \resizebox{\textwidth}{!}{
    \begin{tabular}{lllllllll||l}
\toprule
                                                method &                        LP &                                                                BACE &                                                                  BBBP &                                                               ClinTox &                                                                   HIV &                                         SIDER &                                                                 Tox21 &                                                                Toxcast &                                                                    avg \\
\midrule
                    \cellcolor{jku_grey!10}MolCLR \citep{wang2021molclr} & \cellcolor{jku_grey!10}X &                          \cellcolor{jku_grey!10}${70.43^{\pm{4.6}}}$ &                          \cellcolor{jku_grey!10}${61.07^{\pm{3.9}}}$ &                         \cellcolor{jku_grey!10}${53.64^{\pm{11.7}}}$ &                          \cellcolor{jku_grey!10}${63.87^{\pm{2.6}}}$ &  \cellcolor{jku_grey!10}${56.84^{\pm{7.6}}}$ &                          \cellcolor{jku_grey!10}${63.16^{\pm{4.6}}}$ &                           \cellcolor{jku_grey!10}${59.23^{\pm{8.8}}}$ &                           \cellcolor{jku_grey!10}${61.18^{\pm{5.0}}}$ \\
                              JOAOv2 \citep{wang2022evaluating} &                           &                                                  ${67.40^{\pm{0.7}}}$ &                                                  ${66.40^{\pm{0.9}}}$ &                                                  ${64.50^{\pm{0.9}}}$ &                                                  ${68.40^{\pm{0.5}}}$ &                          ${59.10^{\pm{0.7}}}$ &                                                  ${68.20^{\pm{0.8}}}$ &                                                   ${57.00^{\pm{0.5}}}$ &                                                   ${64.43^{\pm{4.2}}}$ \\
   \cellcolor{jku_grey!10}MolT5-large \citep{edwards2022translation}& \cellcolor{jku_grey!10}X &                          \cellcolor{jku_grey!10}${63.46^{\pm{4.6}}}$ &                          \cellcolor{jku_grey!10}${65.06^{\pm{3.9}}}$ &                          \cellcolor{jku_grey!10}${73.78^{\pm{8.6}}}$ &                          \cellcolor{jku_grey!10}${72.81^{\pm{2.7}}}$ &  \cellcolor{jku_grey!10}${53.20^{\pm{7.4}}}$ &                          \cellcolor{jku_grey!10}${63.63^{\pm{4.4}}}$ &                           \cellcolor{jku_grey!10}${62.98^{\pm{8.5}}}$ &                           \cellcolor{jku_grey!10}${64.99^{\pm{6.4}}}$ \\
                                           Graphormer \citep{ying2021graphormer} &                         X &                                                  ${75.72^{\pm{4.1}}}$ &                                                  ${58.71^{\pm{3.9}}}$ &                                                  ${68.64^{\pm{9.4}}}$ &                                                  ${71.61^{\pm{2.5}}}$ &                          ${58.82^{\pm{7.1}}}$ &                                                  ${65.48^{\pm{4.9}}}$ &                                                   ${62.92^{\pm{8.3}}}$ &                                                   ${65.99^{\pm{5.9}}}$ \\
        \cellcolor{jku_grey!10}GAL 120B  \citep{taylor2022galactica} & \cellcolor{jku_grey!10}  &                          \cellcolor{jku_grey!10}${61.70^{\pm{0.0}}}$ &                          \cellcolor{jku_grey!10}${66.10^{\pm{0.0}}}$ &                          \cellcolor{jku_grey!10}${-}$ &                          \cellcolor{jku_grey!10}${74.50^{\pm{0.0}}}$ &  \cellcolor{jku_grey!10}${-}$ &                          \cellcolor{jku_grey!10}${68.90^{\pm{0.0}}}$ &                           \cellcolor{jku_grey!10}${-}$ &                           \cellcolor{jku_grey!10}${67.80^{\pm{4.6}}}$ \\
                            GNN \citep{wang2022evaluating} &                           &                                                  ${80.90^{\pm{1.9}}}$ &                                                  ${66.40^{\pm{0.9}}}$ &                                                  ${66.30^{\pm{0.6}}}$ &                                                  ${71.40^{\pm{1.2}}}$ &                          ${61.70^{\pm{1.9}}}$ &                                                  ${69.70^{\pm{2.3}}}$ &                                                   ${58.70^{\pm{0.5}}}$ &                                                   ${67.87^{\pm{6.7}}}$ \\
                       \cellcolor{jku_grey!10}Morgan-FP & \cellcolor{jku_grey!10}X &                          \cellcolor{jku_grey!10}${80.94^{\pm{3.5}}}$ &                          \cellcolor{jku_grey!10}${66.09^{\pm{3.8}}}$ &                          \cellcolor{jku_grey!10}${74.99^{\pm{9.5}}}$ &                          \cellcolor{jku_grey!10}${71.31^{\pm{2.7}}}$ &  \cellcolor{jku_grey!10}${59.00^{\pm{8.0}}}$ &                          \cellcolor{jku_grey!10}${64.65^{\pm{4.9}}}$ &                           \cellcolor{jku_grey!10}${59.12^{\pm{8.9}}}$ &                           \cellcolor{jku_grey!10}${68.01^{\pm{7.6}}}$ \\
                        3D Infomax  \citep{stark20223d} &                           &                                                  ${79.42^{\pm{1.9}}}$ &                                                  ${69.10^{\pm{1.1}}}$ &                                                  ${59.43^{\pm{3.2}}}$ &                                                  ${76.08^{\pm{1.3}}}$ &                          ${53.37^{\pm{3.3}}}$ &                                                  ${74.46^{\pm{0.7}}}$ &                                                   ${64.41^{\pm{0.9}}}$ &                                                   ${68.04^{\pm{8.8}}}$ \\
                        \cellcolor{jku_grey!10}KV-PLM  \citep{zeng2022kvplm} & \cellcolor{jku_grey!10}X &                          \cellcolor{jku_grey!10}${79.91^{\pm{3.8}}}$ &                          \cellcolor{jku_grey!10}${69.27^{\pm{3.6}}}$ &                          \cellcolor{jku_grey!10}${73.10^{\pm{9.4}}}$ &                          \cellcolor{jku_grey!10}${69.76^{\pm{2.4}}}$ &  \cellcolor{jku_grey!10}${57.36^{\pm{7.4}}}$ &                          \cellcolor{jku_grey!10}${64.74^{\pm{5.0}}}$ & \cellcolor{jku_grey!10}\cellcolor{jku_yellow!20}${66.32^{\pm{8.1}}}$ &                           \cellcolor{jku_grey!10}${68.64^{\pm{6.5}}}$ \\
                                               MFBERT \citep{abdel2022mfbert} &                         X &                                                  ${71.63^{\pm{4.4}}}$ &                                                  ${71.61^{\pm{3.3}}}$ &                                                 ${77.99^{\pm{10.2}}}$ &                                                  ${71.12^{\pm{2.4}}}$ &                          ${61.14^{\pm{7.4}}}$ &                                                  ${63.95^{\pm{4.8}}}$ &                                                   ${63.72^{\pm{8.3}}}$ &                                                   ${68.74^{\pm{5.5}}}$ \\
           \cellcolor{jku_grey!10}MISU \citep{benjamin2022graph} & \cellcolor{jku_grey!10}  &                          \cellcolor{jku_grey!10}${70.52^{\pm{3.8}}}$ &                          \cellcolor{jku_grey!10}${66.71^{\pm{1.8}}}$ &                          \cellcolor{jku_grey!10}${78.00^{\pm{4.3}}}$ &                          \cellcolor{jku_grey!10}${-}$ &  \cellcolor{jku_grey!10}${59.73^{\pm{0.8}}}$ &                          \cellcolor{jku_grey!10}${76.30^{\pm{0.7}}}$ &                           \cellcolor{jku_grey!10}${62.79^{\pm{0.5}}}$ &                           \cellcolor{jku_grey!10}${69.01^{\pm{6.7}}}$ \\
                                               KV-PLM \cite{zeng2022kvplm} &                           &                                                  ${78.50^{\pm{2.7}}}$ &                                                  ${70.50^{\pm{0.5}}}$ &                                                  ${89.17^{\pm{2.7}}}$ &                                                  ${65.40^{\pm{1.7}}}$ &                          ${59.83^{\pm{0.6}}}$ &                                                  ${72.12^{\pm{1.0}}}$ &                                                   ${55.03^{\pm{1.6}}}$ &                                                  ${70.08^{\pm{10.6}}}$ \\
               \cellcolor{jku_grey!10}BARTSmiles \citep{chilingaryan2022bartsmiles}& \cellcolor{jku_grey!10}X &                          \cellcolor{jku_grey!10}${83.21^{\pm{3.3}}}$ &                          \cellcolor{jku_grey!10}${70.90^{\pm{3.7}}}$ &                          \cellcolor{jku_grey!10}${79.30^{\pm{7.5}}}$ &                          \cellcolor{jku_grey!10}${70.91^{\pm{2.6}}}$ &  \cellcolor{jku_grey!10}${57.66^{\pm{6.9}}}$ &                          \cellcolor{jku_grey!10}${65.16^{\pm{5.0}}}$ &                           \cellcolor{jku_grey!10}${64.98^{\pm{8.6}}}$ &                           \cellcolor{jku_grey!10}${70.30^{\pm{8.1}}}$ \\
                                 MegaMolBART  \citep{liu2022multi} &                           &                                                  ${82.46^{\pm{0.8}}}$ &                                                  ${68.89^{\pm{0.2}}}$ &                                                  ${78.12^{\pm{4.6}}}$ &                                                  ${71.04^{\pm{1.7}}}$ &                          ${59.52^{\pm{1.8}}}$ &                                                  ${73.89^{\pm{0.7}}}$ &                                                   ${63.32^{\pm{0.8}}}$ &                                                   ${71.03^{\pm{7.4}}}$ \\
                 \cellcolor{jku_grey!10}Mc+RDKc-FP & \cellcolor{jku_grey!10}X &                          \cellcolor{jku_grey!10}${81.16^{\pm{3.5}}}$ &                          \cellcolor{jku_grey!10}${69.67^{\pm{3.7}}}$ &                          \cellcolor{jku_grey!10}${84.81^{\pm{4.9}}}$ &                          \cellcolor{jku_grey!10}${72.25^{\pm{2.8}}}$ &  \cellcolor{jku_grey!10}${62.41^{\pm{7.7}}}$ &                          \cellcolor{jku_grey!10}${69.61^{\pm{4.8}}}$ &                           \cellcolor{jku_grey!10}${60.73^{\pm{8.9}}}$ &                           \cellcolor{jku_grey!10}${71.52^{\pm{8.2}}}$ \\
                                               Grover \citep{rong2020grover}r &                         X &                                                  ${78.64^{\pm{3.8}}}$ &                                                  ${67.94^{\pm{4.0}}}$ &                                                  ${89.48^{\pm{4.3}}}$ &                                                  ${77.52^{\pm{2.2}}}$ &                          ${59.63^{\pm{7.8}}}$ &                                                  ${68.16^{\pm{5.4}}}$ &                         \cellcolor{jku_yellow!20}${67.26^{\pm{8.2}}}$ &                         \cellcolor{jku_yellow!20}${72.66^{\pm{9.1}}}$ \\
                         \cellcolor{jku_grey!10}CDDD \citep{winter2019cddd}& \cellcolor{jku_grey!10}X &                          \cellcolor{jku_grey!10}${76.83^{\pm{4.2}}}$ &                          \cellcolor{jku_grey!10}${72.45^{\pm{3.6}}}$ &                          \cellcolor{jku_grey!10}${92.41^{\pm{4.6}}}$ &                          \cellcolor{jku_grey!10}${73.87^{\pm{2.6}}}$ &  \cellcolor{jku_grey!10}${59.72^{\pm{7.1}}}$ &                          \cellcolor{jku_grey!10}${68.55^{\pm{5.0}}}$ &                           \cellcolor{jku_grey!10}${64.85^{\pm{8.2}}}$ & \cellcolor{jku_grey!10}\cellcolor{jku_yellow!20}${72.67^{\pm{9.7}}}$ \\
                                  GraphMVP-C \citep{liu2022multi} &                           &                                                  ${81.20^{\pm{0.9}}}$ &                                                  ${72.40^{\pm{1.6}}}$ &                                                  ${77.50^{\pm{4.2}}}$ &                                                  ${77.00^{\pm{1.2}}}$ &                          ${63.90^{\pm{1.2}}}$ &                                                  ${74.40^{\pm{0.2}}}$ &                                                   ${63.10^{\pm{0.4}}}$ &                         \cellcolor{jku_yellow!20}${72.79^{\pm{6.4}}}$ \\
   \cellcolor{jku_grey!10}GIN-node-pretrain \citep{sun2022does} & \cellcolor{jku_grey!10}  &                          \cellcolor{jku_grey!10}${83.66^{\pm{0.8}}}$ &                          \cellcolor{jku_grey!10}${73.45^{\pm{0.3}}}$ &                          \cellcolor{jku_grey!10}${-}$ &                          \cellcolor{jku_grey!10}${-}$ &  \cellcolor{jku_grey!10}${65.08^{\pm{0.1}}}$ &                          \cellcolor{jku_grey!10}${75.30^{\pm{0.4}}}$ & \cellcolor{jku_grey!10}\cellcolor{jku_yellow!20}${66.50^{\pm{0.1}}}$ & \cellcolor{jku_grey!10}\cellcolor{jku_yellow!20}${72.80^{\pm{6.7}}}$ \\
                              ECFP6||RDKd-FP  &        X                   &                                                  ${80.36^{\pm{0.0}}}$ &                                                  ${71.98^{\pm{0.0}}}$ &                                                  ${87.59^{\pm{0.0}}}$ &                                                  ${73.45^{\pm{0.0}}}$ &                          ${63.02^{\pm{0.0}}}$ &                                                  ${71.41^{\pm{0.0}}}$ &                         \cellcolor{jku_yellow!20}${67.73^{\pm{0.0}}}$ &                         \cellcolor{jku_yellow!20}${73.65^{\pm{7.5}}}$ \\
       \cellcolor{jku_grey!10}MOCO \citep{zhu2022improving} & \cellcolor{jku_grey!10}  &                          \cellcolor{jku_grey!10}${82.60^{\pm{0.3}}}$ &                          \cellcolor{jku_grey!10}${71.60^{\pm{1.0}}}$ &                          \cellcolor{jku_grey!10}${81.60^{\pm{3.7}}}$ &                          \cellcolor{jku_grey!10}${78.30^{\pm{0.4}}}$ &  \cellcolor{jku_grey!10}${61.20^{\pm{0.6}}}$ &                          \cellcolor{jku_grey!10}${76.70^{\pm{0.4}}}$ &                           \cellcolor{jku_grey!10}${64.90^{\pm{0.8}}}$ & \cellcolor{jku_grey!10}\cellcolor{jku_yellow!20}${73.84^{\pm{7.7}}}$ \\
            GIN-sup.-cont.-pt.  \citep{sun2022does} &                           &                        \cellcolor{jku_yellow!20}${86.33^{\pm{0.2}}}$ &                                                  ${74.38^{\pm{0.9}}}$ &                                                  ${-}$ &                                                  ${-}$ &                          ${62.22^{\pm{0.5}}}$ &                                                  ${78.16^{\pm{0.2}}}$ &                         \cellcolor{jku_yellow!20}${68.71^{\pm{0.1}}}$ &                         \cellcolor{jku_yellow!20}${73.96^{\pm{8.2}}}$ \\
 \cellcolor{jku_grey!10}MoleculeSTM-s  \citep{liu2022multi} & \cellcolor{jku_grey!10}  &                          \cellcolor{jku_grey!10}${81.99^{\pm{0.4}}}$ &                          \cellcolor{jku_grey!10}${70.75^{\pm{1.9}}}$ &                          \cellcolor{jku_grey!10}${86.60^{\pm{2.3}}}$ &                          \cellcolor{jku_grey!10}${77.02^{\pm{0.4}}}$ &  \cellcolor{jku_grey!10}${63.70^{\pm{0.8}}}$ &                          \cellcolor{jku_grey!10}${75.71^{\pm{0.9}}}$ &                           \cellcolor{jku_grey!10}${65.17^{\pm{0.4}}}$ & \cellcolor{jku_grey!10}\cellcolor{jku_yellow!20}${74.42^{\pm{7.8}}}$ \\
                                           CLAMP &                         X &                                                  ${84.28^{\pm{3.4}}}$ &                                                  ${68.22^{\pm{3.8}}}$ &                                                  ${75.31^{\pm{7.8}}}$ &                                                  ${76.34^{\pm{2.4}}}$ &                          ${65.15^{\pm{7.5}}}$ &                                                  ${78.23^{\pm{4.2}}}$ &                          \cellcolor{jku_green!20}${74.00^{\pm{7.7}}}$ &                         \cellcolor{jku_yellow!20}${74.50^{\pm{5.9}}}$ \\
  \cellcolor{jku_grey!10}MoleculeSTM-g   \citep{liu2022multi} & \cellcolor{jku_grey!10}  &                          \cellcolor{jku_grey!10}${80.77^{\pm{1.3}}}$ &                          \cellcolor{jku_grey!10}${69.98^{\pm{0.5}}}$ &                          \cellcolor{jku_grey!10}${92.53^{\pm{1.1}}}$ &                          \cellcolor{jku_grey!10}${76.93^{\pm{1.8}}}$ &  \cellcolor{jku_grey!10}${60.96^{\pm{1.1}}}$ &                          \cellcolor{jku_grey!10}${76.91^{\pm{0.5}}}$ &                           \cellcolor{jku_grey!10}${65.05^{\pm{0.4}}}$ & \cellcolor{jku_grey!10}\cellcolor{jku_yellow!20}${74.73^{\pm{9.8}}}$ \\
                                    GEM \citep{fang2022geometry} &                           &                                                  ${85.60^{\pm{1.1}}}$ &                                                  ${72.40^{\pm{0.4}}}$ &                                                  ${90.10^{\pm{1.3}}}$ &                                                  ${-}$ &                          ${67.20^{\pm{0.4}}}$ &                                                  ${78.10^{\pm{0.1}}}$ &                         \cellcolor{jku_yellow!20}${69.20^{\pm{0.4}}}$ &                         \cellcolor{jku_yellow!20}${77.10^{\pm{8.4}}}$ \\
     \cellcolor{jku_grey!10}Uni-Mol \citep{zhou2022uni} & \cellcolor{jku_grey!10}  &                          \cellcolor{jku_grey!10}${85.70^{\pm{0.2}}}$ &                          \cellcolor{jku_grey!10}${72.90^{\pm{0.6}}}$ &                          \cellcolor{jku_grey!10}${91.90^{\pm{1.8}}}$ &                          \cellcolor{jku_grey!10}${80.80^{\pm{0.3}}}$ &  \cellcolor{jku_grey!10}${65.90^{\pm{1.3}}}$ & \cellcolor{jku_grey!10}\cellcolor{jku_green!20}${79.60^{\pm{0.5}}}$ & \cellcolor{jku_grey!10}\cellcolor{jku_yellow!20}${69.60^{\pm{0.1}}}$ & \cellcolor{jku_grey!10}\cellcolor{jku_yellow!20}${78.06^{\pm{8.5}}}$ \\
                                   MMM  \citep{he2022masked} &                           &                        \cellcolor{jku_yellow!20}${86.20^{\pm{2.3}}}$ &                                                  ${75.20^{\pm{0.7}}}$ &                                                  ${90.90^{\pm{2.5}}}$ &                                                  ${-}$ & \cellcolor{jku_green!20}${68.20^{\pm{0.4}}}$ &                        \cellcolor{jku_yellow!20}${79.30^{\pm{0.2}}}$ &                         \cellcolor{jku_yellow!20}${69.90^{\pm{0.2}}}$ &                         \cellcolor{jku_yellow!20}${78.28^{\pm{8.2}}}$ \\
\cellcolor{jku_grey!10}Unified2D3D  \citep{zhu2022unified2d3d} & \cellcolor{jku_grey!10}  & \cellcolor{jku_grey!10}\cellcolor{jku_green!20}${86.80^{\pm{0.6}}}$ & \cellcolor{jku_grey!10}\cellcolor{jku_green!20}${77.40^{\pm{0.6}}}$ & \cellcolor{jku_grey!10}\cellcolor{jku_green!20}${95.40^{\pm{1.1}}}$ & \cellcolor{jku_grey!10}\cellcolor{jku_green!20}${82.20^{\pm{1.0}}}$ &  \cellcolor{jku_grey!10}${67.40^{\pm{0.5}}}$ &                          \cellcolor{jku_grey!10}${75.90^{\pm{0.3}}}$ &                           \cellcolor{jku_grey!10}${-}$ &  \cellcolor{jku_grey!10}\cellcolor{jku_green!20}${80.85^{\pm{8.8}}}$ \\
\bottomrule
\end{tabular}
    }
\end{table*}

\newpage

\subsection{Details on the retrieval and library design study (5.3)}
In this task, molecules from a chemical database must be ranked according to a description of a specific bioassay. Molecules that have been measured as active in the given bioassay should be given a high ranking.
If the top-k ranked molecules contain the known active, then the prediction is considered accurate.
This task is a proxy for how relevant suggestions of a given method are.

We use the PubChem dataset with assay-based temporal split time\_a for molecule retrieval. 
We consider two different set-sizes: 1M an 10k molecules, selected in order of their PubChem Compound-ID (CID). 
The 10k molecule setting is used for computational feasibility and to include Galactica in the comparison. 
To obtain robust estimates of the performance metric, 
we report the average over assays with more than 100 active molecules. 
This results in 190 assays for the 10k molecules setting and 2,543 assays for the 1M molecule setting for testing.

\begin{figure*}[htb]
    \centering
    \includegraphics[width=0.9\columnwidth]{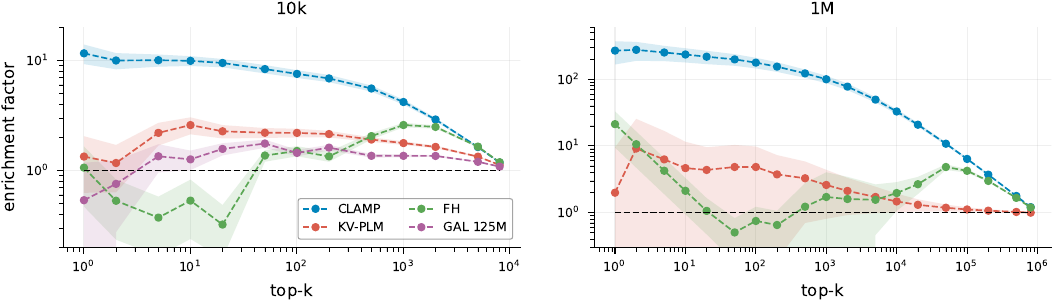}
    \caption{Molecule-retrieval enrichment factor plots. 
    The x-axis displays different $k$ for top-$k$ accuracies and
    on the y-axis the enrichment factor is displayed. Different 
    methods are represented by colored lines. Each dot represents a 
    mean over bioassays and the shaded areas 
    indicate the standard deviation over assays. \textbf{Left:}
    Retrieving from a database of 10k molecules. \textbf{Right:} Retrieval
    from a database of 1M molecules.
    \label{fig:compound_retr}}
\end{figure*}

The enrichment-factor (EF) is used as evaluation metric (see Sec.~\ref{app:metric} for details).
The results of this study in terms of enrichment factor for different top-$k$ accuracies
are shown in Fig.~\ref{fig:compound_retr}.

\newpage

\subsection{Extended experiments}
\label{appsec:representation}
\label{appsec:extended}

\subsubsection{Out-of-domain zero-shot on downstream datasets.}
In this experiment, we evaluate the generalization capabilities of a model in an out-of-domain setting by assessing its robustness to variations in assay description. We aim to test the model's ability to perform well on unseen assay descriptions that are distinct from those in the training dataset.

\textbf{Datasets.} 
The MoleculeNet datasets and Tox21-10k were utilized in this study in a zero-shot manner. The original datasets did not include assay descriptions, thus, we manually created descriptions by integrating information from multiple sources. However, additional efforts are required to refine and optimize these descriptions.
The zero-shot performance is evaluated on the default scaffold test-split, except for Tox21-10k where we use the default test split.

\textbf{Methods compaired.}
We pre-train a CLAMP and FH model on PubChem, removing compounds that are in the test set (as in experiments Sec.~\ref{sec:representation_learning} and further described in Sec.~\ref{appsec:data_overlap}), and train on a random-split.
We benchmark against FH and KV-PLM. Galactica was not included as a comparison model in this study, as the use of pre-trained models on the datasets being tested would introduce a potential source of dataset leakage and lead to an overestimation of performance.

\textbf{Results.}
Results are presented in Tab.~\ref{tab:downstream_zero_shot}. 
Overall, our model demonstrates prominent performance on the HIV dataset, with an AUROC score that surpasses that of linear probing performance. The performance of the FH model on the HIV dataset is also noteworthy, with an AUROC score of 68.21, which is outperformed by CLAMP with an AUROC of 80.67. The zero-shot performance, in this case, is better than the linear-probing performance of 76.34 (see Tab.~\ref{tab:linear_probing_appendix_auroc_default_split}). However, for datasets such as ClinTox and BBBP, the performance of all three models is relatively poor. On average, over datasets, our model outperforms both FH and KV-PLM.

\begin{table*}[h]
    \centering
    \caption{Out of domain zero-shot on downstream datasets. Results are reported in \%. Avg represents the mean over datasets.   \label{tab:downstream_zero_shot}}
    \small
    {\fontsize{10pt}{10pt}
    \begin{tabular}{llllllllll||r}
\toprule
metric &  method&  BACE &  BBBP &   ClinTox &   HIV & SIDER & Tox21 & ToxCast & Tox21-10k &  avg \\
\midrule
   $\Delta$AP&  \cellcolor{jku_grey!10}KVPLM &  \cellcolor{jku_grey!10}${-4.15}$ & \cellcolor{jku_grey!10}\cellcolor{jku_green!20}${08.63}$ & \cellcolor{jku_grey!10}\cellcolor{jku_green!20}${04.54}$ &                          \cellcolor{jku_grey!10}${00.00}$ &                          \cellcolor{jku_grey!10}${01.67}$ &                          \cellcolor{jku_grey!10}${03.20}$ &  \cellcolor{jku_grey!10}${02.61}$ &                          \cellcolor{jku_grey!10}${04.20}$ &                          \cellcolor{jku_grey!10}${02.59}$ \\
                     &        FH & \cellcolor{jku_green!20}${10.78}$ &                                                  ${06.15}$ &                                                  ${01.38}$ &                                                  ${02.89}$ &                                                  ${00.32}$ &                                                  ${01.70}$ & \cellcolor{jku_green!20}${03.66}$ &                                                  ${07.67}$ &                                                  ${04.32}$ \\
&\cellcolor{jku_grey!10}CLAMP &  \cellcolor{jku_grey!10}${07.85}$ &                          \cellcolor{jku_grey!10}${-0.12}$ &                          \cellcolor{jku_grey!10}${03.93}$ & \cellcolor{jku_grey!10}\cellcolor{jku_green!20}${16.24}$ & \cellcolor{jku_grey!10}\cellcolor{jku_green!20}${03.99}$ & \cellcolor{jku_grey!10}\cellcolor{jku_green!20}${05.40}$ &  \cellcolor{jku_grey!10}${02.80}$ & \cellcolor{jku_grey!10}\cellcolor{jku_green!20}${12.02}$ & \cellcolor{jku_grey!10}\cellcolor{jku_green!20}${06.51}$ \\
\midrule
 AUROC &    \cellcolor{jku_grey!10}KVPLM &                          \cellcolor{jku_grey!10}${45.29}$ & \cellcolor{jku_grey!10}\cellcolor{jku_green!20}${56.38}$ & \cellcolor{jku_grey!10}\cellcolor{jku_green!20}${49.94}$ &                          \cellcolor{jku_grey!10}${53.37}$ &                          \cellcolor{jku_grey!10}${47.71}$ &                          \cellcolor{jku_grey!10}${55.73}$ &  \cellcolor{jku_grey!10}${59.07}$ &  \cellcolor{jku_grey!10}${60.62}$ &                          \cellcolor{jku_grey!10}${53.51}$ \\
                &             FH &                                                  ${63.77}$ &                                                  ${56.34}$ &                                                  ${48.03}$ &                                                  ${68.21}$ &                                                  ${49.14}$ &                                                  ${57.15}$ & \cellcolor{jku_green!20}${61.99}$ & \cellcolor{jku_green!20}${71.39}$ &                                                  ${59.50}$ \\
& \cellcolor{jku_grey!10}CLAMP & \cellcolor{jku_grey!10}\cellcolor{jku_green!20}${64.76}$ &                          \cellcolor{jku_grey!10}${47.88}$ &                          \cellcolor{jku_grey!10}${49.03}$ & \cellcolor{jku_grey!10}\cellcolor{jku_green!20}${80.67}$ & \cellcolor{jku_grey!10}\cellcolor{jku_green!20}${54.23}$ & \cellcolor{jku_grey!10}\cellcolor{jku_green!20}${60.58}$ &  \cellcolor{jku_grey!10}${53.83}$ &  \cellcolor{jku_grey!10}${69.70}$ & \cellcolor{jku_grey!10}\cellcolor{jku_green!20}${60.09}$ \\

\bottomrule
\end{tabular}
    }
\end{table*}

\subsubsection{Comparison to bioactivity descriptors}
\label{appsec:hts_benchmarking}
In this experiment, we benchmark learned representation against bioactivity descriptors.
A molecule can be described structurally via e.g. a fingerprint but also assay-measurements can be used to represent a molecule.
The HTS Fingerprint (HTSFP) \citet{petrone2012rethinking} is a bioactivity descriptor using historical activity data to represent a molecules. The HTSFP has the benefit of not requiring any structural information. Additionally, active substances with distinctive mechanisms of action can be found using HTSFPs. However, predictions can only be made for substances that have already been examined in HTS assays, and substances with sparse HTSFPs are frequently removed from the dataset because they may introduce noise into the data and reduce predictive performance.

\textbf{Datasets.}
We reproduce the setting of \citet{laufkotter2019combining}: 582 HTS assays from PubChem are selected, 24 of which are used for testing. The split is termed hts-split. Note that this corresponds to a random assay-split.

\textbf{Methods compared.}
\citet{laufkotter2019combining} suggests the 
bioactivity-structure hybrid (BaSH) fingerprint.
BaSH concatenates HTSFP and a structural FP. They empirically show that the BaSH fingerprint improves performance when compared to the use of the HTSFP fingerprint alone. 
Compared to the implementation of \citet{laufkotter2019combining}, we remove all test and validation assays from the FPs and do not remove only one assay dimension at a time, when testing (leaf-one-out at the task level). Instead of using a Random Forest Model, we train a Logistic Regression model on the FPs, which corresponds to linear probing.

We compare against our method CLAMP: We pre-train a CLAMP model on training assays from the hts-split. For BaSH and HTSFP, missing values get imputed as zeros, which is not the case for our pre-training our method.
We select the best model based on zero-shot validation $\Delta$AP.
Finally, we test the molecular representations on the 24 test-set assays of the hts-split. We perform linear probing for each method for a 5-fold random cross-validation for each test-assays.

\textbf{Results.}
We report the mean 5-fold cross-validation-AUROC over 24 test assays. 
Tab.~\ref{tab:linear_probing_hts} demonstrated that CLAMP outperforms the proposed BaSH fingerprint. The FH baseline is superior to using LP on fingerprint-based molecular descriptors.
With only 1.23 \% of training data in the pretraining set, CLAMP 2048-Shot achieves better results than HTSFP.
CLAMP also addresses its drawbacks. Particularly, the BaSH and HTSFP fingerprint are restricted to molecules with activity data, whereas CLAMP can perform inference on molecules without activity data.

\begin{table*}[ht]
    \centering
    \caption[caption]{Linear probing \%AUROC results for PubChem hts-split. Average over 24 assays. LP stands for linear probing. \% of train corresponds to the amount of training data that has been used.   \label{tab:linear_probing_hts}}
    \begin{tabular}{lclr}
\toprule
 method &  LP & AUROC &  \% of train \\
\midrule
  \cellcolor{jku_grey!10}CLAMP &  \cellcolor{jku_grey!10}x &\cellcolor{jku_grey!10}\cellcolor{jku_green!20}${81.00^{\pm{3.7}}}$ & \cellcolor{jku_grey!10}$100.00$ \\
BaSH \citep{laufkotter2019combining} &  x &  \cellcolor{jku_yellow!20}${79.19^{\pm{4.6}}}$ & $100.00$ \\
 \cellcolor{jku_grey!10}CLAMP 2048-Shot & \cellcolor{jku_grey!10}x &\cellcolor{jku_grey!10}\cellcolor{jku_yellow!20}${78.08}^{\pm{4.1}}$ &  \cellcolor{jku_grey!10} $  1.23 $\\
  HTSFP \citep{laufkotter2019combining} & x & \cellcolor{jku_yellow!20}${77.37^{\pm{4.0}}}$ & $100.00$ \\
\cellcolor{jku_grey!10}CLAMP Zero-Shot \cellcolor{jku_grey!10} &  \cellcolor{jku_grey!10}  & \cellcolor{jku_grey!10}${73.84}^{\pm{4.7}}$ &\cellcolor{jku_grey!10} $0.00$ \\
FH Zero-Shot  &&${73.48}^{\pm{4.8}}$ &$0.00$ \\
\cellcolor{jku_grey!10}ECFP+TT+MACCS &  \cellcolor{jku_grey!10}x & \cellcolor{jku_grey!10}${73.04^{\pm{5.0}}}$ & \cellcolor{jku_grey!10}$100.00$ \\
ECFP4 &x &${71.67^{\pm{4.6}}}$ & $100.00$ \\
\bottomrule
\end{tabular}
\end{table*}

\subsubsection{Few-Shot on FS-Mol}
Few-shot learning is a setting where only a number of training samples are given, the so-called support-set. The support-set can be used to train the model. The model is evaluated on the remaining samples. A support-set-size of 16 is a common choice, also referred to as 16-shot. The zero-shot setting refers to having no support-set molecules, and is particularly challenging.
Specific methods have been suggested for few-shot learning for molecules \citep{stanley2021fs, schimunek2023contextenriched, altae2017low, guo2021few, wang2021property, chen2022meta}.
Schemes include data-augmentation-based approaches, nearest-neighbor approaches and optimization-based or fine-tuning-based methods using meta-optimizers.
We don't employ any of those techniques, and add the support-set to the pre-training set. We do this for CLAMP and FH.

\textbf{Datasets.}
 We use the same setup as in FS-Mol \citep{stanley2021fs}:
 the data is split into a training-, validation- and test-set based on assays. We use the same split and draw a random sample of \emph{support-set-size} number of molecules from the test set.

\begin{figure}
     \centering
     \includegraphics{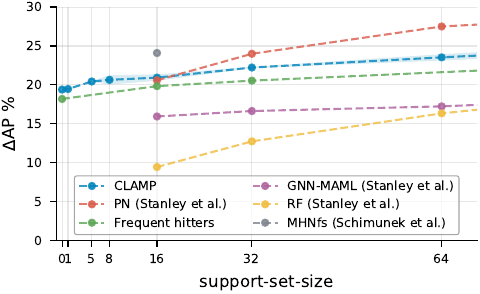}
     \caption{Few- and Zero-shot results for FS-Mol \citep{stanley2021fs}}
     \label{fig:fewshot_fsmol}
 \end{figure}

\textbf{Results.}
Our results can be seen in Figure~\ref{fig:fewshot_fsmol}. The FH method also introduced in \citep{schimunek2023contextenriched}, provides a strong baseline, outperforming other methods like Random-Forest (RF) of few-shot specific methods like GNN-MAML\citep{stanley2021fs} even without a support-set. CLAMP shows impressive zero-and 16-shot performance, but for larger support-set-sizes is outperformed by PN\citep{stanley2021fs} and MHNfs \citep{schimunek2023contextenriched} at 16-shot.

Specific few-shot learning schemes could also be employed on top of the CLAMP representation, which we leaf to future work.

\newpage

\subsubsection{Molecule- and assay-encoder ablation}
\label{appsec:encoder_study}
In this study, we investigated which molecule and text encoder provide the best predictive performance for the CLAMP approach. 
The setting is the same as in the zero-shot transfer experiments.
The experiments were in no way exhaustive. We mainly explored parameters for FS-Mol due to the smaller size.
Tab.~\ref{tab:assay_modes} shows test-set performance for different assay encoders. CLIP concatenated with LSA encoding works well.
Tab.~\ref{tab:compound_modes} shows test-set performance for different molecule encoders.

\begin{table*}[htb]
    \centering
    \caption{Results for different assay encodings.}
    \begin{tabular}{lllll}
\toprule
                                 assay encoding &                                                                  FSMOL & \multicolumn{3}{l}{PubChem} \\
                                            &                                                                default &                                                                    hts &                                        time\_a &                                                               time\_a\_c \\
\midrule
                                  CLIP||LSA &                          \cellcolor{jku_green!20}${19.60^{\pm{0.4}}}$ &                                                   ${07.92 ^{\pm{0.8}}}$ & \cellcolor{jku_green!20}${14.83^{\pm{0.3}}}$ &                         \cellcolor{jku_yellow!20}${10.82^{\pm{1.6}}}$ \\
                \cellcolor{jku_grey!10}LSA & \cellcolor{jku_grey!10}\cellcolor{jku_yellow!20}${19.56^{\pm{0.3}}}$ & \cellcolor{jku_grey!10}\cellcolor{jku_yellow!20}${07.99^{\pm{1.3}}}$ &  \cellcolor{jku_grey!10}${13.37^{\pm{1.1}}}$ & \cellcolor{jku_grey!10}\cellcolor{jku_yellow!20}${11.28^{\pm{0.8}}}$ \\
                                       CLIP &                         \cellcolor{jku_yellow!20}${19.42^{\pm{0.2}}}$ &                                                   &                          ${12.72^{\pm{1.4}}}$ &                     \\
     \cellcolor{jku_grey!10}CLIP||LSA||sT5 & \cellcolor{jku_grey!10}\cellcolor{jku_yellow!20}${19.29^{\pm{0.5}}}$ &  \cellcolor{jku_grey!10}\cellcolor{jku_green!20}${08.24^{\pm{0.3}}}$ &  \cellcolor{jku_grey!10}&  \cellcolor{jku_grey!10}\cellcolor{jku_green!20}${11.95^{\pm{1.4}}}$ \\
                                   category &                                                   ${19.15^{\pm{0.2}}}$ &                                                   &                          &                                                   \\
\cellcolor{jku_grey!10}category||millipore &                           \cellcolor{jku_grey!10}${19.12^{\pm{0.3}}}$ &                           \cellcolor{jku_grey!10}&  \cellcolor{jku_grey!10}&                           \cellcolor{jku_grey!10}\\
                                  millipore &                                                   ${19.07^{\pm{0.5}}}$ &                                                   &                          &                                                   \\
            \cellcolor{jku_grey!10}BioBERT &                           \cellcolor{jku_grey!10}${19.01^{\pm{0.1}}}$ &                           \cellcolor{jku_grey!10}&  \cellcolor{jku_grey!10}&                           \cellcolor{jku_grey!10}\\
                                          constant$^\dagger$ &                                                   ${18.50^{\pm{0.2}}}$ &                                                   ${03.10^{\pm{0.1}}}$ &                          ${10.23^{\pm{0.5}}}$ &                                                   ${10.35^{\pm{0.9}}}$ \\
         \cellcolor{jku_grey!10}GAL 6.7B &\cellcolor{jku_grey!10}${18.50^{\pm{0.2}}}$ & \cellcolor{jku_grey!10} & \cellcolor{jku_grey!10} & \cellcolor{jku_grey!10}  \\
                       
\bottomrule
\end{tabular}
    \label{tab:assay_modes}
    
    \footnotesize{$^\dagger$ a constant as assay-encoding corresponds to FH model} 
\end{table*}

\begin{table*}[htb]
    \centering
    \caption{Results for different molecule encodings. 
    }
    \begin{tabular}{lllll}
\toprule
                                                               molecule encoding &                                                                 FSMOL & \multicolumn{3}{l}{PubChem} \\
                                                                             &                                                               default &                                           hts &                                        time\_a &                                       time\_a\_c \\
\midrule
\cellcolor{jku_grey!10}multiple$^\dagger$ & \cellcolor{jku_grey!10}\cellcolor{jku_green!20}${19.51^{\pm{0.2}}}$ &  \cellcolor{jku_grey!10} &  \cellcolor{jku_grey!10} &   \cellcolor{jku_grey!10}\\
                                                                Morganc+RDKc &                        \cellcolor{jku_yellow!20}${19.35^{\pm{0.2}}}$ & \cellcolor{jku_green!20}${08.35^{\pm{0.3}}}$ & \cellcolor{jku_green!20}${14.77^{\pm{0.3}}}$ &  \cellcolor{jku_green!20}${11.80^{\pm{0.7}}}$ \\
                                                \cellcolor{jku_grey!10}CDDD &                          \cellcolor{jku_grey!10}${18.17^{\pm{1.1}}}$ &  \cellcolor{jku_grey!10}&  \cellcolor{jku_grey!10}&   \cellcolor{jku_grey!10}\\
                                                                      MFBERT &                                                  ${18.43^{\pm{1.2}}}$ &                          &                          &                           \\
                                              \cellcolor{jku_grey!10}Grover &                          \cellcolor{jku_grey!10}${17.86^{\pm{1.1}}}$ &  \cellcolor{jku_grey!10}&  \cellcolor{jku_grey!10}&   \cellcolor{jku_grey!10}\\
                                                                      sprsFP &                                                  ${13.98^{\pm{0.6}}}$ &                          ${04.14^{\pm{3.6}}}$ &                          ${11.28^{\pm{1.3}}}$ & \cellcolor{jku_yellow!20}${11.18^{\pm{0.8}}}$ \\
                                         \cellcolor{jku_grey!10}MolT5-large &                          \cellcolor{jku_grey!10}${16.20^{\pm{2.1}}}$ &  \cellcolor{jku_grey!10}&  \cellcolor{jku_grey!10}&   \cellcolor{jku_grey!10}\\
                                                                      KV-PLM &                                                  ${11.11^{\pm{3.2}}}$ &                          &                          &                           \\
                                       \cellcolor{jku_grey!10}GAL 6.7B &                          \cellcolor{jku_grey!10}${09.25^{\pm{0.9}}}$ &  \cellcolor{jku_grey!10}&  \cellcolor{jku_grey!10}&   \cellcolor{jku_grey!10}\\
\bottomrule
\end{tabular}
    \label{tab:compound_modes}
    
    \footnotesize{$^\dagger$ concatenation of CDDD, Graphormer, Grover, MFBERT, Mc+RDKc.} 
\end{table*}

\subsection{Computational Resources}
\label{appsec:computationl}
The code was run on different servers with diverse Nvidia GPUs, the best of which was an NVIDIA A100-SXM4-80GB, using PyTorch 1.13.0 \citep{paszke2019pytorch}. The total compute run-time was around 170 days and $\sim$800 runs (without linear probing). The minimum GPU-memory experiments were run with 12GB.
Further statement on computational feasibility: Based on our current setup, we would consider an experiment that is 10 times larger than our current setup as computationally infeasible.

\subsection{Broader Impact}
\label{appsec:broader}
We envision that our approach
can also be used to enhance models in other application domains 
with an interface with human language. Furthermore, 
our results reveal drawbacks of multi-modal language models.
Our results also hint at the fact that specialized 
modules for each data modality
whose representations are later combined could be necessary
for good predictive performance. 

\subsection{Social Impact}

Our method has the potential to significantly impact the field of drug discovery and development. By associating molecules with bioassay descriptions, the model can aid in the identification of new drug candidates and provide insight into the potential side effects of existing molecules. This can lead to a more efficient drug development process, potentially reducing time and cost associated with bringing new treatments to market.

However, it is important to note that there are also ethical and societal implications of the CLAMP method. 
Potential biases in the data influence the model's predictions and highlight the need for responsible data management practices.
Additionally, it is important to consider the impact of the technology and take steps to ensure that the benefits of the model are shared equitably. 
As previously demonstrated by \citet{urbina2022dual}, models like CLAMP can not only be used to identify and avoid toxicity but also to suggest highly toxic chemicals.

In conclusion, the CLAMP method has the potential to significantly impact the field of drug discovery and development, but it is crucial to consider the ethical and societal implications of the technology in order to ensure that its benefits are shared equitably. Note that our method would not be used by the general public but by researchers, such as chemists or molecular biologists.

\clearpage
\subsection{List of Acronyms}

\begin{acronym}
 \acro{AUPR}{Area Under Precision Recall curve}
 \acro{AUROC}{Area Under Receiver-Operating-Characterstic curve}
 \acro{AP}{Average Precision}
 \acro{CV}{Computer Vision}
 \acro{CLIP}{Contrastive Language–Image Pre-training}
 \acro{CLAMP}{Contrastive Language–Assay-Molecule Pre-training}
 \acro{ChEMBL}{A biochemical database}
 \acro{ChEBI}{Chemical Entities of Biological Interest}
 \acro{ConVIRT}{Contrastive VIsual Representation Learning from Text}
 \acro{DL}{Deep Learning}
 \acro{DNN}{Deep Neural Network}
 \acro{ECFP}{Extended Connectivity Fingerprint}
 \acro{EF}{Enrichment Factor}
 \acro{FP}{FingerPrint}
  \acro{FH}{Frequent Hitters}
 \acro{GNN}{Graph Neural Network}
 \acro{HTS}{High-Throughput Screening}
 \acro{IUPAC}{International Union of Pure and Applied Chemistry}
 \acro{InChI}{International Chemical Identifier}
 \acro{KV-PLM}{A scientific language model}
 \acro{LSA}{Latent Semantic Analysis}
 \acro{ML}{Machine Learning}
 \acro{NCE}{Noise Contrastive Estimation}
 \acro{NLP}{Natural Language Processing}
 \acro{NN}{Neural Network}
 \acro{PubChem}{A biochemical database}
 \acro{SLM}{Scientific Language Model}
 \acro{SMILES}{Simplified Molecular-Input Line-Entry System}
\end{acronym}

\end{document}